\documentclass[12pt]{amsart}
\usepackage{amssymb}
\usepackage{amsmath}
\usepackage{amsfonts}
\usepackage{graphicx}
\usepackage{color}
\usepackage[onehalfspacing]{setspace}
\usepackage{hyperref}
\usepackage{fix-cm}
\usepackage{epigraph}
\usepackage{natbib}
\usepackage{placeins}
\usepackage{afterpage}

\usepackage[utf8]{inputenc}
\usepackage{charter}
\makeatletter
\def\section{\@startsection{section}{1}
	\z@{0.8\linespacing\@plus\linespacing}{.8\linespacing}{\Large\centering}}

\def\subsection{\@startsection{subsection}{2}
	\z@{.3\linespacing\@plus.3\linespacing}{.3\linespacing}{\large}}

\def\subsubsection{\@startsection{subsubsection}{3}
	\z@{.3\linespacing\@plus.3\linespacing}{-.5em}{\normalfont\bfseries\centering}}
\makeatother

\numberwithin{equation}{section}

\theoremstyle{definition}

\theoremstyle{definition}

\theoremstyle{definition}

\title{}
\begin{document}
	\vspace*{3ex minus 1ex}
	\begin{center}
		\Large \textsc{Measuring the Graph Concordance of
			Locally Dependent Observations}
		\medskip
	\end{center}
	
	\date{%
		\today%
	}
	
	\vspace*{3ex minus 1ex}
	\begin{center}
		Kyungchul Song\\
		\textit{Vancouver School of Economics, University of British Columbia}\\
		\medskip
		
	\end{center}
	
	\thanks{I thank Editor and three referees for valuable comments and suggestions. I am grateful to Colin Cameron, Nathan Canen, Xiaoqi He, Hiro Kasahara, Brian Krauth, Vadim Marmer, Shu Shen, Aaron Smith, and Jungmo Yoon for valuable conversations and comments. I appreciate comments from the participants of Seattle-Vancouver Econometrics Conference, Duke University, University of California at Davis, University of California at San Diego and University of Southern California. I thank Ning Ding for her excellent research assistance. All errors are mine. I acknowledge that this research was supported by Social Sciences and Humanities Research Council of Canada. Corresponding address: Kyungchul Song, Vancouver School of Economics, University of British Columbia, Vancouver, BC, Canada. Email address: kysong@mail.ubc.ca.}
	
	\begin{abstract}
		This paper introduces a simple measure of a concordance
		pattern among observed outcomes along a network, i.e., the
		pattern in which adjacent outcomes tend to be more strongly correlated than
		non-adjacent outcomes. The graph concordance measure can be generally used to quantify the
		empirical relevance of a network in explaining cross-sectional dependence of the outcomes, and as shown in the paper, can also be used to quantify the extent of homophily under certain conditions. When one observes a single large network, it is nontrivial to make inference about the concordance pattern. Assuming dependency graph,  this paper develops a permutation-based confidence interval
		for the graph concordance measure. The confidence interval is valid in
		finite samples when the outcomes are exchangeable, and under the dependency graph assumption together with other regularity conditions, is shown to exhibit asymptotic validity. Monte Carlo simulation results show that the validity of the permutation method is more robust to various graph
		configurations than the asymptotic method.
		\medskip
		
		\medskip
		
	\end{abstract}
	
\maketitle

\section{Introduction}

The role of a network has long received attention in the literature of
social science. When one studies network data and outcomes on the
network (e.g., a friendship nework and test scores of the students), a
primitive empirical task would be to determine whether a network has any
relevance in explaining observed outcomes, i.e., whether it has any
explanatory power, and if so, what aspect of the network is relevant and how
to quantify it taking into account sampling errors properly.

For example, one may ask whether students perform more similarly when they
are friends than when they are not. Or one may ask whether the racial
indicators are more strongly correlated among friends than among
non-friends, that is, whether there is race-based homophily among students.
The observation units do not need to be people or firms; they can be
geographic districts linked along a transportation network or software
products connected by common developers.

This paper introduces a simple parameter which measures the strength of
correlation among linked pairs of outcomes relative to that among unlinked
pairs. This parameter is called the \textit{graph concordance (GC)} in this
paper. Roughly speaking, the GC of a random vector laid on a graph is
defined to be the difference between the average correlation of linked pairs
of outcomes and that of unlinked pairs. A nonzero GC suggests that the graph
induces disparity in correlation between linked pairs of outcomes and
unlinked pairs of outcomes.

The GC can be used as an empirical measure of homophily among individuals in a
large friendship network. When individuals are more likely to befriend people of
the same race, the GC of people's racial backgrounds along the friendship
network will be positive. Indeed, this paper shows that under certain conditions, the GC coincides with the inbreeding homophily of \cite{Currarini/Jackson/Pin:09:Eca} (based on the measure of \cite{Coleman:58:HO}). The GC can be used for other purposes as well.
When pairs of regions with stronger trade links tend to produce more
economic outputs, this suggests a conspicuous role of a trade network in
shaping regional variations in outputs, which will be reflected in the
positive GC of the regional outputs.

Despite the simplicity of GC, developing a formal inference procedure on the
GC\ is a nontrivial task, especially when one observes only a single large complex network. This is because GC is a network-specific parameter that involves a large complex network. This makes contrast with the environment for testing for stochastic dominance or testing for independence where one makes many repeated observations of a random vector of a small dimension.

To deal with this situation, this paper provides a benchmark inference method under the cross-sectional dependence assumption of dependency graph. The dependency graph assumption requires that any two sets of outcomes that have no link between the two sets are independent. While the dependency graph assumption can be restrictive in certain applications, it is flexible in other dimensions. First, the correlations between linked pairs can be heterogeneous across pairs. Second, the local dependence is different from cluster dependence, because the neighborhoods are not disconnected from each other. Therefore, the cross-sectional dependence structure can be very complex, when the network is large and complex.  Third,
one can easily extend this paper's approach to $m$-dependence where any two
sets of outcomes that are $m$ links away are independent.\footnote{The $m$-dependence in times series can be viewed as a special case of an $m$-dependency graph where the graph is given as a line graph along a time ordering.} While it is
preferred to adopt a parsimonious modeling, some network configurations may
allow this extension without obscuring the inference result, such as when there are many components in the large network (e.g. social network studies on households in many villages or students in many schools.)

The main contribution of this paper is to develop a method of permutation-based inference for the GC under the dependency graph assumption. The confidence intervals generated from the method exhibit two-tier validity; when the outcomes are exchangeable, the
inference procedure is valid in finite samples, and when the outcomes fail
to be exchangeable, the inference procedure is shown to be asymptotically
valid under a set of regularity conditions.

The finite sample validity under exchangeability is not surprising. The
main theoretical contribution of this paper is to establish conditions for the networks so that the permutation inference is asymptotically valid even when the outcomes fail to be
exchangeable which is the case with a non-zero GC. This result is crucial
for our purpose because the confidence interval should be valid regardless
of whether the GC\ is zero or not.

Establishing conditions for the network is crucial here. Given the dependency graph assumption, there is no hope of obtaining a meaningful asymptotic distribution for a test statistic from observing a single large network, if the network is too dense. For example, if most nodes are linked to most other nodes, the dependency graph does not produce enough independence. This paper shows that if the maximum number of the neighbors in the network (i.e., the
maximum degree of the network) increases not faster than at a certain polynomial rate, as the number of the nodes in the network increases, the asymptotic validity of permutation inference follows.

The main proof idea of this paper comes from a simple observation as follows.
Since each permutation relabels the observation indices, the cross-sectional
dependence structure and heterogeneity of marginal distributions carry over
under the relabeling. Hence asymptotic derivation can be performed
conditional on a fixed sequence of permutations\ using the central limit theorem for random variables with dependency graphs (e.g. Theorem 2.4 of \cite{Penrose:03:RandomGeometricGraphs}.) The main difficulty in applying it to an estimation problem comes
from lack of a general method to center the permutation test statistic. However,
the permutation test statistic in this paper is approximately centered \textit{regardless of} whether the GC is zero or not, because most permutations reduce adjacent
vertices to non-adjacent ones. This feature enables us to apply the permutation 
method to the estimation problem in this paper.

One might wonder whether we can simply use asymptotic critical values
instead of permutation-based ones, ignoring the finite sample validity
altogether. First, it should be noted that despite the complexity of
theoretical arguments showing its asymptotic validity, the actual
implementation of the permutation inference is pretty simple, as we simply
randomly relabel the cross-sectional units and compute the GC estimator and
its variance estimator the same way as the original test statistic.  More importantly, this paper's Monte Carlo simulation study demonstrates that asymptotic critical values are not as stable as the permutation-based critical values, especially as the network becomes denser. Therefore, this paper's adoption of the permutation approach is not merely a theoretical interest.

To demonstrate the usefulness of this paper's proposal, we apply it to the
social networks data on Indian villages used in Banerjee et al. (2013). The
empirical application is divided into two parts. First, we estimate the GC
of the indicator of whether the household belongs to a minority caste or
tribe along the social network, to see whether there is homophily along the
castes. The main point of this exercise is to quantify the homophily among
the observed households and to provide its confidence interval. It turns out
that the estimated GC\ is around 0.57-0.59 with 95\% confidence intervals
roughly contained in [0.50,0.64]. Thus the GC\ is very high, indicating
strong homophily among a minority caste or tribe.

Second, we search for evidence that the social network plays a role in
explaining microfinancing decisions. The estimated GC\ of the household
decisions on microfinancing from the same sample as before is around
0.11-0.14, and significantly different from zero at
5\%. Therefore, while the GC\ of the microfinancing decisions is much weaker
than the GC\ of the minority castes, it is significantly positive,
indicating the relevance of the social network in explaining the
micro-financing decisions.\medskip

\noindent \textbf{Literature Review:} Defining and estimating a measure of
dependence between variables has drawn interest early on with the beginning
of modern statistics. Among the measures are Pearson's correlation
coefficient, Kendall's $\tau $, and Spearman's $\rho $. (See e.g. Chapter 8 of \cite{Cameron/Trivedi:13:RegressionAnalysis} for various dependence measures.) These measures are
typically between two random variables (or their i.i.d. draws), in contrast
to the GC\ in this paper that measures cross-sectional correlation along a
single large network. 

Closely related to this paper's GC is Moran's spatial autocorrelation test statistic (called $I$ test statistic) that is popularly used in  the literature of spatial modeling (\cite{Moran:50:Biometrika}, \cite{Cliff/Ord:72:GA}.) \cite{Kelejian/Prucha:2001:JOE} established asymptotic theory for Moran's $I$ test statistic in a general set-up. \cite{Pesaran:04:WP} developed a general test for cross-sectional dependence in linear panel models with a short time series dimension. See \cite{Hsiao/Pesaran/Pick:12:OBES} for an extension to limited dependent models. \cite{Robinson:08:JOE} proposed a correlation test that can be applied for testing cross-sectional dependence in a spatial model. The main difference that this paper's contribution makes is two-fold. First, the GC in this paper is proposed as a population quantity to be estimated, not just as a test statistic. As it is an estimation problem, one cannot impose the null hypothesis of, say, no spatial autocorrelation, in developing asymptotic validity of inference. Second, the main focus of the paper is on permutation-based inference instead of asymptotic inference. 

Permutation tests have been among the earliest methods of nonparametric testing in statistics, and have been applied in wide areas of statistics, biometrics and econometrics. They have been mostly used for testing equality of the means or the distributions from different subpopulations, and testing independence between two random variables. \cite{Daniels:44:Biometrika} focused on the permutation-based test on sample correlation
measures between two sets of observations conditional on the samples.
\cite{Friedman/Rafsky:79:AS} considered a two sample problem based on minimum
spanning trees. \cite{Romano:89:AS} investigated bootstrap and permutation tests
and considered, among several others, an example of testing independence.
\cite{Delgado:96:JTSA} developed a permutation test of serial dependence. A textbook treatment of the method and
references are found in \cite{Lehmann/Romano:05:TSH}.

Asymptotic robustness of permutation-based tests when exchangeability fails
has already drawn attention in the literature. \cite{Neuhaus:93:AS} considered a
two-sample problem with random censoring, and found that one can construct a
permutation test that controls the directional error through an asymptotic
pivotal test. \cite{Janssen:97:SPL} extended the approach to other two-sample
testing problems. This approach is substantially generalized by a recent
work by \cite{Chung/Romano:13:AS} who proposed a permutation test in
two-sample problems where the parameter of interest takes a general form. More recently, \cite{Canay/Romano/Shaikh:14:WP} investigated asymptotic validity of randomized tests under approximate symmetry conditions, where the symmetry conditions are imposed
not on the finite sample distribution but on the asymptotic distribution of
the statistic.
\medskip

\noindent \textbf{Organization of the Paper: }The next section introduces
GC, and discusses examples. Section 3 presents the paper's main proposal:
the permutation inference on the GC. The section establishes the asymptotic
validity of the proposed inference method. Section 4 presents and discusses
Monte Carlo simulation studies, and Section 5 illustrates the usefulness of
the proposal through an empirical application on the Indian village data.
Section 6 concludes. The appendix provides heuristics behind the theoretical
results. The full proof of the main theorem and auxiliary results are found
in the supplemental note to this paper.

\section{Graph Concordance of Cross-Sectional Observations}

\subsection{Graph Concordance}

Suppose that we observe random vector $Y=(Y_i)_{i\in N_n},\ Y_i\in 
\mathbf{R}$, with $N_n\equiv \{1,...,n\}$, and a graph (or a network) $%
G_n=(N_n,E_n)$, where $N_n$ indicates the set of \textit{vertices}
(or nodes) and $E_n$ the set of pairs $(i,j)$, $i,j \in N_n$, each pair $(i,j)$
representing an \textit{edge} (or a link) between $i$ and $j$. From here on,
we write $ij$ interchangeably with $(i,j)$. We assume that the graph is 
\textit{undirected}, so that $ij\in E_n$ if and only if $ji\in E_n$. When $ij \in N_n$, we say that $i$ and $j$ are \textit{adjacent}. We
do not allow a loop, i.e., $ii\notin E_n$ for any $i\in N_n$. Let $\tilde{N}_n$ be the set of all the edges possible from $N_n$, i.e., $\tilde{N}_n=\left\{ (i,j)\in N_n^{2}:i\neq j\right\}.$ For each $i\in N_n$, define%
\begin{equation*}
N_n(i)=\{j\in N_n:ij\in E_n\}
\end{equation*}%
to be the (open) neighborhood of vertex $i$, and any vertex in the
neighborhood a \textit{neighbor} of $i$. We let $d_n(i)=|N_n(i)|$ and
call it the \textit{degree} of vertex $i$, i.e., the number of the neighbors
of $i$. We also introduce the closed neighborhood of vertex $i$ which is
defined to be%
\begin{equation*}
\overline N_n(i)=N_n(i)\cup \{i\}.
\end{equation*}

Throughout the paper, we assume that the graph $G_n$ is not a complete
graph, i.e., $E_n$ is a proper subset of $\tilde{N}_n$. This means that
there are pairs of vertices that are not adjacent. We also assume that $%
\overline{N}_n(i)$ is a proper subset of $N_n$ for all $i\in N_n$,
which means that for each vertex $i$, there exists a vertex that is not its
neighbor.

We introduce a parameter that measures the tendency of linked pairs of
outcomes, say, $(Y_i,Y_{j})$'s with $ij\in E_n$, exhibiting a stronger
correlation than unlinked pairs of outcomes, say, $(Y_i,Y_{j})$'s with $%
ij\in \tilde{N}_n\backslash E_n$. First, define
\begin{eqnarray*}
	\overline Y_i &=& \frac{1}{d_n(i)}\sum_{j\in N_n(i)} Y_j, \text{ and } \\
	\overline Y_i^c &=& \frac{1}{|N_n\backslash \overline{N}_n(i)|}\sum_{j\in N_n\backslash 
		\overline{N}_n(i)} Y_j,
\end{eqnarray*}
if $d_n(i) \ge 1$, and $\overline Y_i = 0$, otherwise. Thus $\overline Y_i$ is the average of $Y_j$'s in the neighborhood of $i$. We let 
\begin{eqnarray}
\label{gamma}
\gamma &=&\frac{1}{nv_n^2}\sum_{i\in N_n} Cov(Y_i,\overline Y_i)\text{ and}
\\ \notag
\gamma^{c} &=&\frac{1}{nv_n^2}\sum_{i\in N_n} Cov(Y_i,\overline Y_i^c),
\end{eqnarray}%
where $v_n^{2}=\frac{1}{n}\sum_{i\in N_n}Var(Y_i)$. The quantity $\gamma$ measures
the average correlation of the outcomes $Y_i$ and its neighbors $Y_{j}$
among linked outcomes, and $\gamma^{c}$ measures that among unlinked
outcomes. We define the \textit{graph concordance (GC)} of $Y$ along $G_n$
as: 
\begin{equation*}
C(G_n)=\gamma-\gamma^{c}.
\end{equation*}

When GC is strongly positive, this suggests a conspicuous role played by the
graph in shaping the joint dependence pattern of $Y$. Suppose that
\begin{eqnarray*}
	\sum_{i \in N_n} Var(\overline Y_i) \le \sum_{i \in N_n} Var(Y_i), \text{ and }  \sum_{i \in N_n} Var(\overline Y_i^c) \le \sum_{i \in N_n} Var(Y_i).
\end{eqnarray*}
Then GC takes values in $[-2,2]$ in general, but takes values in $[-1,1]$, when the correlation
between $Y_i$ and $Y_{j}$, $i,j\in N_n$, are nonnegative. For example,
in the extreme case where $G_n$ consists of several complete subgraphs
disconnected from each other, and $Y_i$ and $Y_{j}$ are perfectly
correlated whenever $ij\in E_n$ and uncorrelated whenever $ij\in
N_n\backslash E_n$, the GC becomes 1. We say that $Y$ is \textit{graph
	concordant} if $C(G_n)>0$ and \textit{graph discordant} if $C(G_n)<0$.

Throughout this paper, we regard the graph $
G_n$ as fixed, and pursue conditional inference given $G_n$. This does 
\textit{not} mean that the graph $G_n$ is required to be exogenous for the
validity of inference. The conditional inference merely means that the
coverage probability of the proposed confidence set is expressed in terms of
the conditional probability given $G_n$. When one views the graph $G_n$
as a random graph generated by a certain data generating process, the
conditional validity of the inference means that it is conditionally valid
for \textit{each realization of the graph}. Therefore, the inference is
unconditionally valid as well.

\subsection{Graph Concordance and Inbreeding Homophily}

\cite{Currarini/Jackson/Pin:09:Eca} considered a measure of homophily (due to \cite{Coleman:58:HO}), where the extent of homophily is measured by what they call the inbreeding homophily. Here we show that under certain conditions, graph concordance and inbreeding homophily coincide. To see this, suppose that each individual $i \in N_n$ assumes a type $t \in T$, where $T$ is a finite set so that $i$ is associated with a random variable $D_i \in T$. The inbreeding homophily for type $t$ is defined as follows\footnote{The original definition of the inbreeding homophily measure in \cite{Currarini/Jackson/Pin:09:Eca} involves only the samples, i.e., without the expectation. For inference, we modify their measure as a population quantity here.}
\begin{eqnarray*}
	IH(G_n) = \frac{H - w}{1 - w},
\end{eqnarray*}
where $H = s_n /(s_n + d_n)$, and $s_n$ denotes the average of the probabilities $P\{D_i = t,D_j=t\}$ over $i,j \in N_n$ such that and $ij \in E_n$ and $d_n$ denotes the average of the probabilities $P\{D_i = t,D_j \ne t\}$ over $i,j \in N_n$ such that and $ij \in E_n$, and $w$ the average of $P\{D_i = t\}$ over $i \in N_n$. (See \cite{Currarini/Jackson/Pin:09:Eca} on pages 1007-1008 for motivation.) Letting $Y_i = 1\{D_i = t\}$, we can rewrite $H$ and $w$ as follows: 
\begin{eqnarray*}
	H = \frac{\sum_{i \in N_n} \mathbf{E}[Y_i\overline Y_i]d_n(i)}{\sum_{i \in N_n} \mathbf{E}[Y_i]d_n(i)}, \text{ and } w = \frac{1}{n} \sum_{i \in N_n} \mathbf{E}[Y_i].
\end{eqnarray*}
Now, assume the following three conditions:
\medskip

\noindent (a) $\mathbf{E}[Y_i]$'s are the same across $i \in N_n$.

\noindent (b) $d_n(i)$'s are the same across $i\in N_n$.

\noindent (c) $Y_i$ and $Y_j$ are independent whenever $i$ and $j$ are not linked.
\medskip

The equal degree condition (b) can be removed, if we modify the definition of inbreeding homophily as follows:
\begin{eqnarray*}
	IH'(G_n) = \frac{H' - w}{1 - w},
\end{eqnarray*}
with the modified definition of $H$:
\begin{eqnarray*}
	H' = \frac{\sum_{i \in N_n} \mathbf{E}[Y_i\overline Y_i]}{\sum_{i \in N_n} \mathbf{E}[Y_i]},
\end{eqnarray*}
which can be viewed as the inbreeding homophily with $H$ weighted by the inverse of degree $d_n(i)$. This modified version attenuates the influence of nodes with many links. Condition (c) becomes plausible when $G_n$ is a subgraph of a much larger sparse graph and $Y_i$'s are locally dependent satisfying the law of the large numbers. (See remarks after Theorem 1 in Section 3.1 for details.)

Then we can write
\begin{eqnarray}
\label{equiv}
IH(G_n) = \frac{\sum_{i \in N_n} Cov(Y_i,\overline Y_i)}{\sum_{i \in N_n} Var(Y_i)} = C(G_n). 
\end{eqnarray}
Therefore, under (a)-(c), $IH(G_n) = C(G_n)$, and under (a) and (c), $IH'(G_n) = C(G_n)$. (When (c) is violated, the same statement holds if we replace $C(G_n)$ by $\gamma$ defined in (\ref{gamma}).) Thus the graph concordance $C(G_n)$ can be viewed as a measure of homophily as inbreeding homophily in this case.

\subsection{Residual Graph Concordance}
One may consider graph concordance of $Y_i$ after taking into account variations of observable node characteristics, say, $X_i$. We consider the following linear model:
\begin{eqnarray*}
	Y_i = X_i'\beta_0 + u_i,
\end{eqnarray*} 
where we assume that each $X_i$ is a discrete random vector of a small dimension having a finite support $\mathcal{X}$, and $\mathbf{E}[u_i|(X_j)_{j \in N_n}] = 0$. (We do not necessarily view the linear model as a causal model.) For each $x \in \mathcal{X}$, we define
\begin{eqnarray*}
	N_{n,x} = \{i \in N_n: X_i = x\},
\end{eqnarray*}
and $N_{n,x}(i) = \{j \in N_{n,x}: ij \in E_n\}$, and let
\begin{eqnarray*}
	\overline u_{x,i} &=& \frac{1}{|N_{n,x}(i)|}\sum_{j \in N_{n,x}(i)} u_j, \text{ and }\\
	\overline u_{x,i}^c &=& \frac{1}{|N_{n,x} \backslash \overline N_{n,x}(i)|}\sum_{j \in N_{n,x} \backslash \overline N_{n,x}(i)} u_j,
\end{eqnarray*}
where $\overline N_{n,x}(i) = N_{n,x}(i) \cup \{i\}$. Then we may consider the conditional graph concordance of $u_i$'s as follows:
\begin{eqnarray*}
	C_x(G_n) = \gamma_x - \gamma_x^{c},
\end{eqnarray*}
where 
\begin{eqnarray}
\label{gamma 2}
\gamma_x &=&\frac{1}{nv_{n,x}^2}\sum_{i\in N_n: X_i = x} Cov(u_i,\overline u_{x,i}|X)\text{ and}
\\ \notag
\gamma_x^{c} &=&\frac{1}{nv_{n,x}^2}\sum_{i\in N_n: X_i = x} Cov(u_i,\overline u_{x,i}^c|X),
\end{eqnarray}%
and $v_{n,x}^2=\frac{1}{n}\sum_{i\in N_n: X_i = x}Var(u_i|X)$, $X = [X_1,...,X_n]'$, and $Cov(\cdot,\cdot|X)$ and $Var(\cdot|X)$ denote conditional covariance and conditional variance given $X$. Let us call $C_x(G_n)$ \textit{the residual graph concordance} of $Y_i$ with respect to $G_n$ at $X_i=x$.

\subsection{Examples}

\subsubsection{Homophily on Social Characteristics}

Many studies in sociology and economics are interested in the social
phenomenon called homophily which refers to the tendency of individuals to
associate with others who are similar to themselves in terms of social
characteristics such as race, income, or religion. (For example, see
\cite{Currarini/Jackson/Pin:09:Eca} for an economic model of friendship
formation.) The strength of homophily is a matter of empirical question, and
can be measured by GC, where the observed vector $Y$ is taken to be race,
income, or religion and the graph is the social network among people. Of course, one needs to take care in interpreting the results, because the cross-sectional dependence captured by the GC can arise due to other sources.

\subsubsection{Linear Network Interference on Treatment Outcomes}
\label{sec: network interference}

One may use GC to measure the network interference on treatment outcomes.\footnote{This example is inspired by suggestions by one of the referees.} Let $D_i$ denote the treatment state for individual $i \in N_n$, where individuals are on a friendship network $G_n' = (N_n,E_n')$. The outcome $Y_i$ for individual $i$ is specified as follows:
\begin{eqnarray}
\label{network interference}
Y_i = \beta_0 + \beta_1 \overline{D}_i + \beta_2 D_i + \eta_i,
\end{eqnarray}
where $(D_i,\eta_i)$'s are i.i.d. across $i$'s, $\overline{D}_i = \frac{1}{d_n'(i)}\sum_{j \in N_n'(i)} D_j$, and $N_n'(i)$ denotes the neighborhood of $i$ and $d_n'(i)$ the degree of $i$ in $G_n'$, and $\eta_i$ and $D_i$ are allowed to be correlated. Of frequent interest is whether network interference is present, i.e., whether the outcome $Y_i$ depends on $\overline{D}_i$. (See \cite{Aronow/Samii:15:WP} and \cite{Athey/Eckles/Imbens:15:NBER} and \cite{Leung:16:WPb} for network interference issues in program evaluations, and for references in the related literature.) Suppose further that $D_i$'s are not precisely observed. Then if $\beta_1>0$, the network interference will be revealed through the cross-sectional positive correlation among $Y_i$'s where $Y_i$ and $Y_j$ are positively correlated whenever $N_n'(i) \cap N_n'(j) \ne \varnothing$. Once one defines a graph $G_n = (N_n,E_n)$, where $ij \in E_n$ if and only if $N_n'(i) \cap N_n'(j) \ne \varnothing$, this cross-sectional positive correlation is captured by the positive GC of $Y_i$ along $G_n$.

\subsubsection{Knowledge Spillover and Collaboration Networks}

A\ given set of projects or research papers can form a graph, with an edge
formed by a common developer or an author. Alternatively, a given set of
developers or researchers can form a graph, where an edge is formed between
two vertices if they work on the same project or the same research. (See
\cite{Jackson/Wolinsky:96:JET} for a coauthor model of graph formation. See
also \cite{Fershtman/Gandal:11:RAND} for an empirical analysis of knowledge
spillover along a graph of software developers.) One might be interested in testing whether projects
that are linked through more common developers tend to exhibit higher outcomes than those that are not. Let $N_n$ be the set of the projects, and $Y_i$ the outcome of project $i$, and let a graph $G_n = (N_n,E_n)$ be such that $ij \in E_n$ if and only if projects $i$ and $j$ have more common developers than a benchmark number. If knowledge spillover induces cross-sectional dependence of $Y_i$'s along the graph $G_n$, this cross-sectional dependence can be explored through the GC of $Y_i$'s along the graph $G_n$.

\section{Permutation Inference on Graph Concordance}

\subsection{Point Estimators and Confidence Intervals}

We introduce an estimated version of $C(G_n)$. First, define%
\begin{equation*}
\hat{e}_i=\frac{Y_i-\overline{Y}}{\hat{v}},
\end{equation*}%
where $\overline{Y}=\frac{1}{n}\sum_{j\in N_n}Y_{j}$, and $\hat{v}^{2}=\frac{1}{%
	n}\sum_{i \in N_n}\left( Y_i-\overline{Y}\right) ^{2}$. Let $\hat{a}_i=\frac{1}{%
	d_n(i)}\sum_{j\in N_n(i)}\hat{e}_{j}$ if $d_n(i)\geq 1$ and $\hat{a}%
_i=0$ otherwise, and let%
\begin{equation*}
\hat{a}_i^{c}=\frac{1}{|N_n\backslash \overline{N}_n(i)|}\sum_{j\in
	N_n\backslash \overline{N}_n(i)}\hat{e}_{j}.
\end{equation*}%
Thus $\hat a_i$ is the difference between the average outcome over $i$'s neighbors and the overall average outcome after normalized by $\hat v$. Then the point estimator of the GC is defined as%
\begin{equation*}
\hat{C}(G_n)=\hat{\gamma}-\hat{\gamma}^{c},
\end{equation*}
where%
\begin{equation*}
\hat{\gamma}=\frac{1}{n}\sum_{i\in N_n}\hat{e}_i\hat{a}_i\text{
	and }\hat{\gamma}^{c}=\frac{1}{n}\sum_{i\in N_n}\hat{e}_i\hat{a}%
_i^{c}.
\end{equation*}

As for the confidence intervals, we use the permutation distribution of the test statistic given as follows.
\begin{eqnarray}
T=\frac{\sqrt{n}\{\hat{C}(G_n)-C(G_n)\}}{\hat{\sigma}_{+}},  \label{T}
\end{eqnarray}
where $\hat{\sigma}_{+}$ denotes the scale normalizer which we explain now. A proper scale normalizer should accommodate cross-sectional dependence along the edges. Let us introduce the population versions of $\hat{e}_i$'s, $\hat{a}_i$'s and $\hat{a}_i^c$'s:
\begin{eqnarray*}
	e_i=\frac{Y_i-\mathbf{E}Y_i }{v_n}\text{,}
\end{eqnarray*}
and
\begin{eqnarray*}
	a_i=\frac{1}{d_n(i)}\sum_{j\in N_n(i)}e_{j}\text{ and\ }a_i^{c}=%
	\frac{1}{|N_n\backslash \overline{N}_n(i)|}\sum_{j\in N_n\backslash 
		\overline{N}_n(i)}e_{j}.
\end{eqnarray*} 
We can show that under the stated conditions in Theorem 1 below,%
\begin{eqnarray*}
	\sqrt{n}\left\{ \hat{C}(G_n)-C(G_n)\right\} =\frac{1}{\sqrt{n}}%
	\sum_{i \in N_n}\left( q_i-\mathbf{E}q_i \right) +o_{P}(1),
\end{eqnarray*}
where $q_i=e_i\{a_i-e_i\gamma\}$. Note that $q_i$ is essentially a modified version of $e_i a_i$, where the modification (i.e., subtraction by $e_i^2 \gamma$) is needed for accommodating the first stage estimation error from $\hat{v}$. As for $q_i$, we later assume that for each $d=1,...,d_{mx,n}$, $\mathbf{E}q_i$'s are the same for all $i\in N_n$ having the same degree. Define
\begin{eqnarray*}
	S_n(i) = \{j \in N_n: d_n(j) = d_n(i)\}, \text{ and } s_n(i) = |S_n(i)|,
\end{eqnarray*}
that is, $S_n(i)$ is the set of the vertices which have the same degree as vertex $i$. We construct
\begin{eqnarray*}
	\hat{\sigma}^{2}=\frac{1}{n}\sum_{i_1,i_2\in N_n: \overline N_n(i_1) \sim \overline N_n(i_2)}(\hat{q}_{i_{1}}-\bar{q}_{i_{1}})(\hat{q}_{i_{2}}-\bar{q}_{i_{2}}),
\end{eqnarray*}%
where $\overline N_n(i_1) \sim \overline N_n(i_2)$ means that a vertex from $\overline N_n(i_1)$ is adjacent to a vertex from $\overline N_n(i_2)$, and
\begin{eqnarray}
\hat{q}_i=\hat{e}_i\left\{ \hat{a}_i-\hat{e}_i\hat{\gamma}\right\} \text{ and }\bar{q}_i=\frac{1}{s_n(i)}\sum_{j\in S_n(i)}%
\hat{e}_{j}\left\{ \hat{a}_{j}-\hat{e}_{j}\hat{\gamma}\right\}.
\label{q}
\end{eqnarray}
The location normalization by $\bar{q}%
_{i}$ is later justified by the assumption that $\mathbf{E}q_i$ is the same for all $i$'s having the same degree. (See Assumption 2(ii) and discussions below.) Define 
\begin{eqnarray}
\hat{\sigma}_{+}^{2}=\left\{ 
\begin{array}{l}
\hat{\sigma}^{2}\text{,} \\ 
\hat{\sigma}_{1}^{2},%
\end{array}%
\begin{array}{l}
\text{if }\hat{\sigma}^{2}>0 \\ 
\text{if }\hat{\sigma}^{2}\leq 0%
\end{array}%
\right. ,  \label{sigma3}
\end{eqnarray}%
where $\hat{\sigma}_{1}^{2}=\frac{1}{n}\sum_{i\in N_n}\left( \hat{q}_i-%
\bar{q}_i\right) ^{2}.$ The introduction of $\hat{\sigma}_{1}^{2}$ ensures
positivity of the scale normalizer $\hat{\sigma}_{+}^{2}$ in finite samples.

For critical values, we construct a permutation test statistic as follows.
Define $\Pi _n$ to be the set of permutations on $N_n$. The permutation
test statistic $T_{\pi }$ is obtained by replacing $\hat{e}_i$'s in $T$ by 
$\hat{e}_{\pi (i)}$'s. More specficially, for each $\pi \in \Pi _n$, we let%
\begin{eqnarray*}
	\hat{a}_{i,\pi }=\frac{1}{d_n(i)}\sum_{j\in N_n(i)}\hat{e}_{\pi (j)}%
	\text{ and }\hat{a}_{i,\pi }^{c}=\frac{1}{|N_n\backslash \overline{N}%
		_n(i)|}\sum_{j\in N_n\backslash \overline{N}_n(i)}\hat{e}_{\pi (j)},
\end{eqnarray*}%
and construct%
\begin{eqnarray*}
	\hat{C}_{\pi }(G_n)=\frac{1}{n}\sum_{i\in N_n}\hat{e}_{\pi (i)}\hat{a}%
	_{i,\pi }-\frac{1}{n}\sum_{i\in N_n}\hat{e}_{\pi (i)}\hat{a}_{i,\pi }^{c}.
\end{eqnarray*}%
As for the permutation version of the scale normalizer, we define%
\begin{eqnarray*}
	\hat{q}_{i,\pi }=\hat{e}_{\pi (i)}\left\{ \hat{a}_{i,\pi }-\hat{e}_{\pi (i)}%
	\hat{\gamma}_{\pi }\right\} \text{ and }\bar{q}_{i,\pi }=\frac{1}{%
		s_n(i)}\sum_{j\in S_n(i)}\hat{e}_{\pi (j)}\left\{ \hat{a}_{j,\pi }-%
	\hat{e}_{\pi (j)}\hat{\gamma}_{\pi }\right\} ,
\end{eqnarray*}%
where $\hat{\gamma}_{\pi }=\frac{1}{n}\sum_{i\in N_n}\hat{e}_{\pi (i)}%
\hat{a}_{i,\pi }$. Then we construct $\hat{\sigma}_{\pi }^{2}$ and $\hat{%
	\sigma}_{+,\pi }^{2}$ just as $\hat{\sigma}^{2}$ and $\hat{\sigma}_{+}^{2}$
except that $\hat{q}_{i,\pi }$'s and $\bar{q}_{i,\pi }$'s replace $\hat{q}%
_i$'s and $\bar{q}_i$'s. Define the permutation test statistic as%
\begin{eqnarray}
T_{\pi }=\frac{\sqrt{n}\hat{C}_{\pi }(G_n)}{\hat{\sigma}_{+,\pi }},
\label{Tp}
\end{eqnarray}%
and let%
\begin{eqnarray*}
	c_{\alpha }=\inf \left\{ c\in \mathbf{R}:\frac{1}{|\Pi _n|}\sum_{\pi \in
		\Pi _n}1\left\{ |T_{\pi }|\leq c\right\} >1-\alpha \right\} .
\end{eqnarray*}%
Hence $T_{\pi }$ is computed just as $T$ except that $\hat{e}_i$'s are
replaced by $\hat{e}_{\pi (i)}$'s. Then the permutation-based two-sided
confidence interval for the GC is given as follows:%
\begin{eqnarray}
\mathcal{C}_{\alpha }(G_n)=\left[ \hat{C}(G_n)-c_{\alpha }\hat{\sigma}%
_{+}/\sqrt{n},\ \hat{C}(G_n)+c_{\alpha }\hat{\sigma}_{+}/\sqrt{n}\right] ,
\label{CI}
\end{eqnarray}%
where $\hat{\sigma}_{+}^{2}$ is as defined in (\ref{sigma3}).

It is important that we do not use $\hat a_{\pi(i)}$ in constructing a permutation-based critical value, because we need to create perturbations of the graph structure through permutations. Its motivation is easier to see in the case of the problem of testing for the null of zero graph concordance. In order for the test to work, the resulting graph structure should be totally different from the original graph after a permutation, so that under the null of no graph concordance, both the original test statistic and the permutation test statistic behave similarly, but under the alternative hypothesis where the graph governs the cross-sectional dependence structure, the permutation test statistic should behave differently from the original test statistic, thereby giving power to the test. If the graph structure remains almost invariant after the permutations, the permutation test will not have much power. That is why we use the same neighborhood structure, but only permute the sample index of $\hat e_i$. We cannot use $\hat a_{\pi(i)}$ because then the neighbor structure will also be permuted in tandem with the node indices, leaving the graph structure unchanged and giving no nontrivial power to the test.

\subsection{Inference on Residual Graph Concordance}
Let us see how the previous approach applies to the residual graph concordance (GC). Unlike the GC for $Y_i$'s, the residual GC involves $u_i$'s which are not observed. To deal with this, we first take
\begin{eqnarray*}
	\hat u_i = Y_i - X_i'\hat \beta,
\end{eqnarray*}
where $\hat \beta$ is the usual least squares estimator of $\beta$, i.e., $\hat \beta = (X'X)^{-1}X'y$, with $X$ being the $n \times d$ matrix whose $j$-th row is given by $X_j'$ and $y$ the $n$ dimensional vector whose $j$-th entry is given by $Y_j$. Hence, we have for each $i \in N_{n,x}$,
\begin{eqnarray}
\label{rep}
\hat u_i = u_i - x'(X'X)^{-1}X'u,
\end{eqnarray}
where $u = [u_1,...,u_n]'$. Let us take
\begin{equation*}
\hat{C}_x(G_n)=\hat{\gamma}_{x}-\hat{\gamma}_{x}^{c},
\end{equation*}
and
\begin{equation*}
\hat{\gamma}_x=\frac{1}{|N_{n,x}|}\sum_{i\in N_{n,x}}\hat{e}_{x,i}\hat{a}_{x,i}\text{
	and }\hat{\gamma}_{x}^{c}=\frac{1}{|N_{n,x}|}\sum_{i\in N_{n,x}}\hat{e}_{x,i}\hat{a}%
_{x,i}^{c},
\end{equation*}
with the following definitions:
\begin{eqnarray*}
	\hat{e}_{x,i} = \frac{\hat u_{x,i} - \bar u_x}{\hat v_x}, \quad \hat a_{x,i} = \frac{1}{d_{n,x}(i)}\sum_{j \in N_{n,x}(i)} \hat e_{x,j}, \text{ and } \hat{a}_{x,i}^{c}=\frac{1}{|N_{n,x}\backslash \overline{N}%
		_{n,x}(i)|}\sum_{j\in N_{n,x}\backslash \overline{N}_{n,x}(i)}\hat{e}_{x,j},
\end{eqnarray*}
and $\bar u_x$ denotes the average of $\hat u_{x,i}$'s over $i \in N_{n,x}$ and $\hat v_x$ the average of $(\hat u_i - \bar u_x)$'s over $i\in N_{n,x}$. Thus, we consider the following test statistic:
\begin{eqnarray*}
	T_x = \frac{\sqrt{n}(\hat{C}_x(G_n) - C_x(G_n)}{\hat \sigma_{x,+}},
\end{eqnarray*}
where $\hat \sigma_{x,+}$ is constructed in the same way as we constructed $\hat \sigma_+$ except that we use $\hat u_i$ in place of $Y_i$'s, $N_{n,x}$ in place of $N_n$ and $N_{n,x}(i)$ in place of $N_n(i)$.

Permutation-based inference can proceed similarly as before. Let $\Pi_{n,x}$ be the collection of permutations on $N_{n,x}$. The permutation test statistic $T_{x,\pi}$ is obtained for each $\pi \in \Pi_{n,x}$ by replacing $\hat e_{x,i}$'s in $T$ by $\hat e_{x,\pi(i)}$'s.

Due to (\ref{rep}), we can see that $\hat{C}_x(G_n)$ and $\hat e_{x,i}$ remain the same if we replace $\hat u_i$ by $u_i$ for all $i \in N_{n,x}$. Therefore if $u_i$'s are exchangeable conditional on $X$, the permutation-based inference is valid in finite samples. The asymptotic validity of the permutation inference also follows when $u_i$'s satisfy the conditions specified for $Y_i$'s in the next subsection. 

\subsection{Asymptotic Validity}

When the random vector $Y=(Y_i)_{i=1}^{n}$ is exchangeable, i.e., the
joint distribution of $Y_{\pi }=(Y_{\pi (i)})_{i=1}^{n}$ is the same as that
of $Y=(Y_i)_{i=1}^{n}$ (and consequently, the joint distribution of $(\hat e_{\pi(i)})_{i=1}^n$ is the same as that of $(\hat e_i)_{i=1}^n$) for any permutation $\pi \in \Pi _n$, the
confidence interval is valid in finite samples. However, when $Y$ is
exchangeable, we have $C(G_n')=0$ for any graph $G_n'$ such that for each $i \in N_n$, $d_n'(i) \le n - d_n'(i)$, $d_n'(i)$ denoting the degree of $i$ in $G_n'$. Therefore, this represents a strong form of graph irrelevance of $Y$. For
validity of the confidence intervals, we need to cover the case $%
C(G_n)\neq 0$ as well.

This paper's main result establishes conditions under which the confidence
interval $\mathcal{C}_{\alpha }(G_n)$ is asymptotically valid, even when $%
Y $ fails to be exchangeable. The following definition clarifies which
properties of the graph $G_n$ are relevant for this purpose. Recall that a 
\textit{degree} (denoted by $d_n(i)$) of a vertex $i$ refers to the number
of the edges vertex $i$ has in $G_n$, and the \textit{maximum degree}
(denoted by $d_{mx,n}$)\textit{,} the maximum over the degrees of the
vertices in $N_n$. We define%
\begin{eqnarray*}
	d_{avi,n}=\frac{1}{n}\sum_{i\in N_n:d_n(i)\geq 1}\frac{1}{d_n(i)},
\end{eqnarray*}%
i.e., the average of the inverse degrees. Also define $d_{mx,n,3}$ to be the maximum number of the vertices within three edges from a fixed vertex. We call $d_{mx,n,3}$ \textit{the maximum 3-degree} of graph $G_n$.
\medskip

\noindent \textbf{Definition 1:} For a positive integer $d_{mx,n,3}$, let $\mathcal{G}_n(d_{mx,n,3})$ be the collection of graphs having the maximum 3-degree equal to $d_{mx,n,3}$.\medskip

Let us introduce the population versions of $\hat{e}_i$'s, $\hat{a}_i$'s and $\hat{a}_i^c$'s:
\begin{equation*}
e_i=\frac{Y_i-\mathbf{E}\left[ Y_i\right] }{v_n}\text{,}
\end{equation*}
and
\begin{equation*}
a_i=\frac{1}{d_n(i)}\sum_{j\in N_n(i)}e_{j}\text{ and\ }a_i^{c}=%
\frac{1}{|N_n\backslash \overline{N}_n(i)|}\sum_{j\in N_n\backslash 
	\overline{N}_n(i)}e_{j}.
\end{equation*}
The following assumption is concerned with the nondegeneracy of the limiting
distribution of the test statistic.\medskip

\noindent \textbf{Assumption 1 (Nondegeneracy and Moment Conditions):} There
exist small $c>0$ and large $M>0$ such that the following is satisfied for
all $n\geq 1.$

\noindent (i) $v_n^{2}>c.$

\noindent (ii) For $q_i=e_i\{a_i-e_i\gamma\},$%
\begin{equation}
\mathbf{E}\left[ \left( \frac{1}{\sqrt{n}}\sum_{i \in N_n}\left( q_i-\mathbf{%
	E}q_i \right) \right) ^{2}\right] >c\text{ and\ }\frac{%
	d_{avi,n}}{|\tilde{N}_n|}\sum_{ij\in \tilde{N}_n}\mathbf{E}\left[
e_i^{2}e_{j}^{2}\right] >c.  \label{cd}
\end{equation}

\noindent (iii) max$_{1\leq i\leq n}\mathbf{E}|Y_i|^{8}<M.$\medskip

The first condition in (\ref{cd}) ensures that the variance of the leading
term in the asymptotic linear representation is positive. The second
condition in (\ref{cd}) is a technical, mild condition which essentially requires that $d_{avi,n}$ be bounded away from zero from some large $n$ on. This condition is mostly satisfied by many random graph models (such as Erd\"{o}s-R\'{e}nyi graphs or Barab\'{a}si-Albert graphs) whose degree distribution becomes tight and non-degenerate in the limit. To see this, we let
\begin{equation*}
N_{n,d}=\{i\in N_n:d_n(i)=d\},
\end{equation*}
and rewrite
\begin{eqnarray*}
	d_{avi,n} = \sum_{d=1}^{d_{mx,n}} \frac{1}{d} \frac{|N_{n,d}|}{n}
	\ge \frac{1}{k} \sum_{d=1}^k  \frac{|N_{n,k}|}{n}.
\end{eqnarray*}
for any fixed positive integer $k \le d_{mx,n}$.
Therefore, $d_{avi,n}$ is bounded away from zero, whenever  there is a positive fraction of nodes having degrees less than or equal to $k$ for some fixed $k$.  This condition is violated when all but an asymptotically negligible fraction of nodes in the network becomes either isolated or has a degree that goes to infinity as $n \rightarrow \infty$.\medskip

\noindent \textbf{Assumption 2 (Symmetry in Means):} (i) $\mathbf{E}Y_i$'s
are the same for all $i$'s in $N_n$.

\noindent (ii) For each $d=1,...,d_{mx,n}$, there exists $r_{n,d}\in \mathbf{R}$ such
that $\mathbf{E}q_i=r_{n,d}$ for all $i\in N_{n,d}$.
\medskip

Assumption 2(i) requires that the expected values of $Y_i$'s are the same
across $i$'s, which is weaker than the assumption that $Y_i$'s are
identically distributed. Assumption 2(ii) requires that the expectation of $%
q_i$ be the same for all $i$'s having the same degree. When $i$ and $j$
have the same degree, $q_i$ and $q_{j}$ are sums of the same number of
random variables of the form $e_ie_{k}$ and $e_{j}e_{l}.$ For example,
suppose that $Y_i$'s have the same variance and for each $d$ the correlation between $Y_i$ and $\overline Y_i$ is the same for all $i\in N_{n,d}$. Then Assumption 2(ii) is satisfied.

For example, consider the data generating process for $Y_i$ as in (\ref{network interference}) in the example of treatment outcomes through network interference. Since $(D_i,\eta_i)$'s are i.i.d. across $i$'s, the joint distribution of $((\overline D_j, D_j, \eta_j)_{j \in N_n(i)},D_i,\eta_i)$ is the same across all $i$'s that have the same degree. This means that the joint distribution of $(Y_i,\overline Y_i)$ is the same across all $i$'s that have the same degree. Thus Assumption 2(ii) is satisfied in this case. (The i.i.d. assumption for $(D_i,\eta_i)_{i \in N_n}$ and the linearity in the model also ensure that Assumption 2(i) is satisfied.)

Also, consider the residual graph concordance introduced in a preceding section. Recall that since the sample version of GC remains the same regardless of whether we use $\hat u_i$'s or $u_i$'s, $u_i$ plays the role of $Y_i$, and the subset $N_{n,x}$ of $i$'s such that $X_i = x$ plays the role of $N_n$. Assumption 2 is satisfied as long as the conditional distribution of $(u_i,\bar u_i)$ given $X_i=x$ is the same across $i \in N_{n,x}$ having the same degree.

We introduce a notion of a joint dependence pattern for $Y$ shaped by graph $%
G_n$. We say that $Y=(Y_i)_{i=1}^{n}$ has a \textit{dependency graph} $%
G_n=(N_n,E_n)$, if for any two disjoint subsets $A,A'\subset
N_n$ having no edge in $E_n$ such that one end vertex is in $A$ and the
other end vertex is in $A'$, two random vectors $(Y_i)_{i\in A}$
and $(Y_i)_{i\in A'}$ are independent. (See \cite{Penrose:03:RandomGeometricGraphs},
p.22.) When $Y$ has $G_n$ as a dependency graph, it implies that any two
disjoint sets of $Y_i$'s having no edge between the two sets are
independent. Therefore, $Y_i$ and $Y_{j}$ can be correlated, only when
they are adjacent. The joint dependence of linked pairs $Y_i$ and $Y_{j}$
can be heterogeneous across the linked pairs.\footnote{Under the dependency graph assumption, we have $\gamma^c=0$ and hence we can impose it in the estimator of GC by seting $\hat{\gamma}^c=0$. This does not alter the asymptotic distribution and the variance estimators. We keep the original definition of GC as involving $\gamma^c$ there, because a different inference procedure that does not invoke dependency graph may be possible.} 

Let us define the class of joint distributions of $Y$ to be considered. Let $%
\mathcal{P}_n$ be the collection of joint distributions of $Y$.\medskip

\noindent \textbf{Definition 2:} For $c,M>0$ and a graph $G_n$, we define $%
\mathcal{P}_n(G_n;c,M)$ to be the set of the joint distributions of $%
Y=(Y_i)_{i=1}^{n}$ such that under each $P\in \mathcal{P}_n(G_n;c,M)$,
Assumptions 1 and 2 are satisfied with $(c,M)$ and $G_n$, and $G_n$ is a
dependency graph for $Y$.\medskip

The following theorem is the main result of this paper.\medskip

\noindent \textbf{Theorem 1:} \textit{Suppose that for each }$n\geq 1,\
G_n\in \mathcal{G}_n(d_{mx,n,3})$ \textit{satisfying that as }$n\rightarrow \infty $,%
\begin{equation}
\frac{d_{mx,n,3}^4}{n} \rightarrow 0.  \label{cd6}
\end{equation}

\textit{Then for each }$c,M>0,$\textit{\ and for each sequence }$P_n\in 
\mathcal{P}_n(G_n;c,M),$%
\begin{equation*}
\lim_{n\rightarrow \infty }\ \left\vert P_n\left\{ C(G_n)\in \mathcal{C}%
_{\alpha }(G_n)\right\} -(1-\alpha )\right\vert =0.
\end{equation*}

A notable aspect of Theorem 1 is that despite the fact that the permutation
test statistic does not involve centering explicitly, the confidence
interval is still asymptotically valid regardless of whether $C(G_n)=0$ or
not.

The condition in (\ref{cd6}) allows for the maximum degree $d_{mx,n}$ to
increase at a certain polynomial rate. One should not
interpret the condition (\ref{cd6}) as part of a description of the way the
network forms and grows in reality. The condition should be used to gauge
the finite sample environment in which the validity of inference
justified assuming large $n$ becomes reliable in finite samples.

It is not hard to see that the linear network interference example in Section \ref{sec: network interference} satisfies the dependency graph assumption. The dependency graph assumption is also compatible with the notion of homophily 
among people in the following sense. Homophily refers to the tendency of people being
associated with each other more often when they are of the same social group such as 
race or religion. Suppose that we have two race categories, ${W,B}$, where each individual $i$ is given race indicator $D_i \in \{W,B\}$, and that the observed graph $G_n$ is a subgraph of a very large and sparse graph, say, $G_L=(N_L,E_L)$ where $L$ is much larger than $n$. Also, assume that $(D_i,D_j)$'s have the same joint distribution across $ij \in E_L$ and the same joint distribution across $ij \notin E_L$ such that $i \ne j$. Homophily says that for a person $i$ with race $D_i=t$,
person $j$, if the person is a friend of person $i$, is more likely to be found to be of the same race than of the different race, i.e., for $ij \in E_L$,
\begin{eqnarray*}
	P\{D_j \ne t|D_i=t\} < P\{D_j=t|D_i=t\}. 
\end{eqnarray*}
The dependency graph then imposes that a person $j$ that is not a friend of $i$ is as likely to be of the same race as to be of the different race. In other words, there is no correlation of races between people who are not linked. This latter assumption is plausible when $Y_i$'s are locally dependent so that the law of the large numbers hold. For $t \in \{W,B\}$, let $Y_i(t) = 1\{D_i = t\}$ and
\begin{eqnarray*}
	\overline Y_{i,L}^c(t) = \frac{1}{|N_L \backslash \overline N_L(i)|} \sum_{j \in N_L \backslash \overline N_L(i)} Y_j(t).
\end{eqnarray*}
Then for $t_1,t_2 \in \{W,B\}$, and for $j \in N_L \backslash \overline N_L(i)$,
\begin{eqnarray*}
	P\{D_i=t_1,D_j=t_2\} &=& \mathbf{E}[Y_i(t_1)Y_j(t_2)]\\
	&=& \mathbf{E}[Y_i(t_1)\overline Y_{i,L}^c(t_2)]\\
	&\approx& \mathbf{E}[Y_i(t_1)\mathbf{E}[\overline Y_{i,L}^c(t_2)]]
	= P\{D_i = t_1\} P\{D_j = t_2\},
\end{eqnarray*}
where the approximation above comes from the law of the large numbers. The approximation error will be negligible when $L$ is much larger than $n$.\footnote{This does not mean that the dependency graph assumption is innocuous, because it assumes more than pairwise independence of $Y_i$'s which are not linked.}

The proposal of this paper is still applicable when $Y$ exhibits strong dependence through several common shocks (or aggregate shocks) that affect all the nodes globally, as long as $Y_i$'s have $G_n$ as a dependency graph conditional on the common shocks. In this case, when it comes to measuring the relevance of $G_n$ in explaining the cross-sectional dependence of $Y_i$'s, this paper's view is that our focus should be on the conditional
version of graph concordance given common shocks, where the probability
(both in the definition of graph concordance and the coverage probability of
the confidence set) is now replaced by the conditional probability given the
common shocks. There are two reasons for this view. First, our interest is
not merely in detecting the presence of cross-sectional dependence, but the
presence of a dependence pattern \textit{that is shaped by a given network}.
Therefore, it is better to condition on the source of cross-sectional dependence that
arises for a reason unrelated to the network. Second, as we make only a
single observation of $Y$, there is no hope of recovering unconditional
probability in general that takes into account variations in the common
shocks, because such variations are not observed in the data.\footnote{%
	See Andrews (2005) for examples of linear models where inference based on
	unconditional probability is possible using only cross-sectional
	observations despite the presence of common shocks.}

\subsection{Testing for the Graph Concordance of Cross-Sectional Observations%
}

Suppose that we would like to test whether $C(G_n)>0$ or not. The null and
alternative hypotheses of interest in this paper take the following form:%
\begin{eqnarray*}
	H_{0} :C(G_n)\leq 0,\text{ against } H_{1} :C(G_n)>0.
\end{eqnarray*}

Similarly as before, we define 
\begin{equation*}
T_{1}=\frac{\sqrt{n}\hat{C}(G_n)}{\hat{\sigma}_{+}},
\end{equation*}%
which is a version of $T$ with the restriction $C(G_n)=0$ imposed. For
critical values, we use $T_{\pi }$ defined previously, and find a critical
value $c_{\alpha ,1}$ as follows:%
\begin{equation*}
c_{\alpha ,1}=\inf \left\{ c\in \mathbf{R}:\frac{1}{|\Pi _n|}\sum_{\pi \in
	\Pi _n}1\left\{ T_{\pi }\leq c\right\} >1-\alpha \right\} ,
\end{equation*}%
which is a one-sided version of $c_{\alpha }$. Then we can perform the test
by rejecting the null hypothesis if and only if $T_{1}>c_{\alpha ,1}$.

The null hypothesis is weaker than the exchangeability of $Y$. Hence the permutation test does not preserve finite sample validity in all the cases of the null hypothesis. The following theorem shows that the permutation test
controls the size asymptotically under the null hypothesis.

Let $\mathcal{P}_{n,0}(G_n)$ be the collection of the probabilities under which $C(G_n)\leq 0$, and let $\mathcal{P}_{n,00}(G_n)$ be the collection of the probabilities under which $C(G_n)=0$. Also, let $\mathcal{P}_{n,1}(G_n;b)$ be the collection of the	probabilities under which $C(G_n) \ge b.$ \medskip

\noindent \textbf{Theorem 2:} (i) \textit{Suppose that the conditions of Theorem
	1 hold. Then for each }$c,M>0,$\textit{\ and for each sequence }$P_n\in 
\mathcal{P}_n(G_n;c,M)\cap \mathcal{P}_{n,0}(G_n),$
\begin{equation*}
\underset{n\rightarrow \infty }{\text{limsup }} P_n\left\{ T_{1}>c_{\alpha
	,1}\right\} \leq \alpha.
\end{equation*}

\textit{Furthermore, for each sequence }$P_n\in \mathcal{P}%
_n(G_n;c,M)\cap \mathcal{P}_{n,00}(G_n),$ 
\begin{equation*}
\lim_{n\rightarrow \infty }\ \left\vert P_n\left\{ T_{1}>c_{\alpha
	,1}\right\} -\alpha \right\vert =0,
\end{equation*}

\noindent (ii) \textit{For each }$c,M>0$,\textit{ and for each sequence }$P_n\in 
\mathcal{P}_n(G_n;c,M)\cap \mathcal{P}_{n,1}(G_n;b)$\textit{ with }$b>0$,
\begin{eqnarray*}
	\liminf_{n\rightarrow \infty} P_n\left\{ T_{1}>c_{\alpha
		,1}\right\} = 1.
\end{eqnarray*}
\medskip

The proof of Theorem 2 comes from the arguments used for proving Theorem 1.
Details are omitted for brevity.

\section{Monte Carlo Simulation Studies}

\subsection{Data Generating Process}

For the graph in the simulation study, we consider two classes of random
graphs, each having size $n\in \{300,600\}$. The first class is an Erd\"{o}%
s-R\'{e}nyi (E-R) random graph, where each pair of the vertices form an edge with
equal probability $p_n=\lambda /(n-1)$, where $\lambda $ is chosen from $%
\{1,3,5\}$. Each vertex from this random graph has neighbors of size $%
\lambda $ on average. The distribution of the size of the neighborhood is
approximately a Poisson distribution with parameter $\lambda $ when $n$ is
large. The second class is a Barab\'{a}si-Albert (B-A) random graph of preferential
attachment. To generate this random graph, we first began with an Erd\"{o}%
s-R\'{e}nyi random graph of size $20$ with $\lambda =1$. Then we let the graph
grow by adding each vertex sequentially and let the vertex form edges with $%
m $ other existing vertices. (We chose $m$ from $\{1,2,3\}$.) The
probability of a new vertex forming an edge with an existing vertex is
proportional to the number of the neighbors of the existing vertex. We keep
adding new vertices until the size of the graph becomes $n$. We generate the graphs once and fix them when we generate the outcomes. The graph characteristics used in the simulation study are summarized in Table 1.

As for the data generating process for the outcomes, we first generate $%
\{Y_i\}_{i=1}^{n}$ i.i.d. from $N(0,1)$ under the null hypothesis. Under
the alternative hypothesis, we generate $Y_i$ as follows. We first
generate $\{Y_i^{\ast }\}_{i=1}^{n}$ i.i.d. from $N(0,1)$. Let $%
E=\{e_{1},...,e_{S}\}$ be the set of edges in the graph generated as
previously. We remove redundant edges from $E$ (i.e., remove $ji$ with $j<i$%
) and let $M$ be two-column matrix whose entries are of the form $%
[i_{s},j_{s}]$ for $e_{s}=i_{s}j_{s}$. Let $M$ be sorted on the first column
so that $i_{s}\leq i_{s+1}$.\medskip

\noindent \textsc{Step 1:}\textbf{\ }For $s=1$, such that $e_{1}=i_{1}j_{1}$%
, we draw $Z_{1}\sim N(0,1)$ and set%
\begin{equation*}
(Y_{i_{1}},Y_{j_{1}})=\sqrt{1-c^{2}}\times (Y_{i_{1}}^{\ast
},Y_{j_{1}}^{\ast })+c\times Z_{1},
\end{equation*}%
where $c$ is a parameter that determines the strength of the stochastic dependence of linked outcomes. (When $c$ is away from zero, the variable $Z_1$ serving as a common factor to $%
Y_{i_{1}}$ and $Y_{j_1}$ induces correlation between $Y_{i_{1}}$ and $%
Y_{j_{2}}$ and this correlation increases in $c$ other things being equal.)
We replace $(Y_{i_{1}}^{\ast },Y_{j_{1}}^{\ast })$ by $(Y_{i_{1}},Y_{j_{1}})$%
, and redefine the series $\{Y_i^{\ast }\}_{i=1}^{n}$.\medskip

\noindent \textsc{Step }$s$\textbf{: }For $s>1$ such that $%
e_{s}=(i_{s},j_{s})$, we draw $Z_{s}\sim N(0,1)$ and set%
\begin{equation*}
(Y_{i_{s}},Y_{j_{s}})=\sqrt{1-c^{2}}\times (Y_{i_{s}}^{\ast
},Y_{j_{s}}^{\ast })+c\times Z_{s}.
\end{equation*}%
We replace $(Y_{i_{s}}^{\ast },Y_{j_{s}}^{\ast })$ by $(Y_{i_{s}},Y_{j_{s}})$%
, and redefine the series $\{Y_i^{\ast }\}_{i=1}^{n}$.\medskip

\begin{table}[t]
	\caption{The Degree Characteristics of the Graphs and the True GCs Used in
		the Simulation Study}
	\begin{center}
		\begin{tabular}{rrrrl|rrrl}
			\hline\hline
			&  & E-R & Graph &  &  & B-A & Graph &  \\ \cline{2-9}
			& $n=$ & \multicolumn{1}{l}{$300$} & \multicolumn{1}{|r}{$n=$} & $600$ & $n=$
			& \multicolumn{1}{l}{$300$} & \multicolumn{1}{|r}{$n=$} & $600$ \\ 
			\cline{2-9}
			\multicolumn{1}{c}{} & \multicolumn{1}{c}{$\lambda =1$} & \multicolumn{1}{c}{%
				$\lambda =5$} & \multicolumn{1}{|c}{$\lambda =1$} & \multicolumn{1}{c|}{$%
				\lambda =5$} & \multicolumn{1}{|c}{$m=1$} & \multicolumn{1}{c}{$m=3$} & 
			\multicolumn{1}{|c}{$m=1$} & \multicolumn{1}{c}{$m=3$} \\ \hline
			\multicolumn{1}{c}{max. deg.} & \multicolumn{1}{|c}{5} & \multicolumn{1}{c}{
				12} & \multicolumn{1}{|c}{6} & \multicolumn{1}{c|}{13} & \multicolumn{1}{|c}{
				22} & \multicolumn{1}{c}{42} & \multicolumn{1}{|c}{40} & \multicolumn{1}{c}{
				70} \\ 
			\multicolumn{1}{c}{ave. deg.} & \multicolumn{1}{|c}{0.960} & 
			\multicolumn{1}{c}{4.807} & \multicolumn{1}{|c}{0.920} & \multicolumn{1}{c|}{
				5.133} & \multicolumn{1}{|c}{1.927} & \multicolumn{1}{c}{5.620} & 
			\multicolumn{1}{|c}{1.950} & \multicolumn{1}{c}{5.810} \\ \hline
			\multicolumn{1}{c}{true GC ($c=0.3$)} & \multicolumn{1}{|c}{0.056} & 
			\multicolumn{1}{c}{0.073} & \multicolumn{1}{|c}{0.052} & \multicolumn{1}{c|}{
				0.072} & \multicolumn{1}{|c}{0.078} & \multicolumn{1}{c}{0.068} & 
			\multicolumn{1}{|c}{0.078} & \multicolumn{1}{c}{0.068} \\ 
			\multicolumn{1}{c}{true GC ($c=0.6$)} & \multicolumn{1}{|c}{0.233} & 
			\multicolumn{1}{c}{0.171} & \multicolumn{1}{|c}{0.210} & \multicolumn{1}{c|}{
				0.159} & \multicolumn{1}{|c}{0.260} & \multicolumn{1}{c}{0.163} & 
			\multicolumn{1}{|c}{0.262} & \multicolumn{1}{c}{0.167} \\ \hline
		\end{tabular}%
	\end{center}
	\par
	\medskip 
	\parbox{6.2in}{\small
		
		Notes: The table gives the network characteristics of the graph that was used for the simulation study.
		The simulation study was based on a single generation of the random graphs.
		The E-R represents Erd\"{o}s-R\'{e}nyi Random Graph (E-R graph) 
		with probability equal to $p=\lambda /(n-1)$, where $\lambda $ is chosen from ${1,3,5}$ 
		and the B-A represents Barab\'{a}si-Albert random graph (B-A graph) 
		of preferential attachment, where the parameter $m$ refers to
		the number of links each new node forms with other existing nodes.
		The true GCs were computed from one million simulations using the single realization of the random graphs.
		
		\medskip \medskip \medskip}
\end{table}
\medskip

This algorithm generates $\{Y_i\}_{i=1}^{n}$ with positive graph
concordance when $c>0$. When $c=0$, this series is i.i.d. from $N(0,1)$. As
for the parameter $c$, we consider values of $0,\ 0.3,$ and $0.6.$ Note that 
$c$ does not necessarily represent the correlation between $Y_i$ and $%
Y_{j} $ for $ij\in E_n$, because we keep adding $c\times Z_{s}$ to $Y_i$ as
we run along the neighbors of vertex $i$.

For the confidence intervals, the E-R graphs and B-A graphs were generated
once, and based on these graphs, the true GCs were computed using one
million Monte Carlo simulations. The graph characteristics and the true GCs
are presented in Table 1.

The number of Monte Carlo simulations in the investigation of the finite
sample properties was set to be 5,000 and the number of random permutations
used to construct critical values was set to be 1,000.

\begin{table}[t]
	\caption{The Empirical Coverage Probability of Permutation-Based Confidence
		Intervals of the GC\ at 95\% Nominal Level}
	\begin{center}
		\begin{tabular}{cccccccc}
			\hline\hline
			&  &  & E-R &  &  & B-A &  \\ \cline{3-8}
			$c$ &  & $\lambda =1$ & $\lambda =3$ & $\lambda =5$ & $m=1$ & $m=2$ & $m=3$
			\\ \hline
			0 & $n\ =300$ & \multicolumn{1}{|c}{0.9470} & 0.9512 & 0.9466 & 0.9528 & 
			0.9472 & 0.9496 \\ 
			& $n\ =600$ & \multicolumn{1}{|c}{0.9516} & 0.9534 & 0.9542 & 0.9476 & 0.9534
			& 0.9500 \\ \hline
			0.3 & $n\ =300$ & \multicolumn{1}{|c}{0.9518} & 0.9488 & 0.9496 & 0.9460 & 
			0.9534 & 0.9476 \\ 
			& $n\ =600$ & \multicolumn{1}{|c}{0.9494} & 0.9480 & 0.9508 & 0.9502 & 0.9514
			& 0.9416 \\ \hline
			0.6 & $n\ =300$ & \multicolumn{1}{|c}{0.9456} & 0.9462 & 0.9502 & 0.9488 & 
			0.9528 & 0.9498 \\ 
			& $n\ =600$ & \multicolumn{1}{|c}{0.9496} & 0.9500 & 0.9456 & 0.9528 & 0.9564
			& 0.9578 \\ \hline
		\end{tabular}%
	\end{center}
	\par
	\medskip 
	\parbox{6.2in}{\small
		
		Notes: The E-R represents Erd\"{o}s-R\'{e}nyi Random Graph (E-R graph) 
		with probability equal to $p=\lambda /(n-1)$, where $\lambda $ is chosen from ${1,3,5}$ 
		and the B-A represents Barab\'{a}si-Albert random graph (B-A graph) 
		of preferential attachment, where the parameter $m$ refers to
		the number of the links each new node forms with other existing nodes.
		
		\medskip \medskip \medskip}
\end{table}

\begin{table}[t]
	\caption{The Mean Length of Permutation-Based Confidence Intervals of the
		GC\ at 95\% Nominal Level}
	\begin{center}
		\begin{tabular}{cccccccc}
			\hline\hline
			&  &  & E-R &  &  & B-A &  \\ \cline{3-8}
			$c$ &  & $\lambda =1$ & $\lambda =3$ & $\lambda =5$ & $m=1$ & $m=2$ & $m=3$
			\\ \hline
			0 & $n\ =300$ & \multicolumn{1}{|c}{0.2283} & 0.2040 & 0.1979 & 0.2515 & 
			0.2221 & 0.2031 \\ 
			& $n\ =600$ & \multicolumn{1}{|c}{0.1503} & 0.1428 & 0.1234 & 0.1749 & 0.1434
			& 0.1350 \\ \hline
			0.3 & $n\ =300$ & \multicolumn{1}{|c}{0.2276} & 0.2022 & 0.1952 & 0.2506 & 
			0.2211 & 0.2007 \\ 
			& $n\ =600$ & \multicolumn{1}{|c}{0.1498} & 0.1415 & 0.1214 & 0.1741 & 0.1429
			& 0.1326 \\ \hline
			0.6 & $n\ =300$ & \multicolumn{1}{|c}{0.2189} & 0.1876 & 0.1847 & 0.2482 & 
			0.2101 & 0.1930 \\ 
			& $n\ =600$ & \multicolumn{1}{|c}{0.1458} & 0.1322 & 0.1161 & 0.1774 & 0.1401
			& 0.1280 \\ \hline
		\end{tabular}%
	\end{center}
	\par
	\medskip 
	\parbox{6.2in}{\small
		
		Notes: The mean length of the confidence intervals is computed from 5,000 Monte Carlo simulations.
		
		\medskip}
\end{table}

\begin{table}[t]
	\caption{The Empirical Coverage Probability of Confidence Intervals of the
		GC\ from Asymptotic Normal Distribution at 95\% Nominal Level}
	\begin{center}
		\begin{tabular}{cccccccc}
			\hline\hline
			&  &  & E-R &  &  & B-A &  \\ \cline{3-8}
			$c$ &  & $\lambda =1$ & $\lambda =3$ & $\lambda =5$ & $m=1$ & $m=2$ & $m=3$
			\\ \hline
			0 & $n\ =300$ & \multicolumn{1}{|c}{0.9292} & 0.8912 & 0.8228 & 0.9112 & 
			0.8262 & 0.7354 \\ 
			& $n\ =600$ & \multicolumn{1}{|c}{0.9416} & 0.9330 & 0.8786 & 0.9224 & 0.8732
			& 0.7914 \\ \hline
			0.3 & $n\ =300$ & \multicolumn{1}{|c}{0.9326} & 0.8860 & 0.8334 & 0.9072 & 
			0.8212 & 0.7404 \\ 
			& $n\ =600$ & \multicolumn{1}{|c}{0.9376} & 0.9282 & 0.8730 & 0.9264 & 0.8614
			& 0.7938 \\ \hline
			0.6 & $n\ =300$ & \multicolumn{1}{|c}{0.9268} & 0.8884 & 0.8300 & 0.9080 & 
			0.8236 & 0.7518 \\ 
			& $n\ =600$ & \multicolumn{1}{|c}{0.9414} & 0.9284 & 0.8668 & 0.9322 & 0.8848
			& 0.8018 \\ \hline
		\end{tabular}%
	\end{center}
	\par
	\medskip 
	\parbox{6.2in}{\small
		
		Notes: The confidence intervals from asymptotic normal distribution are not stable.
		For example, see the case of E-R graphs with $\lambda = 5$ and $c = 0$, where the coverage
		probability is only 0.8228 (with $n = 300$) and 0.8786 (with $n = 600$), when the nominal
		coverage probability is 0.95. Also see the case of B-A graphs 
		with $m = 3$ with $c=0$, where the coverage probability is only 0.7354  (with $n = 300$) and 0.7914 (with $n = 600$).
		
		\medskip}
\end{table}

\begin{table}[t]
	\caption{The Mean Length of Confidence Intervals of the
		GC from Asymptotic Normal Distribution at 95\% Nominal Level}
	\begin{center}
		\begin{tabular}{cccccccc}
			\hline\hline
			&  &  & E-R &  &  & B-A &  \\ \cline{3-8}
			$c$ &  & $\lambda =1$ & $\lambda =3$ & $\lambda =5$ & $m=1$ & $m=2$ & $m=3$
			\\ \hline
			0 & $n\ =300$ & \multicolumn{1}{|c}{0.2111} & 0.1620 & 0.1205 & 0.2125 & 
			0.1300 & 0.0889 \\ 
			& $n\ =600$ & \multicolumn{1}{|c}{0.1439} & 0.1303 & 0.0895 & 0.1586 & 0.1006
			& 0.0702 \\ \hline
			0.3 & $n\ =300$ & \multicolumn{1}{|c}{0.2104} & 0.1606 & 0.1191 & 0.2119 & 
			0.1294 & 0.0876 \\ 
			& $n\ =600$ & \multicolumn{1}{|c}{0.1435} & 0.1292 & 0.0881 & 0.1579 & 0.1000
			& 0.0689 \\ \hline
			0.6 & $n\ =300$ & \multicolumn{1}{|c}{0.2023} & 0.1490 & 0.1126 & 0.2096 & 
			0.1227 & 0.0844 \\ 
			& $n\ =600$ & \multicolumn{1}{|c}{0.1396} & 0.1206 & 0.0841 & 0.1606 & 0.0981
			& 0.0667 \\ \hline
		\end{tabular}%
	\end{center}
	\par
	\medskip 
	\parbox{6.2in}{\small
		
		Notes: The mean length of the confidence intervals is computed from 5,000 Monte Carlo simulations.
		
		\medskip}
\end{table}

\subsection{Results and Discussions}

The empirical coverage probabilities of permutation-based confidence
intervals are reported in Table 2. The probabilities were computed for
nominal level $95\%$. When $c=0$ or $c=0.3,$ the finite sample coverage
probabilities are close to 0.95. However, when $c=0.6,$ the coverage
probabilities are slightly lower than 0.95 when the graph becomes more dense.

Table 3 reports the mean length of the confidence intervals from 5,000
simulations. The lengths of the confidence intervals are not comparable
directly across different graphs and different $c$'s, because the GCs are
different. However, we can see that there is a substantial reduction in the
mean length of the confidence intervals when we increase the sample size
from 300 to 600, indicating more information from more samples. The
reduction in the mean length is most substantial when the graph has fewer
edges.

Finally, we investigate the alternative method based on asymptotic normal
distribution of the test statistic. The results are reported in Table 4. Using asymptotic critical values leads to coverage probabilities
lower than the nominal level of 0.95. This under-coverage becomes more
severe when the graph becomes denser. For example, when $n=300$ and a B-A
graph with $m=3$ were used, the asymptotic coverage probability in the case of $c=0$ is only
0.7354, while that of the permutation confidence interval is 0.9496. This illustrates the merit of the permutation-based approach over the asymptotic method.

\section{Empirical Application:\ Social Networks in Indian Villages}

\subsection{Data}

The data used here is originated from the data used in \cite{Banerjee/Chandrasekhar/Duflo/Jackson:13:Science}. In 2006, the social network data were collected for 75 rural
villages in Karnataka, a state in southern India. The data contained
household information such as caste, village leader indicator, and various
amenities (e.g. roofing material, and type of latrine, etc.) The social
network data were collected along 12 dimensions in terms of visiting each
other, kinship, borrowing or lending money, rice, or kerosene, giving or
exchanging advice, or going to the place of prayer together. In 2007, a
microfinancing institution, Bharatha Swamukti Samsthe (BSS), began
operations in 43 of these villages, and collected data on who joined the
microfinancing program. (See \cite{Banerjee/Chandrasekhar/Duflo/Jackson:13:Science} for details.)

Out of 75 villages in the original sample, we selected villages that have at
least one minority caste/tribe household. The resulting villages contain a
total of 10,176 households. As for the definitions of the social nework, we
considered three definitions. The graph $G_{n,ALL}$ is defined to be the
social network where two households are linked if and only if any of the 12
dimensions in the social network data (as mentioned before) occurred between
the households. The graph $G_{n,EE}$ is defined to be the social nework
where two households are linked if and only if material exchanges
(borrowing/lending rice, kerosene or money) occurred, and the graph $%
G_{n,SC} $ is defined to be the social network where two households are
linked if and only if some social activities (such as seeking advice, or
going to the same temple or church, etc.) occurred.

\subsection{Application 1: Homophily along Castes/Tribes}

\cite{Jackson:14:JEP} observed homophily of households from the same Indian village
data, when the social network is defined based on borrowing/lending of
kerosene and rice. Similarly in \cite{Jackson:14:JEP}, we define a minority
caste/tribe household to be the household reported to be scheduled caste or
scheduled tribe, which is considered for affirmative action by the Indian
government (\cite{Jackson:14:JEP}). Here we use the GC\ of this minority indicators
along with social networks as a measure of homophily and provide formal
inference results.

\begin{table}[t]
	\caption{Social Network Characteristics from Data on Indian Villages}
	\label{table-mc1}
	\begin{center}
		\begin{tabular}{c|cccc}
			\hline\hline
			Networks & Size & Maximum Degree & Average Degree & Median Degree \\ \hline
			$G_{n,EE}$ & 10,176 & 56 & 5.7545 & 5 \\ 
			$G_{n,SC}$ & 10,176 & 87 & 8.4754 & 7 \\ 
			$G_{n,ALL}$ & 10,176 & 90 & 9.1824 & 8 \\ \hline
		\end{tabular}%
	\end{center}
	\par
	\medskip 
	\parbox{6.2in}{\small 
		
		Notes: The network $G_{n,EE}$ is defined based on observed material exchanges between households
		(such as borrowing or lending rice, kerosene or money). The network $G_{n,SC}$ is defined based on 
		social activities (such as seeking advice or going to the same temple or church). 
		The network $G_{n,ALL}$ is the union of the two networks $G_{n,EE}$ and $G_{n,SC}$.
		
		\medskip \medskip \medskip }
\end{table}

\begin{table}[t]
	\caption{The Estimated GC of Minority Castes/Tribes along the Social
		Networks }
	\begin{center}
		\begin{tabular}{c|ccccc}
			\hline\hline
			Networks & GC &  & Confidence & Intervals &  \\ \cline{4-6}
			&  &  & 99\% & 95\% & 90\% \\ \hline
			$G_{n,EE}$ & 0.5657 &  & [0.5141,0.6174] & [0.5271,0.6043] & [0.5331,0.5984]
			\\ 
			$G_{n,SC}$ & 0.5936 &  & [0.5385,0.6488] & [0.5512,0.6361] & [0.5583,0.6290]
			\\ 
			$G_{n,ALL}$ & 0.5813 &  & [0.5231,0.6395] & [0.5370,0.6256] & [0.5446,0.6180]
			\\ \hline
		\end{tabular}%
	\end{center}
	\par
	\medskip 
	\parbox{6.2in}{\small 
		
		Notes: Recall that the GC in the case of nonnegative pairwise concordance measures takes values in $[-1,1]$.
		The estimated GC's are positive significantly, indicating strong evidence for homophily 
		among households in the minority groups.
		
		\medskip \medskip \medskip }
\end{table}

The results are shown in Table 7. The estimated GC\ is from 0.5657 to
0.5936. The GC\ is very high, indicating strong homophily along the castes.
This strong homophily is robust to alternative definitions of the social
networks. In particular, the estimated GC\ is significantly away from zero,
regardless of whether the social network is that of economic exchanges or
that of social activities. The permutation-based confidence intervals are
narrow around the point estimates, supporting the strong homophily.

\subsection{Application 2: Graph Concordance of Microfinancing Decisions}

\cite{Banerjee/Chandrasekhar/Duflo/Jackson:13:Science} investigated the role of the social networks in
microfinancing decisions by households, modeling the microfinancing
decisions using a logit model. We search for evidence that the social network plays any role in
explaining the cross-sectional dependence of microfinancing decisions, without
modeling any parametric/nonparametric functional relationships among the
variables. The social network can play a role in affecting the
cross-sectional pattern of microfinancing decisions in two different ways.
First, the social network can serve as a channel for information diffusion
from one household to another. Second, the edges in the social network
merely reflect the common social context that the linked households share
and affect the mirofinancing decisions. A necessary implication in either
way would be that the GC of microfinancing decisions along the social
network should be positive. Otherwise, the role of the social network in
either way may lack empirical support. (The network statistics are in Table 8.)

\begin{table}[t]
	\caption{Social Network Characteristics from Data on Indian Villages}
	\begin{center}
		\begin{tabular}{cc|c|c|c|c}
			\hline\hline
			Data Set & Networks & Size & Maximum Degree & Average Degree & Median Degree
			\\ \hline
			& $G_{n,EE}$ & 7,537 & 56 & 5.80 & 5 \\ 
			A & $G_{n,SC}$ & 7,537 & 87 & 8.65 & 7 \\ 
			& $G_{n,ALL}$ & 7,537 & 90 & 9.39 & 8 \\ \hline
			& $G_{n,EE}$ & 5,915 & 56 & 5.88 & 5 \\ 
			B & $G_{n,SC}$ & 5,915 & 71 & 8.73 & 7 \\ 
			& $G_{n,ALL}$ & 5,915 & 80 & 9.47 & 8 \\ \hline
		\end{tabular}%
	\end{center}
	\par
	\medskip 
	\parbox{6.2in}{\small 
		
		Notes: The data set is from villages that have microfinancing data.
		First, Data Set A consists of households in the villages that have at least one minority household.
		Second, Data Set B consists of households in Data A that are in villages whose partiticpation rate
		in microfinancing is at least 10 percent. As before, the network $G_{n,EE}$ is defined based on 
		observed material exchanges between households
		(such as borrowing or lending rice, kerosene or money). The network $G_{n,SC}$ is defined based on 
		social activities (such as seeking advice or going to the same temple or church). 
		The network $G_{n,ALL}$ is the union of the two networks $G_{n,EE}$ and $G_{n,SC}$.
		
		\medskip}
\end{table}

\begin{table}[t]
	\caption{The Estimated GC of Microfinancing Decisions along the Social
		Networks}
	\begin{center}
		\begin{tabular}{cc|cccc}
			\hline\hline
			Data Set & Networks & \ \ \ \ GC \ \ \ \  & Confidence & Intervals &  \\ 
			\cline{4-6}
			&  &  & 99\% & 95\% & 90\% \\ \hline
			& $G_{n,EE}$ & 0.1386 & [0.1006,0.1765] & [0.1105,0.1666] & [0.1156,0.1615]
			\\ 
			A & $G_{n,SC}$ & 0.1302 & [0.0974,0.1630] & [0.1055,0.1549] & [0.1091,0.1513]
			\\ 
			& $G_{n,ALL}$ & 0.1257 & [0.0940,0.1574] & [0.1002,0.1512] & [0.1043,0.1471]
			\\ \hline
			& $G_{n,EE}$ & 0.1345 & [0.0872,0.1818] & [0.1022,0.1668] & [0.1087,0.1603]
			\\ 
			B & $G_{n,SC}$ & 0.1225 & [0.0881,0.1569] & [0.0957,0.1493] & [0.0999,0.1451]
			\\ 
			& $G_{n,ALL}$ & 0.1182 & [0.0832,0.1532] & [0.0924,0.1440] & [0.0964,0.1401]
			\\ \hline
		\end{tabular}%
	\end{center}
	\par
	\medskip 
	\parbox{6.2in}{\small
		
		Notes: Across different data sets and different definitions of social networks, the GC 
		of microfinancing decisions of households is positive along the social networks. It is interesting
		to see that GC is strongest when we considered $G_{n,EE}$, i.e., the social network defined 
		in terms of material exchanges among households in the minority groups.
		
		\medskip}
\end{table}

The estimation results of GC are reported in Table 9. The estimated GC\ ranges from 0.1182 to 0.1386. The confidence intervals of all the GCs are substantially away from zero. This suggests that the social network plays a role in explaining the micro-financing decisions.
Interestingly, the GC is strongest when we considered $G_{n,EE}$, i.e., the
social network defined in terms of material exchanges among households in the minority  groups.

\section{Conclusion}

This paper proposes a new measure that gauges the relevance of a given graph
in explaining the cross-sectional dependence of an observed random vector laid on the graph. Examples include homophily on social networks and network interference of treatments. This paper develops a method of permutation inference for the measure, and shows that this method exhibits two-tier validity.

The inference procedure does not require a complex set of conditions or a delicate choice of tuning parameters. Furthermore, the procedure does not presume any
functional relationships among observables, and hence is suitable for a
diagnostic purpose in investigating the role of a graph in explaining the cross-sectional dependence of an observed random vector.

\section{Appendix}

\subsection{A Matrix Formulation of Permutation-Based Inference}

The section gives a reformulation of test statistics and critical values using
vectors and matrices. While this formulation is less illuminating, it is
convenient for programming the inference procedure using programs such as
Matlab, because the programs are known to run faster when the algorithms are
properly vectorized.

Define $A_n$ to be the $n \times n$ adjacency matrix whose $(i,j)$-th entry is given by 
$1$ if $ij\in E_n$ and $0$ otherwise. Introduce the degree vector $\mathbf{%
	d}_n=A_n\mathbf{1}$, where $\mathbf{1}$ denotes the $n\times 1$ vector
of ones. Let $\mathbf{d}_{n,inv}$ be an $n\times 1$ vector whose $i$-th
entry is $d_n(i)^{-1}$ if $d_n(i)\geq 1$ and $0$ otherwise, where $d_n(i)$ is
the $i$-th entry of $\mathbf{d}_n$. Also let $\mathbf{d}_{n,inv}^{c}$ be
an $n\times 1$ vector whose $i$-th entry is $(n-1-d_n(i))^{-1}$.

Let $\mathbf{\hat{e}}$ be the $n$-dimensional vector whose $i$-th entry is
given by $\hat{e}_i$. Define 
\begin{equation*}
\mathbf{\hat{e}}_{A}=(A_n\mathbf{\hat{e})\odot d}_{n,inv}\text{ and\ }%
\mathbf{\hat{e}}_{A^{c}}=(A_n^{c}\mathbf{\hat{e})\odot d}_{n,inv}^{c},
\end{equation*}%
where $A_n^{c}$ is the $n\times n$ matrix whose $(i,j)$-th entry is given
by $1\{a_{ij}=0\}-1\{i=j\}$, $a_{ij}$ is the $(i,j)$-th entry of $A_n$, and%
$\ \odot $ denotes the Hadarmard product. Then we can write
\begin{equation*}
\hat{C}(G_n)=\mathbf{1}^{\top }(\mathbf{\hat{e}}\odot (\mathbf{\hat{e}}%
_{A}-\mathbf{\hat{e}}_{A^{c}}))/n.
\end{equation*}

Now, let us turn to $\hat{\sigma}_{+}^{2}.$ First, define the matrix\ $M_{0}=%
\mathbf{d}_n\mathbf{1}^{\top }-\mathbf{1d}_n^{\top }$ and let $M$ be the
matrix whose $(i,j)$-th entry is 1 if the $(i,j)$-th entry of $M_{0}$ is
zero, and 0 otherwise. Define%
\begin{equation*}
\mathbf{s}_n=M\mathbf{1,}
\end{equation*}%
so that the $i$-th entry of $\mathbf{s}_n$ is equal to the number of the
vertices having the same degree as vertex $i$. Define $\mathbf{\hat{e}}%
_{\tau }$ to be the subvector of $\mathbf{\hat{e}}$ that collects only the
entries having the corresponding entries of $\mathbf{s}_n$ greater than or
equal to 1. We similarly define the subvector $\mathbf{\hat{e}}_{A,\tau }$
of $\mathbf{\hat{e}}_{A}$ using $\mathbf{s}_n$. Define $M_{\tau }$ to be
the submatrix of $M$ that selects only entries $(i,j)$ of $M$ such that the $%
i$-th and the $j$-th entries of $\mathbf{s}_n$ are greater than or equal
to 1. Let $\mathbf{s}_{\tau }$ be the subvector of $\mathbf{s}_n$ that
selects only the entries that are greater than or equal to 1. Then let $%
n_{\tau }$ be the number of entries in $\mathbf{s}_{\tau }.$ Let $\mathbf{s}%
_{\tau ,inv}$ be the $n_{\tau }\times 1$ vector whose $i$-th entry is given
by the inverse of the $i$-th entry of $\mathbf{s}_{\tau }.$

Now, define$\ A_n^{\ast }$ to be the $n\times n$ matrix whose $(i,j)$-th
entry is 1 if the $(i,j)$-entry of $I_n+A_n+A_n^{2}+A_n^3$ is positive, and
0 otherwise. Let%
\begin{equation*}
\hat{\gamma}=\mathbf{1}^{\top }\left( \mathbf{\hat{e}}\odot \mathbf{\hat{%
		e}}_{A}\right) /n,
\end{equation*}%
and define%
\begin{eqnarray*}
	\mathbf{\hat{q}} &=&\mathbf{\hat{e}}\odot (\mathbf{\hat{e}}_{A}-\hat{\gamma}
	\mathbf{\hat{e})}-M_{\tau }(\mathbf{\hat{e}}_{\tau }\odot (\mathbf{\hat{e%
	}}_{A,\tau }-\hat{\gamma}\mathbf{\hat{e}}_{\tau }\mathbf{)})\odot 
	\mathbf{s}_{\tau ,inv}, \\
	\hat{\sigma}^{2} &=&\mathbf{1}^{\top}((\mathbf{\hat{q}\hat{q}}^{\top
	})\odot A_n^{\ast })\mathbf{1},\text{ and} \\
	\hat{\sigma}_{1}^{2} &=&\mathbf{\hat{q}}^{\top }\mathbf{\hat{q}}/n.
\end{eqnarray*}%
Then we construct%
\begin{equation*}
\hat{\sigma}_{+}^{2}=\hat{\sigma}^{2}1\left\{ \hat{\sigma}^{2}>0\right\} +%
\hat{\sigma}_{1}^{2}1\left\{ \hat{\sigma}^{2}\leq 0\right\} .
\end{equation*}
Once this procedure is programmed, the permutation version is simply
obtained by permuting the indices of $\mathbf{\hat{e}}$, and use the vector
instead of $\mathbf{\hat{e}}$ in the above computations.\medskip \medskip
\medskip \medskip

\subsection{Asymptotic Analysis through Conditioning on Permutations}

Once we fix a permutation $\pi \in \Pi _n$, the permuted observations
inherit their dependence structure and heterogeneity of marginals under the
relabeling of observations through $\pi$. When the relabeling of the observations does not alter the
limiting properties of the test statistic, the asymptotic analysis can be
performed by considering a fixed permutation asymptotics first and then by
taking care of the randomness of the permutation.

We introduce some notation. Let $\mathcal{X}_n\subset \mathbf{R}^{m}$ be a
measurable set from which random vector $X_n$ takes values. Let $\mathbb{T}%
_n$ be a finite collection of transforms $\tau _n:\mathcal{X}%
_n\rightarrow \mathcal{X}_n$, and $t_n:\mathcal{X}_n\rightarrow 
\mathbf{R}^{m}$ be a Borel measurable map. Suppose that $T_{1n}\in \mathbb{T}%
_n$ and $T_{2n}\in \mathbb{T}_n$ are independent and identically
distributed, and independent of $X_n$, where $T_{1n}$ is drawn from the
uniform distribution on $\mathbb{T}_n$.

The theorem below is Theorem 3.2 of \cite{Hoeffding:52:AMS} (adapted to 
notation here.)\medskip

\noindent \textbf{Theorem A1 (Hoeffding)}: \textit{Suppose that }$F$\textit{%
	\ is the CDF of a random vector in }$\mathbf{R}^{m}$\textit{\ and that for any continuity points }$t_1,t_2\in 
\mathbf{R}^{m}$\textit{\ of }$F,$
\begin{equation*}
\left\vert P\left\{ \left[ 
\begin{array}{c}
t_n(T_{1n}X_n) \\ 
t_n(T_{2n}X_n)%
\end{array}%
\right] \leq t\right\} -P\left\{ \left[ 
\begin{array}{c}
Q \\ 
Q'%
\end{array}%
\right] \leq t\right\} \right\vert =o(1),\textnormal{  } t =(t_1',t_2')',
\end{equation*}%
\textit{as }$n\rightarrow \infty $\textit{, where }$Q$ \textit{and} $%
Q'$\textit{\ are independent and identifically distributed as }$F$.

\textit{Then for any }$\eta >0$\textit{\ and for any continuity points }$t_1,t_2\in 
\mathbf{R}^{m}$\textit{\ of }$F,$
\begin{equation*}
P\left\{ \left\vert P\left\{ t_n(T_nX_n)\leq t|X_n\right\}
-F(t)\right\vert >\eta \right\} \rightarrow 0,\textnormal{  } t =(t_1',t_2')',
\end{equation*}%
\textit{as }$n\rightarrow \infty $.\medskip

Let $q_n:\mathcal{X}_n\rightarrow \mathbf{R}^{m}$ be a Borel measurable
map and let $\lambda _n:\mathcal{X}_n\rightarrow \mathbf{R}^{m}$ be an $%
m $-dimensional positive definite matrix-valued map. Define for $\tau_n \in \mathbb{T}_n$,
\begin{equation*}
h_{\tau_n}=Var\left( q_n(T_nX_n)|T_n = \tau_n\right) \text{.}
\end{equation*}%
We introduce the following lemma.\medskip

\noindent \textbf{Lemma A1}: \textit{Suppose that }$T_n$\textit{\ is
	independent of }$X_n$\textit{\ and is drawn from the uniform distribution
	on }$\mathbb{T}_n$.\textit{\ Suppose further that for each }$\varepsilon
\in (0,1)$, \textit{there exist a sequence of subsets }$\mathbb{T}%
_n(\varepsilon )\subset \mathbb{T}_n$\textit{\ and }$\mathbb{\tilde{T}}%
_n(\varepsilon )\subset \mathbb{T}_n\times \mathbb{T}_n$ \textit{such
	that }$|\mathbb{T}_n(\varepsilon )|/|\mathbb{T}_n|\rightarrow 1$\textit{%
	\ and }$|\mathbb{\tilde{T}}_n(\varepsilon )|/|\mathbb{T}%
_n|^{2}\rightarrow 1$ \textit{as }$n\rightarrow \infty $, \textit{and a
	continuous, strictly increasing function }$\varphi :[0,\infty )\rightarrow \lbrack
0,\infty )$\textit{\ (not depending on }$n$\textit{)} \textit{such that }$%
\varphi (0)=0$\textit{, where the sets }$\mathbb{T}_n(\varepsilon )$%
\textit{\ and }$\mathbb{\tilde{T}}_n(\varepsilon )$ \textit{and the
	function }$\varphi $ \textit{satisfy the following conditions (C1) and (C2)}.
\medskip

\noindent (C1) \textit{For any fixed sequences }$(\tau _{1n},\tau _{2n})\in 
\mathbb{\tilde{T}}_n(\varepsilon )$,
\begin{equation*}
\sup_{t\in \mathbf{R}^{2m}}\left\vert P\left\{ \left[ 
\begin{array}{c}
h_{\tau _{1n}}^{-1/2}q_n(\tau _{1n}X_n) \\ 
h_{\tau _{2n}}^{-1/2}q_n(\tau _{2n}X_n)%
\end{array}%
\right] \leq t\right\} -\Phi _{2m}(t)\right\vert \leq \varphi (\varepsilon
)+o(1),
\end{equation*}%
\textit{as }$n\rightarrow \infty $\textit{, where }$\Phi _{2m}$\textit{\ is
	the CDF of }$N(0,I_{2m}).$

\noindent (C2) \textit{For any fixed sequence }$\tau _n\in \mathbb{T}%
_n(\varepsilon )$,%
\begin{equation}
\left\Vert h_{\tau _n}^{-1/2}\left( \lambda _n^{1/2}(\tau
_nX_n)-h_{\tau _n}^{1/2}\right) \right\Vert \leq \varphi (\varepsilon
)+o_{P}(1),  \label{ineq}
\end{equation}%
\textit{as }$n\rightarrow \infty $.\medskip

\textit{Then for any }$\eta >0$, \textit{%
	as }$n\rightarrow \infty ,$%
\begin{equation}
\sup_{t\in \mathbf{R}^{m}} P\left\{ \left\vert P\left\{ \lambda
_n^{-1/2}(T_nX_n)q_n(T_nX_n)\leq t|X_n\right\} -\Phi_m
(t)\right\vert >\eta \right\} \rightarrow 0,  \label{cv7}
\end{equation}%
\textit{where }$\Phi_m $\textit{\ is the CDF of\ }$N(0,I_m).$\medskip

\noindent \textbf{Proof:} Define $\Delta _n(\tau _n)=h_{\tau
	_n}^{-1/2}\{\lambda _n^{1/2}(\tau _nX_n)-h_{\tau _n}^{1/2}\}$ and
write%
\begin{eqnarray}
&&\{\lambda _n^{-1/2}(\tau _nX_n)-h_{\tau _n}^{-1/2}\}q_n(\tau
_nX_n)  \label{cv5} \\
&=&\{\lambda _n^{-1/2}(\tau _nX_n)-h_{\tau _n}^{-1/2}\}h_{\tau
	_n}^{1/2}h_{\tau _n}^{-1/2}q_n(\tau _nX_n).  \notag
\end{eqnarray}%
Define $1_n(X_n) =1\{|| h_{\tau_n}^{-1/2}q_n(\tau_n X_n)|| \le 1/\varphi^{1/4}(\varepsilon) \}$. Since 
\begin{equation*}
\{\lambda _n^{-1/2}(\tau _nX_n)-h_{\tau _n}^{-1/2}\}h_{\tau
	_n}^{1/2}=-(\Delta _n(\tau _n)+I_{2m})^{-1}\Delta _n(\tau _n),
\end{equation*}%
Condition (C2) implies that for any $\tau_n \in \mathbb{T}
_n(\varepsilon)$,
\begin{equation*}
\left\Vert \{\lambda _n^{-1/2}(\tau _nX_n)-h_{\tau
	_n}^{-1/2}\}q_n(\tau _nX_n)\right\Vert 1_n(X_n) \leq \varphi^{3/4}
(\varepsilon )+o_{P}(1),
\end{equation*}%
as $n\rightarrow \infty .$ Therefore, for
any sequence of pairs $(\tau _{1n},\tau _{2n})$ from $\bar{\mathbb{T}}%
_n(\varepsilon) \equiv \mathbb{T}_n^2(\varepsilon) \cap \tilde{\mathbb{T}}_n(\varepsilon)$, and for any $t\in \mathbf{R}^{2m},$
\begin{eqnarray*}
	&& P\left\{ Z^*_{2n}(\tau _{1n},\tau _{2n}) \leq t-\varphi^{3/4}(\varepsilon) \mathbf{1}_{2m} \right\}-b_n(\varepsilon;\tau_{1n})-b_n(\varepsilon;\tau_{2n})\\
	&&\le P\left\{ Z_{2n}(\tau _{1n},\tau _{2n})\leq t\right\}\\
	&& \le P\left\{ Z^*_{2n}(\tau _{1n},\tau _{2n}) \leq t+\varphi^{3/4}(\varepsilon) \mathbf{1}_{2m} \right\}+b_n(\varepsilon;\tau_{1n})+b_n(\varepsilon;\tau_{2n}),
\end{eqnarray*}
where $\mathbf{1}_{2m}$ is the $2m$-dimensional vector of ones, $b_n(\varepsilon;\tau_n) \equiv P\{|| h_{\tau_n}^{-1/2}q_n(\tau_n X_n)|| > 1/\varphi^{1/4}(\varepsilon) \}$,
\begin{eqnarray*}
	Z^*_{2n}(\tau _{1n},\tau _{2n}) &\equiv& \left[ 
	\begin{array}{c}
		h_{\tau_{1n}}^{-1/2}q_n(\tau _{1n}X_n) \\ 
		h_{\tau_{2n}}^{-1/2}q_n(\tau _{2n}X_n)%
	\end{array}%
	\right], \textnormal{ and }\\
	Z_{2n}(\tau _{1n},\tau _{2n}) &\equiv& \left[ 
	\begin{array}{c}
		\lambda _n^{-1/2}(\tau _{1n}X_n)q_n(\tau _{1n}X_n) \\ 
		\lambda _n^{-1/2}(\tau _{2n}X_n)q_n(\tau _{2n}X_n)%
	\end{array}
	\right].
\end{eqnarray*}
Since $\mathbf{E}[|| h_{\tau_n}^{-1/2}q_n(\tau_n X_n)||^2] = m$, using Chebyshev's inequality, we deduce that for all $\tau_n \in \mathbb{T}_n(\varepsilon)$, $
b_n(\varepsilon;\tau_n) \le m \varphi^{1/2}(\varepsilon).$
Hence by (C1),
\begin{eqnarray*}
	&& \Phi(t-\varphi^{3/4}(\varepsilon) \mathbf{1}_{2m})-2m\varphi^{1/2}(\varepsilon)-\varphi(\varepsilon)+o(1)\\
	&& \le P\left\{ Z_{2n}(\tau _{1n},\tau _{2n})\leq t\right\}\\
	&& \le \Phi(t+\varphi^{3/4}(\varepsilon) \mathbf{1}_{2m})+2m\varphi^{1/2}(\varepsilon)+\varphi(\varepsilon)+o(1),
\end{eqnarray*}
as $n \rightarrow \infty$, or, for some constant $C_2>0$,
\begin{equation}
\left\vert P\left\{ Z_{2n}(\tau _{1n},\tau _{2n})\leq t\right\} -\Phi
_{2m}(t)\right\vert \leq C_2\varphi^{1/2}(\varepsilon )+o(1),  \label{bd9}
\end{equation}%
as $n\rightarrow \infty $. (Here the terms $o(1)$ are uniform over $t \in \mathbf{R}^{2m}$.)

Now, note that by Jensen's inequality,%
\begin{eqnarray}
\label{bd78} \quad \quad \left\vert P\left\{ Z_{2n}(T_{1n},T_{2n})\leq t\right\} -\Phi
_{2m}(t)\right\vert   
\leq \mathbf{E}\left\vert P\left\{ Z_{2n}(T_{1n},T_{2n})\leq
t|T_{1n},T_{2n}\right\} -\Phi _{2m}(t)\right\vert.
\end{eqnarray}%
Since $T_{1n}$ and $T_{2n}$ are drawn from the uniform distribution on $%
\mathbb{T}_n$ independently of $X_n$, the last expectation is equal to%
\begin{eqnarray*}
	&&\frac{1}{|\mathbb{T}_n|^{2}}\sum_{\tau _{1,}\tau _{2}\in \mathbb{T}%
		_n}\left\vert P\left\{ Z_{2n}(\tau _{1},\tau _{2})\leq t\right\} -\Phi
	_{2m}(t)\right\vert \\
	&\leq &\max_{\tau _{1},\tau _{2}\in \bar{\mathbb{T}}_n(\varepsilon)}\left\vert P\left\{ Z_{2n}(\tau _{1},\tau _{2})\leq t\right\} -\Phi
	_{2m}(t)\right\vert +\frac{|\mathbb{T}_n|^{2}-\left\vert \bar{\mathbb{T}}
		_n(\varepsilon)\right\vert }{|\mathbb{T}_n|^2}.
\end{eqnarray*}%
The last term is $o(1)$ as $n\rightarrow \infty $. Since $\bar{\mathbb{T}}_n(\varepsilon)$ is a finite set, there exists a sequence $(\tau
_{1n},\tau _{2n})$ that achieves the maximum above, and by (\ref{bd9}), the absolute value on the left hand side of (\ref{bd78}) vanishes, as we send $n\rightarrow \infty $ and $\varepsilon
\rightarrow 0$, fulfilling the condition
of Theorem A1. Hence (\ref{cv7}) follows. $\blacksquare $\medskip

The convergence in Condition (C1) usually follows because for each fixed
permutation, the dependence structure and marginal heterogeneity are
inherited under the relabeling of the observations. The main requirement in
Condition (C1) is that the correlation between $q_n(\tau _{1n}X_n)$ and $%
q_n(\tau _{2n}X_n)$ is small for most $\tau _{1n}$'s and $\tau _{2,n}$%
's. As for Condition (C2), the matrix $h_{\tau _n}$ is usually the
conditional variance of $q_n(T_nX_n)$ given $T_n=\tau _n$ such
that for each fixed $\tau _n\in \mathbb{T}_n,$ $h_{\tau _n}^{-1/2}q_n(\tau _nX_n)\overset{d}{\rightarrow }N(0,I_{m})$, 
and the matrix $\lambda _n(\tau _nX_n)$ represents its consistent
estimator for each fixed $\tau _n\in \mathbb{T}_n$. Then Lemma A1 says
that the permutation distribution of $\lambda
_n^{-1/2}(T_nX_n)q_n(T_nX_n)$ is asymptotically the same as $%
N(0,I_{m})$.

\subsection{A Synopsis of the Proof of Theorem 1}

In this subsection, we provide a synopsis of the proof of Theorem 1. The
full proofs of the results are found in the supplemental note of this paper
(attached to this paper at the end).

First, define%
\begin{equation*}
C_{\pi }(G_n)=\frac{1}{n}\sum_{i\in N_n}\mathbf{E}\left[ e_{\pi
	(i)}a_{i,\pi }|\pi \right] -\frac{1}{n}\sum_{i\in N_n}\mathbf{E}\left[
e_{\pi (i)}a_{i,\pi }^{c}|\pi \right] ,
\end{equation*}%
where $a_{i,\pi }=\frac{1}{d_n(i)}\sum_{j\in N_n(i)}e_{\pi (j)}$ if $%
d_n(i)\geq 1$, and $a_{i,\pi }=0$\ otherwise, and 
\begin{equation*}
a_{i,\pi }^{c}=\frac{1}{|N_n\backslash \overline{N}_n(i)|}\sum_{j\in
	N_n\backslash \overline{N}_n(i)}e_{\pi (j)}.
\end{equation*}%
Also let%
\begin{eqnarray*}
	\zeta _n = \frac{1}{\sqrt{n}}\sum_{i\in N_n}\left( q_i-\mathbf{E}%
	[q_i]\right) \text{, and } 
	\zeta _{n,\pi } = \frac{1}{\sqrt{n}}\sum_{i\in N_n}\left( q_{i,\pi }-%
	\mathbf{E}[q_{i,\pi }|\pi ]\right) .
\end{eqnarray*}%
Then we can show the following%
\begin{eqnarray*}
	\sqrt{n}\{\hat{C}(G_n)-C(G_n)\} &=& \zeta _n+o_{P}(1), \text{ and }\\
	\sqrt{n}\{\hat{C}_{\pi }(G_n)-C_{\pi }(G_n)\} &=& \zeta _{n,\pi }+o_{P}(1),
\end{eqnarray*}
uniformly over $\pi \in \Pi_n$. Furthermore, after some algebra, we can show that 
\begin{equation*}
\hat{\sigma}_n^{2}=Var\left( \zeta _n\right) +o_{P}(1),
\end{equation*}
Once asymptotic linear representation is obtained, we turn to asymptotic
normality. It is not hard to see that $q_i$'s have as dependency graph the
graph $G_{n,2}=(N_n,E_{n,3})$, where $ij\in E_{n,3}$ if and only if $i$
and $j$ are within three edges from each other. Using the Central Limit
Theorem for a sum of random variables having a dependency graph (e.g.
Theorem 2.4 of \cite{Penrose:03:RandomGeometricGraphs}), we can show that%
\begin{equation*}
T\overset{d}{\rightarrow }N(0,1).
\end{equation*}%
The convergence rate in the normal approximation is $O(d_{mx,n,3}/n^{1/4})$ which explains the rate requirement in (\ref{cd6}).

From here upon, we focus on $T_{\pi }$. For this, we use Lemma A1. We take $X_n\ $and $q_n(X_n)$ in
Lemma A1 to be $Y$ and $\zeta _{n,\pi }$ respectively, and take $\mathbb{T}%
_n$ to be $\Pi _n$ with the identification: $\tau _nX_n=(Y_{\pi
	(i)})_{i\in N_n}$. Also take $\lambda _n(X_n)$ and $h_{\tau _n}^{2}$
in Lemma A1 to be $\hat{\sigma}_{\pi }^{2}$ and $h_{n,\pi }^{2}\equiv
Var(\zeta _{n,\pi }|\pi )$.

For asymptotic normality of $\zeta _{n,\pi }$, we first show that $h_{n,\pi }^{2}$ is asymptotically nondegenerate
for most permutations $\pi $'s\footnote{When we say a statement holds ``for most permutations" here, it means that the statement holds for all but an asymptotically negligible fraction of permutations.}, that is,
\begin{equation}
\frac{1}{\left\vert \Pi _n\right\vert }\sum_{\pi \in \Pi _n}1\left\{
h_{n,\pi }^{2}>\frac{c}{2} \right\} \rightarrow 1 \text{, as }%
n\rightarrow \infty \text{.}  \label{ineq1}
\end{equation}%
Then we apply the Central Limit Theorem to $\zeta _{n,\pi }$. For this, note
that $q_{i,\pi }$'s have as dependency graph the graph $G_{n,3,\pi}=(N_n,E_{n,3,\pi })$, where $ij\in E_{n,3,\pi }$ if and only if $\pi (i)$and $\pi(j)$ are within two edges from each other. Therefore we can apply
the Berry-Esseen bound in Theorem 2.4 of \cite{Penrose:03:RandomGeometricGraphs} \textit{for each
	fixed sequence of permutations }$\pi \in \Pi _n$. The bound is uniform over $\pi \in \Pi _n$, giving this convergence in distribution uniform over $\pi \in \Pi _n$. Note that $Cov(\zeta _{n,\pi _{1}},\zeta
_{n,\pi _{2}}|\pi _{1},\pi _{2})$ is small for most permutations if $%
Cov(q_{i,\pi _{1},}q_{i,\pi _{2}}|\pi _{1},\pi _{2})$ is so. The latter
asymptotic negligibility follows, because for most permutations $\pi _{1}$
and $\pi _{2}$, $q_{i,\pi _{1}}$ and $q_{i,\pi _{2}}$ are independent.

Unlike $T,$ the permutation test statistic $T_{\pi }$ does not involve a
centering by $C_{\pi }(G_n)$. (Compare (\ref{T}) and (\ref{Tp}).)
Centering by $C_{\pi }(G_n)$ is not possible because it is unknown.
However, the centering is not needed, because $C_{\pi }(G_n)$ is
asymptotically negligible for most $\pi $'s, regardless of whether $%
C(G_n)=0$ or not. Indeed for any $\varepsilon >0,$%
\begin{equation}
\frac{1}{\left\vert \Pi _n\right\vert }\sum_{\pi \in \Pi _n}1\left\{
\left\vert \sqrt{n}C_{\pi }(G_n)\right\vert >\varepsilon \right\} =o(1)%
\text{, as }n\rightarrow \infty \text{.}  \label{ineq2}
\end{equation}%
This result comes from the fact that for most permutations, the correlation
between $e_{\pi (i)}$ and $a_{i,\pi }$ (or $a_{i,\pi }^{c}$) is zero under
the local dependence set-up when $d_{mx,n}$ is small relative to $n$. Once
we take $\Pi _n(\varepsilon )$ to be the set of permutations satisfying
the inequalities in the indicators of (\ref{ineq1}) and (\ref{ineq2}), we
obtain (C1) of Lemma A1.

To show (C2) of Lemma A1, we need to deal with $\hat{\sigma}_{\pi }^{2}$. We
show that $\hat{\sigma}_{\pi }^{2}$ is close to $h_{n,\pi }^{2}$ for most $%
\pi $'s in large samples. For this, we first define%
\begin{equation}
\label{q_j}
q_{i,\pi }^{\ast }=\frac{1}{|S_n(i)|}\sum_{j\in S_n(i)}\mathbf{E}\left[
q_{j,\pi }|\pi \right] ,
\end{equation}
and let\ $\eta _{i,\pi }=q_{i,\pi }-\mathbf{E}\left[ q_{i,\pi }|\pi \right] $
and $\tilde{\eta}_{i,\pi }=q_{i,\pi }-q_{i,\pi }^{\ast }$. Let $\sigma
_{n,\pi }^{2}$ be the population version of $\hat{\sigma}_{\pi }^{2}\ $%
(conditional on fixed $\pi $) defined by%
\begin{equation*}
\sigma _{n,\pi }^{2}=\frac{1}{n}\sum_{i_{1}\in N_n}\mathbf{E}\left[ \tilde{%
	\eta}_{i_{1},\pi }^{2}|\pi \right] +\frac{1}{n}\sum_{i_{1}\in
	N_n}\sum_{i_{2}\in N_{n,3}(i_{1})}\mathbf{E}\left[ \tilde{\eta}_{i_{1},\pi
}\tilde{\eta}_{i_{2},\pi }|\pi \right] .
\end{equation*}%
Then after some algebra, we can show that for most permutations,%
\begin{equation*}
\hat{\sigma}_{\pi }^{2}=\sigma _{n,\pi }^{2}+o_{P}(1).
\end{equation*}%
It remains to show that $\sigma _{n,\pi }^{2}$ is close to $h_{n,\pi }^{2}$.
First we write%
\begin{eqnarray}
h_{n,\pi }^{2}-\tilde{h}_{n,\pi }^{2} &=&\frac{1}{n}\sum_{i_{1}\in
	N_n}\sum_{i_{2}\in N_n\backslash \{i_{1}\}}\mathbf{E}\left[ \eta
_{i_{1},\pi }\eta _{i_{2},\pi }|\pi \right] \text{ and}  \label{eq56} \\
\sigma _{n,\pi }^{2}-\tilde{\sigma}_{n,\pi }^{2} &=&\frac{1}{n}%
\sum_{i_{1}\in N_n}\sum_{i_{2}\in N_{n,3}(i_{1})}\mathbf{E}\left[ \tilde{%
	\eta}_{i_{1},\pi }\tilde{\eta}_{i_{2},\pi }|\pi \right] ,  \notag
\end{eqnarray}%
where $\tilde{h}_{n,\pi }^{2}=\frac{1}{n}\sum_{i_{1}\in N_n}\mathbf{E}\left[ \eta
_{i_{1},\pi }^{2}|\pi \right] \text{ and\ }\tilde{\sigma}_{n,\pi }^{2}=\frac{%
	1}{n}\sum_{i_{1}\in N_n}\mathbf{E}\left[ \tilde{\eta}_{i_{1},\pi }^{2}|\pi %
\right].$ Again the double sums on the right hand side in (\ref{eq56}) are
asymptotically negligible because $\eta _{i_{1},\pi }$ and $\eta _{i_{2},\pi
}$ are uncorrelated for most permutations $\pi $. It remains to compare $%
\tilde{h}_{n,\pi }^{2}$ and $\tilde{\sigma}_{n,\pi }^{2}$. Observe that%
\begin{eqnarray}
\mathbf{E}\left[ \eta _{i,\pi }^{2}-\tilde{\eta}_{i,\pi }^{2}|\pi \right]
&=&\left( q_{i,\pi }^{\ast }-\mathbf{E}\left[ q_{i,\pi }|\pi \right] \right) 
\mathbf{E}\left[ 2q_{i,\pi }-\mathbf{E}\left[ q_{i,\pi }|\pi \right]
-q_{i,\pi }^{\ast }|\pi \right]  \label{ineq6} \\
&=&\left( q_{i,\pi }^{\ast }-\mathbf{E}\left[ q_{i,\pi }|\pi \right] \right)
(\mathbf{E}\left[ q_{i,\pi }|\pi \right] -q_{i,\pi }^{\ast })  \notag \\
&=&-\left( q_{i,\pi }^{\ast }-\mathbf{E}\left[ q_{i,\pi }|\pi \right]
\right) ^{2}.  \notag
\end{eqnarray}%
However, for most $\pi $'s, $e_{\pi (i)}$ and $a_{i,\pi }$ are independent.
Hence for such permutations, we have 
\begin{equation*}
\mathbf{E}[q_{i,\pi}|\pi] = \mathbf{E}\left[ e_{\pi (i)}|\pi \right] \mathbf{E}\left[ a_{i,\pi }|\pi %
\right] =\mathbf{E}\left[ e_{\pi (i)}|\pi \right] \frac{1}{d_n(i)}%
\sum_{j\in N_n(i)}\mathbf{E}\left[ e_{\pi (j)}|\pi \right] =0,
\end{equation*}%
because $\mathbf{E}\left[ e_i\right] =0$ for all $i\in N_n,$ and hence $
q_{i,\pi }^*=0$ from (\ref{q_j}). Therefore, $\tilde{h}_{n,\pi }^{2}$ and $\tilde{\sigma}_{n,\pi }^{2}$ are close to each other for most permutations.
Thus by Lemma A1, we obtain that for all $t\in \mathbf{R},$%
\begin{equation*}
P\left\{ \sqrt{n}\hat{C}_{\pi }/\hat{\sigma}_{\pi} \leq t|Y\right\}
=\Phi(t) + o_{P}(1).
\end{equation*}

\bibliographystyle{econometrica}
\bibliography{information_matching_networks_A2}
\pagebreak

\begin{center}
	\Large Supplemental Note for ``Measuring the Graph Concordance of Locally Dependent Observations" \medskip
	
	\normalsize
	
	Kyungchul Song\medskip
	
	\textit{Vancouver School of Economics, University of British Columbia}\medskip
	\medskip
\end{center}

\noindent {\Large 1. Introduction}\medskip

This supplemental note collects the mathematical proofs of the results in the paper ``Measuring the Graph Concordance of Locally Dependent Observations" by the author. The next section begins by introducing notation. In Section 2.1., preliminary results are provided. Then we move to the asymptotic linear representation of the graph concordance estimator and its permuted version in Section 2.2. From this, we prove asymptotic normality of the graph concordance estimator and its permuted version in Section 2.3. Section 2.4 is devoted to the consistency of asymptotic variance estimators and Section 2.5 to that of permuted versions. Using the results so far, and checking the conditions of Lemma A1 in the main paper, we obtain the proof of Theorem 1.\medskip \medskip
	
\noindent {\Large 2. Proofs}\medskip
	
	All the asymptotic statements in this section assume that $P\in \mathcal{P}
	_n(G_n;c,M)$ for all $n\geq 1$ for a given sequence of graphs $G_n,$
	and for given $(c,M)\in (0,\infty )^{2}$. All the auxiliary results from here on are under Assumptions 1 and 2. Throughout the proofs, the notation $\pi \in \Pi _n$ represents a random permutation with uniform distribution on $\Pi _n$ whenever it appears inside expectation or probability, and the notation $C$ an absolute constant that does not depend
	on $n\mathcal{\ }$or $P$.
	
	We introduce notation for uniform convergence in $\pi \in \Pi _n$. For any
	nonstochastic sequence of finite dimensional vectors $b_{n,P}(\pi )$, we say
	that $b_{n,P}(\pi )=o(1),$ $\Pi _n$-unif., or $b_{n,P}(\pi )=O(1),$ $\Pi
	_n$-unif., if, respectively, 
	\begin{equation*}
	\max_{\pi \in \Pi _n}||b_{n,P}(\pi )||=o(1)\text{ or }\max_{\pi \in \Pi
		_n}||b_{n,P}(\pi )||=O(1).
	\end{equation*}%
	For any sequence of random vectors of the form $g_{n,\pi }(Y)\in \mathbf{R}%
	^{d}$ with $g_{n,\pi }$ being an $\mathbf{R}^{d}$-valued Borel meansurable
	map on $\mathbf{R}^{n}$ indexed by $\pi \in \Pi _n$, and any nonstochastic
	positive sequence of numbers, $\{a_n\}_{n\geq 1}$, we write%
	\begin{eqnarray*}
		g_{n,\pi }(Y) &=&o_{P}(a_n),\text{ }\Pi _n\text{-unif., and} \\
		g_{n,\pi }(Y) &=&O_{P}(a_n),\text{ }\Pi _n\text{-unif.,}
	\end{eqnarray*}%
	if, respectively, 
	\begin{eqnarray*}
		\underset{n\rightarrow \infty }{\text{limsup}}\max_{\pi \in \Pi
			_n}P\left\{ \frac{\left\Vert g_{n,\pi }(Y)\right\Vert }{a_n}>\varepsilon
		|\pi \right\} &=&0\text{ for all }\varepsilon >0\text{, and} \\
		\underset{M_{1}\rightarrow \infty }{\text{lim}}\ \underset{n\rightarrow
			\infty }{\text{limsup}}\max_{\pi \in \Pi _n}P\left\{ \frac{\left\Vert
			g_{n,\pi }(Y)\right\Vert }{a_n}>M_{1}|\pi \right\} &=&0.
	\end{eqnarray*}%
	Furthermore, we write for a random vector $Z\in \mathbf{R}^{d}$,%
	\begin{equation*}
	g_{n,\pi }(Y)\overset{d}{\rightarrow }Z,\ \Pi _n\text{-unif.,}
	\end{equation*}%
	if for any continuity point $t\in \mathbf{R}^{d}$ of the CDF\ of $Z$,%
	\begin{equation*}
	\underset{n\rightarrow \infty }{\text{limsup}}\max_{\pi \in \Pi
		_n}\left\vert P\left\{ g_{n,\pi }(Y)\leq t|\pi \right\} -P\left\{ Z\leq
	t\right\} \right\vert =0.
	\end{equation*}
	
	Let $d_{av,n}$ denote the average degree of $G_n$, i.e., 
	\begin{equation*}
	d_{av,n}=\frac{1}{n}\sum_{i\in N_n}d_n(i).
	\end{equation*}%
	Certainly, we have $d_{av,n}\leq d_{mx,n}$. Also, recall the fact that $%
	|E_n|=nd_{av,n}$. (Note that according to our definition of $E_n$, we
	have $ij\in E_n$ if and only if $ji\in E_n$.) For any two subsets $%
	A_{1},A_{2}\subset N_n$, we say that $A_{1}$ and $A_{2}$ are \textit{%
		adjacent} if there exist one vertex from $A_{1}$ and another vertex from $%
	A_{2}$ that are adjacent. We write $i\sim j$ to mean that $i$ and $j$ are
	adjacent. We say that $i_{1}j_{1}$ and $i_{2}j_{2}$ (or $\{i_{1},j_{1}\}$
	and $\{i_{2},j_{2}\}$) are \textit{adjacent }if one of the vertices $i_{1}$
	and $j_{1}$ is adjacent to one of the vertices $i_{2}$ and $j_{2}$, and we
	write $i_{1}j_{1}\sim i_{2}j_{2}$.\medskip
	
	\noindent {\large 2.1. Preliminary Results}\medskip
	
	We begin with a lemma that shows that the average degree $d_{av,n}\ $is
	bounded away from zero.\medskip
	
	\noindent \textbf{Lemma B1:}\textit{\ There exists }$c_{1}>0$\textit{\ such
		that}%
	\begin{equation*}
	d_{av,n}>c_{1}\text{\textit{\ for all }}n\geq 1.
	\end{equation*}%
	\medskip
	
	\noindent \textbf{Proof:}\textit{\ }Suppose to the contrary that for any $%
	\varepsilon >0$, there exists $n$ such that $d_{av,n}<\varepsilon $. Then%
	\begin{equation*}
	d_{avi,n}=\frac{1}{n}\sum_{i\in N_n:d_n(i)\geq 1}\frac{1}{d_n(i)}\leq 
	\frac{1}{n}\sum_{i\in N_n:d_n(i)\geq 1}d_n(i)\leq d_{av,n}<\varepsilon
	.
	\end{equation*}%
	Since the choice of $\varepsilon >0$ is arbitrary, this violates the second condition in Assumption 1(ii). $\blacksquare $%
	\medskip
	
	\noindent \textbf{Lemma B2:}\textit{\ For any sequence of nonstochastic real
		Borel measurable functions }$\{g_i\}_{i=1}^{n}$\textit{\ on the real line
		such that\ for each }$n\geq 1,$%
	\begin{equation*}
	\max_{i \in N_n}Var(g_i(Y_i))<\infty ,
	\end{equation*}%
	\textit{we have}%
	\begin{equation}
	Var\left( \frac{1}{n}\sum_{i\in N_n}g_i(Y_i)\right) =O\left( \frac{%
		d_{av,n}}{n}\right) \text{.}  \label{st1}
	\end{equation}
	
	\textit{Furthermore, suppose that }$d_{av,n}/n\rightarrow 0$\textit{, as }$%
	n\rightarrow \infty $. \textit{Then for }$1\leq \lambda \leq 4,$%
	\begin{equation}
	\frac{1}{n}\sum_{i\in N_n}\left\vert \hat{e}_i-e_i\right\vert
	^{\lambda }=O_{P}\left( \frac{d_{av,n}^{\lambda /2}}{n^{\lambda /2}}\right) 
	\text{.}  \label{st2}
	\end{equation}%
	\medskip
	
	\noindent \textbf{Proof:}\textit{\ }As for (\ref{st1}), observe that%
	\begin{equation*}
	Var\left( \frac{1}{n}\sum_{i \in N_n}g_i(Y_i)\right) = \frac{1}{n^2}\sum_{i \in N_n} Var\left(g_i(Y_i)\right) + \frac{1}{n^{2}}%
	\sum_{ij\in E_n}Cov(g_i(Y_i),g_{j}(Y_{j}))=O\left( \frac{|E_n|}{n^{2}%
	}\right) .
	\end{equation*}%
	Since $|E_n|=nd_{av,n},$ the desired result follows.
	
	We turn to (\ref{st2}). Write%
	\begin{equation*}
	\sqrt{n}\left( \hat{v}^{2}-v_n^{2}\right) =\frac{1}{\sqrt{n}}%
	\sum_{i \in N_n} \left( Y_i^{2}-\mathbf{E}Y_i^{2}\right) -\sqrt{n}%
	\left( \overline{Y}^{2}-(\mathbf{E}Y_i)^{2}\right),
	\end{equation*}%
	and note that%
	\begin{equation*}
	\sqrt{n}\left( \overline{Y}^{2}-(\mathbf{E}Y_i)^{2}\right) =2\sqrt{n}\left( 
	\overline{Y}-\mathbf{E}Y_i\right) \mathbf{E}Y_i+\sqrt{n}\left( \overline{Y}-\mathbf{E}Y_i\right) ^{2}.
	\end{equation*}%
	The expected value of the last term is $O(d_{av,n}/\sqrt{n})$ by (\ref{st1}%
	) and by Assumption 1(iii). Since $e_i^2 - \mathbf{E}e_i^2 = (Y_i^2 - \mathbf{E}Y_i^2 - 2(Y_i - \mathbf{E}Y_i)\mathbf{E}Y_i)/v_n^2$, we find that
	\begin{equation}
	\mathbf{E}\left\vert \sqrt{n}\left( \hat{v}^{2}-v_n^{2}\right) -\frac{%
		v_n^{2}}{\sqrt{n}}\sum_{i \in N_n}\left( e_i^{2}-\mathbf{E}%
	e_i^{2}\right) \right\vert =O\left( \frac{d_{av,n}}{\sqrt{n}}\right) .
	\label{exp78}
	\end{equation}%
	By this and (\ref{st1}),%
	\begin{equation}
	\mathbf{E}\left\vert \hat{v}^{2}-v_n^{2}\right\vert \leq \mathbf{E}%
	\left\vert \frac{v_n^{2}}{n}\sum_{i \in N_n}\left( e_i^{2}-\mathbf{E}%
	e_i^{2}\right) \right\vert +O\left( \frac{d_{av,n}}{n}\right) =O\left( 
	\frac{d_{av,n}^{1/2}}{\sqrt{n}}\right) .  \label{bd61}
	\end{equation}%
	By the mean-value expansion, for some $\alpha _n\in \lbrack 0,1],$%
	\begin{equation}
	\hat{v}-v_n=\sqrt{\hat{v}^{2}}-\sqrt{v_n^{2}}=\frac{1}{2}\frac{\hat{v}%
		^{2}-v_n^{2}}{\sqrt{v_n^{2}+(1-\alpha _n)(\hat{v}^{2}-v_n^{2})}}%
	=O_{P}\left( \frac{d_{av,n}^{1/2}}{\sqrt{n}}\right) ,  \label{c}
	\end{equation}%
	where the last rate follows by the condition $d_{av,n}/n\rightarrow 0$, (\ref{bd61}), and Assumption 1(i). Using the second equality above, we write
	\begin{equation}
	\sqrt{n}v_n\left( \frac{1}{v_n}-\frac{1}{\hat{v}}\right) =\frac{1}{2}\frac{%
		\sqrt{n}(\hat{v}^{2}-v_n^{2})}{v_n^{2}+o_{P}(1)}=\frac{1}{2\sqrt{n}}%
	\sum_{i \in N_n}\left( e_i^{2}-\mathbf{E}e_i^{2}\right) +O_{P}\left( \frac{%
		d_{av,n}}{\sqrt{n}}\right) ,  \label{AL2}
	\end{equation}%
	by (\ref{exp78}).
	
	Since $\mathbf{E}[Y_i]$ is the same for all $%
	i\in N_n$ (Assumption 2(i)), if we let $\varepsilon _i=Y_i-\mathbf{E}%
	Y_i,$ we can write%
	\begin{equation}
	\hat{e}_i-e_i=-\frac{\bar{\varepsilon}}{\hat{v}}+\left( \frac{1}{\hat{v}}%
	-\frac{1}{v_n}\right) \varepsilon _i,  \label{dec67}
	\end{equation}%
	where $\bar{\varepsilon}=\frac{1}{n}\sum_{j\in N_n}\varepsilon _{j}$.
	Hence we bound%
	\begin{equation}
	\frac{1}{n}\sum_{i\in N_n}\left\vert \hat{e}_i-e_i\right\vert
	^{\lambda }\leq \frac{2^{\lambda -1}|\bar{\varepsilon}|^{\lambda }}{|\hat{v}%
		|^{\lambda }}+2^{\lambda -1}\left\vert \frac{1}{v_n}-\frac{1}{\hat{v}}%
	\right\vert ^{\lambda }\frac{1}{n}\sum_{i\in N_n}\left\vert \varepsilon
	_i\right\vert ^{\lambda }.  \label{bd671}
	\end{equation}%
	Since $\bar{\varepsilon}=O_{P}(d_{av,n}^{1/2}/n^{1/2})$ by (\ref{st1}), the
	leading term is $O_{P}(d_{av,n}^{\lambda /2}/n^{\lambda /2})$ by (\ref{c}).
	By (\ref{AL2}) and (\ref{st1}), we have%
	\begin{equation}
	\left\vert \frac{1}{v_n}-\frac{1}{\hat{v}}\right\vert ^{\lambda
	}=O_{P}\left( \frac{d_{av,n}^{\lambda /2}}{n^{\lambda /2}}\right) .
	\label{bd67}
	\end{equation}%
	Applying this to (\ref{bd671}), we obtain the desired result. $\blacksquare $%
	\medskip
	
	\noindent \textbf{Lemma B3: }(i)(a)\textit{\ }%
	\begin{equation*}
	Var\left( \frac{1}{\sqrt{n}}\sum_{i\in N_n}e_ia_i\right)
	=O(d_{mx,n}^2d_{av,n})\text{\textit{.}}
	\end{equation*}%
	\noindent (b)\textit{\ }%
	\begin{equation}
	Var\left( \frac{1}{\sqrt{n}}\sum_{i\in N_n}a_i\right) =O(d_{mx,n}^2d_{av,n})\text{ 
		\textit{and} }Var\left( \frac{1}{\sqrt{n}}\sum_{i\in N_n}a_i^{c}\right)
	=O(d_{av,n})\text{\textit{.}}  \label{rateV}
	\end{equation}
	
	\noindent (ii)(a)\textit{\ }%
	\begin{equation}
	Var\left( \frac{1}{\sqrt{n}}\sum_{i\in N_n}e_{\pi (i)}a_{i,\pi }|\pi
	\right) =O(d_{mx,n}^2d_{av,n}),\text{ }\Pi _n\text{\textit{-unif.}}
	\label{rateV4}
	\end{equation}%
	\noindent (b)\textit{\ }%
	\begin{eqnarray}
	Var\left( \frac{1}{\sqrt{n}}\sum_{i\in N_n}a_{i,\pi }|\pi \right)
	&=&O(d_{mx,n}^2d_{av,n}),\ \Pi _n\text{\textit{-unif}, \textit{and}}  \label{rateV2} \\
	Var\left( \frac{1}{\sqrt{n}}\sum_{i\in N_n}a_{i,\pi }^{c}|\pi \right)
	&=&O(d_{av,n}),\text{ }\Pi _n\text{\textit{-unif.}}  \notag
	\end{eqnarray}%
	\medskip
	
	\noindent \textbf{Proof:} (i)(a) Let $N_{n,3}(i)$ be the set of the vertices which is connected to $i$ within three edges (with $i$ excluded from the set). We write%
	\begin{equation*}
	Var\left( \frac{1}{\sqrt{n}}\sum_{i\in N_n}e_ia_i\right) \leq \frac{1}{%
		n}\sum_{i\in N_n}\mathbf{E}(e_i^{2}a_i^{2})+\frac{1}{n}\sum_{i\in
		N_n}\sum_{j\in N_{n,3}(i)}\sqrt{\mathbf{E}(e_i^{2}a_i^{2})\mathbf{E}%
		(e_{j}^{2}a_{j}^{2})},
	\end{equation*}%
	because $e_ia_i$ and $e_{j}a_{j}$ are uncorrelated if $j$ is more than
	two edges away from $i$. If $d_n(i)=0$, $\mathbf{E}(e_i^{2}a_i^{2})=0,$
	and if $d_n(i)\geq 1,$ 
	\begin{equation*}
	\mathbf{E}(e_i^{2}a_i^{2})\leq \frac{1}{d_n(i)}\sum_{j\in N_n(i)}%
	\mathbf{E}[e_i^{2}e_{j}^{2}]\leq \max_{i,j \in N_n}\mathbf{E}%
	[e_i^{2}e_{j}^{2}]<C,
	\end{equation*}%
	by Assumption 1(iii) for some $C>0$. Hence for this constant $C>0,$%
	\begin{equation}
	Var\left( \frac{1}{\sqrt{n}}\sum_{i\in N_n}e_ia_i\right) \leq C+\frac{C%
	}{n}\sum_{i\in N_n}d_n(i)d_{mx,n}^2=O\left( d_{mx,n}^2d_{av,n}\right) .
	\label{var1}
	\end{equation}
	
	\noindent (b) Write
	\begin{eqnarray}
	&& Var\left( \frac{1}{\sqrt{n}}\sum_{i\in N_n}a_i\right) \\
	&\leq& \frac{1}{n}%
	\sum_{i_{1}\in N_n:d_n(i_{1})\geq 1}\sum_{i_{2}\in N_{n,3}(i_1)\cup\{i\}:d_n(i_{2})\geq 1}\sum_{j_{1}\in
		N_n(i_{1})}\sum_{j_{2}\in N_n(i_{2})}\frac{|\mathbf{E}%
		(e_{j_{1}}e_{j_{2}})|}{d_n(i_{1})d_n(i_{2})}\\
	&\leq&
	\frac{1}{n}%
	\sum_{i_{1}\in N_n} d_{mx,n}^2(d_n(i_1)+1)=O(d_{mx,n}^2(d_{av,n}+1))=O(d_{mx,n}^2d_{av,n}),  \label{bd50}
	\end{eqnarray}
	where the last equality uses Lemma B1. As for the second statement in (\ref{rateV}),
	\begin{eqnarray}
	&&Var\left( \frac{1}{\sqrt{n}}\sum_{i\in N_n}a_i^{c}\right)  \label{bd51}
	\\
	&\leq &\frac{1}{n(n-d_{mx,n}-1)^{2}}\sum_{i_{1},i_{2}\in
		N_n}\sum_{j_{1}\in N_n\backslash \overline{N}_n(i_{1})}\sum_{j_{2}\in
		N_n\backslash \overline{N}_n(i_{2})}|\mathbf{E}(e_{j_{1}}e_{j_{2}})|  \notag
	\\
	&\leq &\frac{C}{n(n-d_{mx,n}-1)^{2}}\sum_{i_{1},i_{2}\in
		N_n}\sum_{j_{1}\in N_n\backslash \overline{N}_n(i_{1})}\left(
	d_n(j_{1})+1\right) ,  \notag
	\end{eqnarray}%
	because $\mathbf{E}(e_{j_{1}}e_{j_{2}})=0$ if $j_{2}\in N_n\backslash \bar{%
		N}_n(j_{1})$. The last term is $O(d_{av,n})$.\medskip
	
	\noindent (ii)(a) Similarly, we bound
	\begin{eqnarray*}
		&&Var\left( \frac{1}{\sqrt{n}}\sum_{i\in N_n}e_{\pi (i)}a_{i,\pi }|\pi
		\right) \\
		&\leq &\frac{1}{n}\sum_{i\in N_n}\mathbf{E}(e_{\pi (i)}^{2}a_{i,\pi
		}^{2}|\pi )+\frac{1}{n}\sum_{i\in N_n}\sum_{j\in N_{n,3}(i)}\sqrt{\mathbf{E%
		}(e_{\pi (i)}^{2}a_{i,\pi }^{2}|\pi )\mathbf{E}(e_{\pi (j)}^{2}a_{j,\pi
	}^{2}|\pi )}.
\end{eqnarray*}%
Both terms are $O(d_{mx,n}d_{av,n})$, $\Pi _n$-unif., because if $%
d_n(i)=0 $, $\mathbf{E}(e_{\pi (i)}^{2}a_{i,\pi }^{2})=0,$ and if $%
d_n(i)\geq 1,$ 
\begin{eqnarray}
\mathbf{E}(e_{\pi (i)}^{2}a_{i,\pi }^{2}|\pi ) &\leq &\frac{1}{d_n(i)}%
\sum_{j\in N_n(i)}\mathbf{E}[e_{\pi (i)}^{2}e_{\pi (j)}^{2}|\pi ]
\label{bd725} \\
&\leq &\frac{1}{d_n(i)}\sum_{j\in N_n(i)}\max_{i \in N_n}
\mathbf{E}[e_i^{4}]\leq C.  \notag
\end{eqnarray}%
\noindent (b) Note that by\ Assumption 1(iii),
\begin{equation}
\left\vert \mathbf{E}[e_{\pi (i_{2})}e_{\pi (j_{2})}|\pi ]\right\vert \leq
\max_{i,j \in N_n}\left\vert \mathbf{E}[e_ie_{j}]\right\vert \leq C,
\label{bd11}
\end{equation}%
and hence following the same arguments in (\ref{bd50})\ and (\ref{bd51}), we
obtain the desired result. $\blacksquare $\medskip

Recall the definition:%
\begin{equation*}
q_i=e_i(a_i-\gamma e_i)\text{ and\ }q_{i,\pi }=e_{\pi
	(i)}(a_{i,\pi }-\gamma_{\pi}e_{\pi (i)})\text{,}
\end{equation*}%
where $\gamma=\frac{1}{n}\sum_{i\in N_n}\mathbf{E}\left(
a_ie_i\right) $ and $\gamma_{\pi}=\frac{1}{n}\sum_{i\in N_n}%
\mathbf{E}\left( a_{i,\pi }e_{\pi (i)}|\pi \right) .$\medskip

\noindent \textbf{Lemma B4:}(i)\textit{\ }%
\begin{equation*}
Var\left( \frac{1}{\sqrt{n}}\sum_{i\in N_n}q_i\right)
=O(d_{mx,n}^2d_{av,n})\text{\textit{.}}
\end{equation*}

\noindent (ii)\textit{\ }%
\begin{equation*}
Var\left( \frac{1}{\sqrt{n}}\sum_{i\in N_n}q_{i,\pi }|\pi \right)
=O(d_{mx,n}^2d_{av,n}),\text{ }\Pi _n\text{\textit{-unif.}}
\end{equation*}%
\medskip

\noindent \textbf{Proof:}(i) Note that%
\begin{equation}
Var\left( \frac{1}{\sqrt{n}}\sum_{i\in N_n}q_i\right) \leq 2Var\left( 
\frac{1}{\sqrt{n}}\sum_{i\in N_n}e_ia_i\right) +2\gamma^{2}Var\left( \frac{1}{\sqrt{n}}\sum_{i\in N_n}e_i^{2}\right) .
\label{bd781}
\end{equation}%
The leading term is $O(d_{mx,n}^2d_{av,n})$ by Lemma B3(i)(a). Since%
\begin{equation}
|\gamma|=\left\vert \frac{1}{n}\sum_{i\in N_n}\mathbf{E}\left[
e_ia_i\right] \right\vert \leq \frac{1}{n}\sum_{i\in N_n:d_n(i)\geq
	1}\frac{1}{d_n(i)}\sum_{j\in N_n(i)}\left\vert \mathbf{E}\left[
e_ie_{j}\right] \right\vert \leq C,  \label{gm}
\end{equation}%
the last term in (\ref{bd781}) is $O(d_{av,n})$ by Lemma B2.\medskip\ 

\noindent (ii) Since%
\begin{equation}
|\gamma_{\pi}|\leq \frac{1}{n}\sum_{i\in N_n}\left\vert \mathbf{E}%
(e_{\pi (i)}a_{i,\pi }|\pi )\right\vert \leq \max_{i,j \in N_n}\left\vert \mathbf{E}[e_ie_{j}]\right\vert \leq C,  \label{gmp}
\end{equation}%
we obtain the desired rate using (\ref{rateV4}) and the same arguments in (%
\ref{bd781}). $\blacksquare $\medskip

\noindent \textbf{Lemma B5:} \textit{Suppose that for each }$i\in N_n$%
\textit{, let }$B_n(i)\subset N_n$ \textit{be a nonempty subset and }$%
\psi _n(i)$ \textit{a positive number. Then the following holds for }$%
1\leq \lambda \leq 4$\textit{.}

\noindent (i)%
\begin{equation}
\frac{1}{|\Pi _n|}\sum_{\pi \in \Pi _n}\sum_{i\in N_n}\frac{\psi
	_n(i)}{|B_n(i)|}\sum_{j\in B_n(i)}\left\vert \mathbf{E}\left[ e_{\pi
	(j)}a_{j,\pi }|\pi \right] \right\vert ^{\lambda }=O\left( \frac{d_{av,n}}{n}%
\sum_{i\in N_n}\psi _n(i)\right) .  \label{c1}
\end{equation}%
\noindent (ii) 
\begin{equation}
\frac{1}{|\Pi _n|}\sum_{\pi \in \Pi _n}\sum_{i\in N_n}\frac{\psi
	_n(i)}{|B_n(i)|}\sum_{j\in B_n(i)}\left\vert \mathbf{E}\left[ q_{j,\pi
}|\pi \right] \right\vert ^{\lambda }=O\left( \frac{d_{av,n}}{n}\sum_{i\in
N_n}\psi _n(i)\right) \text{.}  \label{c2}
\end{equation}%
\medskip

\noindent \textbf{Proof:} (i) Using the definition of $a_{i,\pi }$, we bound
the left hand side of (\ref{c1}) by%
\begin{eqnarray*}
	&&\sum_{i\in N_n:d_n(i)\geq 1}\frac{\psi _n(i)}{|B_n(i)|}\sum_{j\in
		B_n(i):d_n(j) \ge 1}\frac{1}{d_n(j)}\sum_{k\in N_n(j)}\frac{1}{|\Pi _n|}\sum_{\pi
		\in \Pi _n}\left\vert \mathbf{E}\left[ e_{\pi (j)}e_{\pi (k)}|\pi \right]
	\right\vert ^{\lambda } \\
	&\leq &\sum_{i\in N_n}\frac{\psi _n(i)}{n(n-1)}\sum_{ij\in \tilde{N}%
		_n}\left\vert \mathbf{E}\left[ e_ie_{j}\right] \right\vert ^{\lambda }.
\end{eqnarray*}%
Since $\mathbf{E}\left[ e_ie_{j}\right] =\mathbf{E}\left[ e_i\right] 
\mathbf{E}\left[ e_{j}\right] =0$ for all $ij\in \tilde{N}_n\backslash
E_n$, we have%
\begin{eqnarray*}
	\frac{1}{n-1}\sum_{ij\in \tilde{N}_n}\left\vert \mathbf{E}\left[ e_ie_{j}%
	\right] \right\vert ^{\lambda } &=&\frac{1}{n-1}\sum_{ij\in E_n}\left\vert 
	\mathbf{E}\left[ e_ie_{j}\right] \right\vert ^{\lambda }\\ 
	&=& O\left( \frac{|E_n|}{n}\right) =O\left( d_{av,n}\right).
\end{eqnarray*}

\noindent (ii) First, observe that
\begin{eqnarray}
\label{ave gamma}
\frac{1}{|\Pi _n|}\sum_{\pi \in \Pi _n} 
|\gamma_{\pi}|^\lambda 
\le 
\frac{1}{|\Pi _n|}\sum_{\pi \in \Pi _n} \frac{1}{n} \sum_{i\in N_n}
\left| \mathbf{E}\left[ e_{\pi (i)}a_{i,\pi }|\pi \right] \right|^{\lambda } = O \left(\frac{d_{av,n}}{n} \right),	
\end{eqnarray}
by (i). The left hand side of (\ref{c2}) is
bounded by
\begin{eqnarray*}
	&&\frac{2^{\lambda -1}}{|\Pi _n|}\sum_{\pi \in \Pi _n}\sum_{i\in N_n}%
	\frac{\psi _n(i)}{|B_n(i)|}\sum_{j\in B_n(i)}\left\vert \mathbf{E}%
	\left[ e_{\pi (j)}a_{j,\pi }|\pi \right] \right\vert ^{\lambda } \\
	&&+\frac{C2^{\lambda -1}\left( \max_{i \in N_n}\mathbf{E}\left[ e_i^{2}
		\right] \right) ^{\lambda }}{|\Pi _n|}\sum_{\pi \in \Pi _n} \left|\gamma_{\pi} \right|^{\lambda} \sum_{i\in
		N_n} \psi _n(i) .
\end{eqnarray*}%
The desired result follows from (\ref{ave gamma}). $\blacksquare $\medskip

\noindent \textbf{Lemma B6:} \textit{Suppose that } $d_{av,n}d_{mx,n}^2 = O(n^{3/2})$. \textit{Then the
	following holds.}

\noindent (i)%
\begin{equation*}
\frac{1}{|\Pi _n|}\sum_{\pi \in \Pi _n}\frac{1}{n}\sum_{i_{1}\in
	N_n}\sum_{i_{2}\in N_n\backslash \{i_{1}\}}\left\vert Cov\left(
q_{i_{1},\pi },q_{i_{2},\pi }|\pi \right) \right\vert =O\left( \frac{d_{av,n}}{\sqrt{n}}\right) \text{.}
\end{equation*}

\noindent (ii) 
\begin{equation*}
\frac{1}{|\Pi _n|^{2}}\sum_{\pi _{1}\in \Pi _n}\sum_{\pi _{2}\in \Pi
	_n}\frac{1}{n}\sum_{i_{1},i_{2}\in N_n}\left\vert Cov(q_{i_{1},\pi
	_{1}},q_{i_{2},\pi _{2}}|\pi _{1},\pi _{2})\right\vert =O\left( \frac{%
	d_{mx,n}d_{av,n}}{n}\right) .
\end{equation*}%
\medskip

\noindent \textbf{Proof:} (i) First, we bound%
\begin{equation}
\frac{1}{|\Pi _n|}\sum_{\pi \in \Pi _n}\frac{1}{n}\sum_{i_{1}\in
	N_n}\sum_{i_{2}\in N_n\backslash \{i_{1}\}}\left\vert Cov\left(
q_{i_{1},\pi },q_{i_{2},\pi }|\pi \right) \right\vert \leq A_{1n}+A_{2n},
\label{st8}
\end{equation}%
where%
\begin{eqnarray*}
	A_{1n} &=&\frac{1}{|\Pi _n|}\sum_{\pi \in \Pi _n}\frac{1}{n}%
	\sum_{i_{1}\in N_n}\sum_{i_{2}\in (N_n\setminus \{i_1\})\backslash
		N_{n,2}(i_{1})}\left\vert Cov\left( q_{i_{1},\pi },q_{i_{2},\pi }|\pi
	\right) \right\vert \text{ and} \\
	A_{2n} &=&\frac{1}{|\Pi _n|}\sum_{\pi \in \Pi _n}\frac{1}{n}%
	\sum_{i_{1}\in N_n}\sum_{i_{2}\in N_{n,2}(i_{1})}\left\vert Cov\left(
	q_{i_{1},\pi },q_{i_{2},\pi }|\pi \right) \right\vert .
\end{eqnarray*}
and $N_{n,2}(i)$ denotes the set of the vertices which is connected to $i$ within two edges (with $i$ excluded from the set).
We deal with $A_{1n}$ first. For $\mathbf{i}=(i_{1}j_{1},i_{2}j_{2})$, define%
\begin{eqnarray}
\xi _{1}(\mathbf{i}) &=&Cov(e_{i_{1}}e_{j_{1}},e_{i_{2}}e_{j_{2}}),
\label{chsi} \\
\xi _{2}(\mathbf{i}) &=&Cov(e_{i_{1}}e_{j_{1}},e_{i_{2}}^{2}),
\notag \\
\xi _{3}(\mathbf{i}) &=&Cov(e_{i_{2}}e_{j_{2}},e_{i_{1}}^{2}),%
\text{ and}  \notag \\
\xi _{4}(\mathbf{i}) &=&Cov(e_{i_{1}}^{2},e_{i_{2}}^{2}). 
\notag
\end{eqnarray}%
For $k=1,...,4$, let $\xi _{k,\pi }(\mathbf{i})$ be $\xi _{k}(\mathbf{i})$
except that $\mathbf{i}$ is replaced by $\pi (\mathbf{i})$, where
\begin{eqnarray*}
	\pi (\mathbf{i})=(\pi(i_1)\pi(j_1),\pi(i_2)\pi(j_2)).
\end{eqnarray*}
We bound
\begin{equation}
A_{1n}\leq \sum_{k=1}^{4}B_{kn},  \label{bd54}
\end{equation}
where%
\begin{equation}
B_{kn}=\frac{1}{n}\sum_{i_{1}\in N_n}\sum_{i_{2}\in (N_n\setminus\{i_1\})\backslash
	N_{n,2}(i_{1})}\sum_{j_{1}\in N_n(i_{1})}\sum_{j_{2}\in N_n(i_{2})}\frac{%
	1}{|\Pi _n|}\sum_{\pi \in \Pi _n}\frac{|\gamma_{k,\pi}|\left\vert \xi _{k,\pi }(\mathbf{i%
	})\right\vert }{d_n(i_{1})d_n(i_{2})},  \label{A}
\end{equation}
where $\gamma_{1,\pi} = 1, \gamma_{2,\pi} =\gamma_{3,\pi} = \gamma_{\pi}$, and $\gamma_{4,\pi} = \gamma_{\pi}^2$. Define
\begin{eqnarray*}
	S_{n,1} &=&\left\{ (i_{1}j_{1},i_{2}j_{2})\in \tilde{N}_n\times \tilde{N}%
	_n:\{i_{1},j_{1}\}\cap \{i_{2},j_{2}\}=\varnothing \right\} \text{ and} \\
	S_{n,2} &=&\left\{ (i_{1}j_{1},i_{2}j_{2})\in S_{n,1}:i_{2}=j_{2}\right\} .
\end{eqnarray*}%
Let 
\begin{eqnarray*}
	R_{n,1} &=&(E_n\times E_n)\cap S_{n,1}\text{ and} \\
	R_{n,2} &=&(E_n\times N_n^2)\cap S_{n,2}.
\end{eqnarray*}%
Let $H_{n,1}$ and $H_{n,2}$ be the collection of $\mathbf{i}$ in $S_{n,1}$
and $S_{n,2}$ respectively such that $\{i_{1},j_{1}\}\sim \{i_{2},j_{2}\}$
(i.e., the two pairs are adjacent to each other) and there is no vertex in $%
\mathbf{i}$ that is non-adjacent to the other vertices in $\mathbf{i}$. For $%
\mathbf{i}\in (S_{n,1}\backslash H_{n,1})\cup (S_{n,2}\backslash H_{n,2}),$ $%
\xi _{k}(\mathbf{i})=0$ for all $k=1,...,4$. Furthermore, note that%
\begin{equation}
|H_{n,1}|=O\left( |E_n|^2\right) =O(n^2 d_{av,n}^2),
\label{H1}
\end{equation}%
which is the number of ways that one chooses an edge $i_{1}i_{2}\in E_n$
and chooses $j_1j_2 \in E_n$. (The number of the ways in this case dominates (up to a constant) the number of all the other ways that $\mathbf{i} \in H_{n,1}$.) Also, observe that%
\begin{equation}
|H_{n,2}|=O\left( |E_n|d_{mx,n}\right) =O(nd_{mx,n}d_{av,n}),  \label{H2}
\end{equation}%
which is the number of ways that one chooses an edge $i_{1}i_{2}$ $\in E_n$
and chooses $j_{1}$ from its neighborhood. Hence note that%
\begin{eqnarray*}
	B_{1n} &\le& \frac{1}{n}\sum_{\mathbf{i}\in R_{n,1}}\frac{1}{|\Pi _n|}%
	\sum_{\pi \in \Pi _n}\left\vert \xi _{1,\pi }(\mathbf{i})\right\vert \leq 
	\frac{1}{n}\frac{\left\vert R_{n,1}\right\vert }{|S_{n,1}|}\sum_{\mathbf{i}%
		\in S_{n,1}}\left\vert \xi _{1}(\mathbf{i})\right\vert \\
	&=& \frac{1}{n}\frac{\left\vert R_{n,1}\right\vert }{|S_{n,1}|}\sum_{\mathbf{i%
		}\in H_{n,1}}\left\vert \xi _{1}(\mathbf{i})\right\vert \leq \frac{C}{n}%
	\frac{\left\vert R_{n,1}\right\vert |H_{n,1}|}{|S_{n,1}|} \\
	&=& O\left( \frac{(nd_{av,n})^{2}n^2 d_{av,n}^2}{n^{5}}\right)
	=O\left( \frac{d_{av,n}^{4}}{n}\right) .
\end{eqnarray*}%
The second inequality follows because all four elements of each $\mathbf{i} \in S_{n,1}$ are distinct, and the second equality follows because $\left\vert R_{n,1}\right\vert \leq
\left\vert E_n\right\vert ^{2}=(nd_{av,n})^{2}$. By Cauchy-Schwarz inequality and (\ref{ave gamma}),
\begin{eqnarray*}
	B_{2n} &\leq &\sqrt{\frac{1}{|\Pi _n|}%
		\sum_{\pi \in \Pi _n}|\gamma_\pi|} \sqrt{\frac{1}{n}\frac{\left\vert R_{n,2}\right\vert }{|S_{n,2}|}%
	\sum_{\mathbf{i}\in S_{n,2}}\left\vert \xi _{2}(\mathbf{i})\right\vert} \leq 
	O\left( \frac{\sqrt{d_{av,n}}}{\sqrt{n}} \right) \sqrt{\frac{C}{n}\frac{\left\vert R_{n,2}\right\vert |H_{n,2}|}{|S_{n,2}|}} \\
	&=&O\left( \frac{\sqrt{d_{av,n}}}{\sqrt{n}} \times \sqrt{\frac{(n^{2}d_{av,n})nd_{mx,n}d_{av,n}}{n^{4}}}\right) =O\left( 
	\frac{d_{mx,n}^{1/2}d_{av,n}^{3/2}}{n}\right) ,
\end{eqnarray*}
because $\left\vert R_{n,2}\right\vert \leq n|E_n|$ which is the number
ways to choose $i_{1}j_{1}\in E_n$ and choose $i_{2}=j_{2}$ from $N_n$.
We obtain the same rate for $B_{3n}$ using symmetry.

Finally, since for $i'j'\in \tilde{N%
}_n\backslash E_n$, $\xi _{4}(i'i',j'j')=0$, we have 
\begin{eqnarray*}
	B_{4n} &\leq & \sqrt{\frac{1}{|\Pi _n|}%
		\sum_{\pi \in \Pi _n}|\gamma_\pi|^2} \sqrt{\frac{1}{n}\sum_{i_{1}\in
		N_n}\sum_{i_{2}\in N_n\backslash \{i_{1}\}}\frac{1}{|\tilde{N}%
		_n|}\sum_{i'j'\in \tilde{N}_n}\left\vert \mathbf{E}\left[ \xi _{4}(i'i',j'j')%
	\right] \right\vert} \\
 &=& O\left( \frac{\sqrt{d_{av,n}}}{\sqrt{n}} \times  \sqrt{\frac{ n-1 }{|\tilde{N}_n|}\sum_{i'j'\in E_n}\left\vert \mathbf{E}\left[ \xi
	_{4}(i'i',j'j')\right] \right\vert} \right) = O\left( \frac{d_{av,n}}{\sqrt{n}}\right) .
\end{eqnarray*}%
Therefore, from (\ref{bd54}), we conclude that%
\begin{equation}
A_{1n}=O\left(\frac{d_{av,n}}{\sqrt{n}}\right) .  \label{A1}
\end{equation}

Let us turn to $A_{2n}$. Similarly as in (\ref{bd54}), we can bound%
\begin{equation*}
A_{2n}\leq \sum_{k=1}^{4}B_{kn}',
\end{equation*}%
where%
\begin{equation*}
B_{kn}'=\frac{1}{n}\sum_{i_{1}\in N_n}\sum_{i_{2}\in
	N_{n,2}(i_{1})}\sum_{j_{1}\in N_n(i_{1})}\sum_{j_{2}\in N_n(i_{2})}\frac{%
	1}{|\Pi _n|}\sum_{\pi \in \Pi _n}\frac{|\gamma_{k,\pi}|\left\vert \xi _{k,\pi }(\mathbf{i%
	})\right\vert }{d_n(i_{1})d_n(i_{2})}.
\end{equation*}%
Observe that
\begin{equation*}
B_{1n}'\leq \frac{C}{n}\sum_{i_{1}\in N_n}\sum_{i_{2}\in
	N_{n,2}(i_{1})}\frac{1}{|\Pi _n|}\sum_{\pi \in \Pi _n}1\{(\pi (i_{1})\pi
(j_{1}),\pi(i_{2})\pi (j_{2})) \in H_{n,1}\}.
\end{equation*}%
However, 
\begin{equation*}
\frac{1}{|\Pi _n|}\sum_{\pi \in \Pi _n}1\{(\pi (i_{1})\pi
(j_{1}),\pi(i_{2})\pi (j_{2})) \in H_{n,1}\}\leq \frac{C(nd_{av,n})^2(n-4)!}{n!},
\end{equation*}
for some absolute constant $C>0$. Therefore, for some absolute constant $C'>0$,
\begin{eqnarray*}
	B_{1n}' &\leq & \frac{C'}{n}\sum_{i_{1}\in N_n}\sum_{i_{2}\in
		N_{n,2}(i_{1})}\frac{(nd_{av,n})^2(n-4)!}{n!} \\
	&\leq &\frac{C'}{n}\sum_{i_{1}\in N_n}d_n(i_{1})\frac{%
		n^2d_{mx,n} d_{av,n}^2 (n-4)!}{n!}=O\left( \frac{d_{mx,n}d_{av,n}^3}{n^2}\right) .
\end{eqnarray*}
Let us turn to $B_{2n}'$ which we bound by
\begin{eqnarray*}
	\frac{1}{n}\sum_{i_{1}\in N_n}\sum_{i_{2}\in
			N_{n,2}(i_{1})}\frac{1}{|\Pi _n|}\sum_{\pi \in \Pi _n}1\{(\pi (i_{1})\pi
		(j_{1}),\pi(i_{2})\pi (i_{2})) \in H_{n,2}\}.
\end{eqnarray*}
Note that
\begin{eqnarray*}
	\frac{1}{|\Pi _n|}\sum_{\pi \in \Pi _n}1\{(\pi (i_{1})\pi
	(j_{1}),\pi(i_{2})\pi (i_{2})) \in H_{n,2}\}
	\leq \frac{Cnd_{mx,n}d_{av,n}(n-3)!}{n!}.
\end{eqnarray*}
Hence
\begin{eqnarray*}
	\frac{1}{n}\sum_{i_{1}\in N_n}\sum_{i_{2}\in
		N_{n,2}(i_{1})}\frac{1}{|\Pi _n|}\sum_{\pi \in \Pi _n}1\{(\pi (i_{1})\pi
	(j_{1}),\pi(i_{2})\pi (i_{2})) \in H_{n,2}\}
	=O\left(\frac{d_{mx,n}^2 d_{av,n}^2}{n^2} \right)
\end{eqnarray*}
Therefore, we conclude that
\begin{equation}
B_{2n}'=O\left( \frac{d_{mx,n}^2 d_{av,n}^2}{n^2} \right).  \label{A2}
\end{equation}
We obtain the same rate for $B_{3n}'$. Let us consider $B_{4n}'$ which we bound by
\begin{eqnarray*}
	\frac{C}{n}\sum_{i_{1}\in N_n}\sum_{i_{2}\in
		N_{n,2}(i_{1})}\frac{1}{|\Pi _n|}\sum_{\pi \in \Pi _n}1\{(\pi (i_{1})\pi
	(i_{1}),\pi(i_{2})\pi (i_{2})) \in E_n\}.
\end{eqnarray*}
Note that
\begin{eqnarray*}
	\frac{1}{|\Pi _n|}\sum_{\pi \in \Pi _n}1\{(\pi (i_{1})\pi
	(i_{1}),\pi(i_{2})\pi (i_{2})) \in E_n\} = O\left(\frac{n d_{av,n} (n-2)!}{n!} \right).
\end{eqnarray*}
Hence
\begin{eqnarray*}
	B_{4n}' & \le & \frac{1}{n}\sum_{i_{1}\in N_n}\sum_{i_{2}\in
		N_{n,2}(i_{1})}\frac{1}{|\Pi _n|}\sum_{\pi \in \Pi _n}1\{(\pi (i_{1})\pi
	(i_{1}),\pi(i_{2})\pi (i_{2})) \in E_n\} \\
	&=& O\left(\frac{n d_{mx,n} d_{av,n}^2 (n-2)!}{n!} \right) 
	= O\left(\frac{d_{mx,n} d_{av,n}^2}{n^2} \right).
\end{eqnarray*}
Combining (\ref{A1}) and (\ref{A2}) with (\ref{st8}), we obtain the desired
result.\medskip

\noindent (ii) For $\mathbf{i}=(i_{1}j_{1},i_{2}j_{2})$, define $\xi _{k}(%
\mathbf{i})$, $k=1,...,4,$ as in the proof of (i). Also let%
\begin{eqnarray}
\xi _{1,\pi _{1},\pi _{2}}(\mathbf{i}) &=&Cov(e_{\pi_{1}(i_{1})}e_{\pi
	_{1}(j_{1})},e_{\pi _{2}(i_{2})}e_{\pi _{2}(j_{2})}|\pi _{1},\pi _{2}),
\label{chsipi} \\
\xi _{2,\pi _{1},\pi _{2}}(\mathbf{i}) &=&\gamma _{\pi_{1}} Cov(e_{\pi
	_{1}(i_{1})}e_{\pi _{1}(j_{1})},e_{\pi _{2}(i_{2})}^{2}|\pi _{1},\pi _{2}), 
\notag \\
\xi _{3,\pi _{1},\pi _{2}}(\mathbf{i}) &=&\gamma _{\pi_{2}} Cov(e_{\pi
	_{2}(i_{2})}e_{\pi _{2}(j_{2})},e_{\pi _{1}(i_{1})}^{2}|\pi _{1},\pi _{2}),%
\text{ and}  \notag \\
\xi _{4,\pi _{1},\pi _{2}}(\mathbf{i}) &=&\gamma _{\pi_{1}} \gamma
_{\pi_{2}} Cov(e_{\pi _{1}(i_{1})}^{2},e_{\pi _{2}(i_{2})}^{2}|\pi
_{1},\pi _{2}).  \notag
\end{eqnarray}%
Observe that%
\begin{equation}
\frac{1}{n}\sum_{i_{1},i_{2}\in N_n}\frac{1}{|\Pi _n|^{2}}\sum_{\pi
	_{1}\in \Pi _n}\sum_{\pi _{2}\in \Pi _n}\left\vert Cov(q_{i_{1},\pi
	_{1}},q_{i_{2},\pi _{2}}|\pi _{1},\pi _{2})\right\vert \leq
\sum_{k=1}^{4}F_{k,n},  \label{bd783}
\end{equation}%
where%
\begin{equation*}
F_{k,n}=\frac{1}{n}\sum_{i_{1},i_{2}\in N_n}\sum_{j_{1}\in
	N_n(i_{1})}\sum_{j_{2}\in N_n(i_{2})}\frac{1}{|\Pi _n|^{2}}\sum_{\pi
	_{1}\in \Pi _n}\sum_{\pi _{2}\in \Pi _n}\frac{\left\vert \xi _{k,\pi
		_{1},\pi _{2}}(\mathbf{i})\right\vert }{d_n(i_{1})d_n(i_{2})}.
\end{equation*}%
Similarly as before, define%
\begin{eqnarray*}
	S_{n,1}' &=&\tilde{N}_n\times \tilde{N}_n\text{ and} \\
	S_{n,2}' &=&\left\{ (i_{1}j_{1},i_{2}j_{2})\in S_{n,1}^{\prime
	}:i_{2}=j_{2}\right\} .
\end{eqnarray*}%
Let $H_{n,1}'$ and $H_{n,2}'$ be the collection of $%
\mathbf{i}$ in $S_{n,1}'$ and $S_{n,2}'$ respectively such
that $\{i_{1},j_{1}\}\sim \{i_{2},j_{2}\}$ and there is no vertex in $%
\mathbf{i}$ that is non-adjacent to the other vertices in $\mathbf{i}$. Then
it is not hard to see that the bounds for $H_{n,1}$ and $H_{n,2}$ in (\ref%
{H1})\ and (\ref{H2}) apply to $H_{n,1}'$ and $H_{n,2}'$
as well. Also, for $\mathbf{i}\in (S_{n,1}'\backslash
H_{n,1}')\cup (S_{n,2}'\backslash H_{n,2}'),$ $%
\xi _{k}(\mathbf{i})=0$ for all $k=1,...,4$.

Hence%
\begin{eqnarray*}
	F_{1,n} &=&\frac{n}{|\tilde{N}_n|^{2}}\sum_{i_{1}j_{1}\in \tilde{N}_n}\sum_{i_{2}j_{2}\in \tilde{N}_n}\left\vert \xi _{1}(\mathbf{i}
	)\right\vert \leq \frac{Cn|H_{n,1}'|}{|\tilde{N}_n|^{2}} \\
	&=&O\left( \frac{n}{|\tilde{N}_n|^{2}}\times n^2 d_{av,n}^2 \right)
	=O\left( \frac{d_{av,n}^2}{n}\right) .
\end{eqnarray*}

Let us turn to $F_{2,n}$:
\begin{equation*}
F_{2,n}=\frac{n}{n|\tilde{N}_n|}\sum_{i_{1}j_{1}\in \tilde{N}%
	_n}\sum_{i_{2}\in N_n}\left\vert \xi _{2}(i_1 j_1,i_2 i_2)\right\vert \leq 
\frac{C|H_{n,2}'|}{|\tilde{N}_n|}=O\left( \frac{d_{mx,n}d_{av,n}}{%
	n}\right) .
\end{equation*}%
We obtain the same rate for $F_{3,n}$ using symmetry. Finally, as for $%
F_{4,n}$, note that%
\begin{eqnarray*}
	F_{4,n} &\leq &\frac{1}{n}\sum_{i_{1},i_{2}\in N_n}\frac{1}{|\Pi _n|^{2}}%
	\sum_{\pi _{1}\in \Pi _n}\sum_{\pi _{2}\in \Pi _n}\gamma_{\pi
		_{1}} \gamma _{\pi _{2}} \left\vert Cov(e_{\pi
		_{1}(i_{1})}^{2},e_{\pi _{2}(i_{2})}^{2}|\pi _{1},\pi _{2})\right\vert \\
	&\leq &Cn\left( \frac{1}{|\Pi _n|}\sum_{\pi \in \Pi _n}\gamma_{\pi
	}\right) ^{2}=O\left( \frac{d_{av,n}^{2}}{n}\right) ,
\end{eqnarray*}%
by (\ref{ave gamma}). Combining these results with (\ref{bd783}), we obtain the
desired result. $\blacksquare $\medskip \medskip

\noindent {\large 2.2. Asymptotic Linear Representation}\medskip

\noindent \textbf{2.2.1 The First Order Analysis of Estimation Errors}\medskip

\noindent \textbf{Lemma B7:}\textit{\ Suppose that }$d_{av,n}/\sqrt{n}=O(1),$%
\textit{\ as }$n\rightarrow \infty $.\textit{\ Then the following holds.}

\noindent (i)%
\begin{equation}
\frac{1}{\sqrt{n}}\sum_{i\in N_n}\{e_ia_i^{c}-\mathbf{E}%
[e_ia_i^{c}]\}=O_{P}\left( \frac{d_{av,n}}{\sqrt{n}}\right) .
\label{eq79}
\end{equation}

\noindent (ii)%
\begin{eqnarray}
\frac{1}{\sqrt{n}}\sum_{i\in N_n}\mathbf{E}\left[ e_{\pi (i)}a_{i,\pi
}^{c}|\pi \right] &=&O\left( \frac{d_{av,n}}{\sqrt{n}}\right) ,\ \Pi _n%
\text{\textit{-unif., and}}  \label{eq81} \\
\frac{1}{\sqrt{n}}\sum_{i\in N_n}\{e_{\pi (i)}a_{i,\pi }^{c}-\mathbf{E}%
\left[ e_{\pi (i)}a_{i,\pi }^{c}|\pi \right] \} &=&O_P\left( \frac{d_{av,n}}{%
	\sqrt{n}}\right) ,\ \Pi _n\text{\textit{-unif}.}  \notag
\end{eqnarray}%
\medskip

\noindent \textbf{Proof:}\textit{\ }(i) When $i$ and $j$ are not adjacent in 
$G_n$, $\mathbf{E}[e_ie_{j}]=\mathbf{E}[e_i]\mathbf{E}[e_{j}]=0$. Hence%
\begin{equation}
\mathbf{E}\left[ e_ia_i^{c}\right] =\frac{1}{n-1-d_n(i)}\sum_{j\in
	N_n\backslash \overline{N}_n(i)}\mathbf{E}\left[ e_i\right] \mathbf{E}\left[
e_{j}\right] =0.  \label{zer}
\end{equation}%
Write the left hand side of (\ref{eq79}) as%
\begin{equation}
\frac{1}{\sqrt{n}}\sum_{i\in N_n}e_ia_i^{c}=\frac{1}{\sqrt{n}}%
\sum_{i\in N_n}\frac{e_i}{n-1-d_n(i)}\sum_{j\in N_n\backslash \overline{N}%
	_n(i)}e_{j}=A_{1n}-A_{2n},  \label{dec8}
\end{equation}%
where%
\begin{eqnarray*}
	A_{1n} &=&\frac{1}{\sqrt{n}}\sum_{i\in N_n}\frac{ne_i}{n-1-d_n(i)}%
	\left( \frac{1}{n}\sum_{j\in N_n}e_{j}\right) \text{ and} \\
	A_{2n} &=&\frac{1}{\sqrt{n}}\sum_{i\in N_n}\frac{e_i}{n-1-d_n(i)}%
	\sum_{j\in \overline{N}_n(i)}e_{j}.
\end{eqnarray*}%
As for $A_{1n}$, note that%
\begin{eqnarray*}
	\mathbf{E}\left( \frac{1}{\sqrt{n}}\sum_{i\in N_n}\frac{ne_i}{%
		n-1-d_n(i)}\right) ^{2} &\leq &\left( \frac{n}{n-d_{mx,n}-1}\right) ^{2}%
	\frac{1}{n}\sum_{i\in N_n}\mathbf{E}e_i^{2} \\
	&&+\left( \frac{n}{n-d_{mx,n}-1}\right) ^{2}\frac{1}{n}\sum_{ij\in
		E_n}\left\vert \mathbf{E}e_ie_{j}\right\vert .
\end{eqnarray*}%
The leading term on the right hand side is $O(1)$ and the last term is $%
O(|E_n|/n)=O(d_{av,n})$. Therefore the left hand side term above is $%
O\left( 1+d_{av,n}\right) =O(d_{av,n})$ by Lemma B1. Using this and (\ref%
{st1}), we conclude that 
\begin{equation*}
A_{1n}=O_{P}\left( \frac{d_{av,n}}{\sqrt{n}}\right) .
\end{equation*}%
As for $A_{2n}$, the expected value of its absolute value is bounded by%
\begin{equation*}
\frac{1}{\sqrt{n}}\sum_{i\in N_n}\frac{1}{n-1-d_n(i)}\sum_{j\in \overline{N}%
	_n(i)}\mathbf{E}\left\vert e_ie_{j}\right\vert \leq \frac{C}{\sqrt{n}}%
\sum_{i\in N_n}\frac{d_n(i)+1}{n-1-d_{mx,n}}=O\left( \frac{d_{av,n}}{%
	\sqrt{n}}\right) .
\end{equation*}%
Hence we obtain the desired rate.\medskip

\noindent (ii) Write%
\begin{equation*}
\frac{1}{\sqrt{n}}\sum_{i\in N_n}\mathbf{E}\left[ e_{\pi (i)}a_{i,\pi
}^{c}|\pi \right] =\frac{1}{\sqrt{n}}\sum_{i\in N_n}\frac{1}{n-1-d_n(i)}%
\sum_{j\in N_n\backslash \overline{N}_n(i)}\mathbf{E}\left[ e_{\pi (i)}e_{\pi
	(j)}|\pi \right] .
\end{equation*}%
The last term is bounded by%
\begin{equation*}
\frac{1}{\sqrt{n}}\frac{1}{n-1-d_{mx,n}}\sum_{i\in N_n}\sum_{j\in
	N_n\backslash \overline{N}_n(i):\pi (j)\in N_n(\pi (i))}\left\vert \mathbf{E%
}[e_{\pi (i)}e_{\pi (j)}|\pi ]\right\vert ,
\end{equation*}%
because if $\pi (j)\notin N_n(\pi (i))$, $\mathbf{E}[e_{\pi (i)}e_{\pi
	(j)}|\pi ]=0$. By (\ref{bd11}), 
\begin{eqnarray*}
	\sum_{i\in N_n}\sum_{j\in N_n\backslash \overline{N}_n(i):\pi (j)\in
		N_n(\pi (i))}\left\vert \mathbf{E}[e_{\pi (i)}e_{\pi (j)}|\pi ]\right\vert
	&\leq &C\sum_{i\in N_n}d_n(\pi (i)) \\
	&=&C\sum_{i\in N_n}d_n(i)=O\left( nd_{av,n}\right) .
\end{eqnarray*}%
This establishes the first statement.

Let us turn to the second statement. Similarly as before,%
\begin{equation*}
\frac{1}{\sqrt{n}}\sum_{i\in N_n}e_{\pi (i)}a_{i,\pi }^{c}=A_{1n,\pi
}-A_{2n,\pi },
\end{equation*}%
where%
\begin{eqnarray*}
	A_{1n,\pi } &=&\frac{1}{\sqrt{n}}\sum_{i\in N_n}\frac{ne_{\pi (i)}}{%
		n-1-d_n(i)}\left( \frac{1}{n}\sum_{j\in N_n}e_{\pi (j)}\right) \text{ and%
	} \\
	A_{2n,\pi } &=&\frac{1}{\sqrt{n}}\sum_{i\in N_n}\frac{e_{\pi (i)}}{%
		n-1-d_n(i)}\sum_{j\in \overline{N}_n(i)}e_{\pi (j)}.
\end{eqnarray*}%
As for $A_{1n,\pi }$, note that $\sum_{j\in N_n}e_{\pi (j)}=\sum_{j\in
	N_n}e_{j}$ and%
\begin{eqnarray*}
	\mathbf{E}\left[\left( \frac{1}{\sqrt{n}}\sum_{i\in N_n}\frac{ne_{\pi (i)}}{%
		n-1-d_n(i)} \right) ^{2} |\pi\right] &\leq &\left( \frac{n}{n-d_{mx,n}-1}\right) ^{2}%
	\frac{1}{n}\sum_{i\in N_n}\mathbf{E}[e_{\pi (i)}^{2}|\pi ] \\
	&&+\left( \frac{n}{n-d_{mx,n}-1}\right) ^{2}\frac{1}{n}\sum_{ij\in
		E_n}\left\vert \mathbf{E[}e_{\pi (i)}e_{\pi (j)}|\pi ]\right\vert .
\end{eqnarray*}%
Since $\mathbf{E}[e_{\pi (i)}^{2}|\pi ]\leq \max_{i \in N_n}\mathbf{E}%
[e_i^{2}]\leq C$ and $\left\vert \mathbf{E[}e_{\pi (i)}e_{\pi (j)}|\pi
]\right\vert \leq \max_{i,j \in N_n}|\mathbf{E[}e_ie_{j}]|$, both terms on the
right hand side are $O(d_{av,n})$, $\Pi _n$-unif., similarly as before. After using the same arguments as in the proof of (i), we obtain the desired result. $\blacksquare $\medskip

\noindent \textbf{Lemma B8:}\textit{\ Suppose that }$d_{av,n}/\sqrt{n}%
\rightarrow 0,$\textit{\ as }$n\rightarrow \infty $.\textit{\ Then the
	following holds.}

\noindent (i)%
\begin{equation}
\frac{1}{\sqrt{n}}\sum_{i\in N_n}\{\hat{e}_i\hat{a}%
_i^{c}-e_ia_i^{c}\}=O_{P}\left( \frac{d_{av,n}}{\sqrt{n}}\right)
.  \label{eq78}
\end{equation}

\noindent (ii)%
\begin{equation*}
\frac{1}{\sqrt{n}}\sum_{i\in N_n}\left\{ \hat{e}_{i,\pi }\hat{a}_{i,\pi
}^{c}-e_{\pi (i)}a_{i,\pi }^{c}\right\} =O_{P}\left( \frac{d_{av,n}}{%
\sqrt{n}}\right) ,\ \Pi _n\text{\textit{-unif.}}
\end{equation*}%
\medskip

\noindent \textbf{Proof:}\textit{\ } (i) We write the sum as%
\begin{eqnarray}
\frac{1}{\sqrt{n}}\sum_{i\in N_n}\left\{ \hat{e}_i\hat{a}%
_i^{c}-e_ia_i^{c}\right\} &=&\frac{1}{\sqrt{n}}\sum_{i\in
	N_n}e_i\left\{ \hat{a}_i^{c}-a_i^{c}\right\} +\frac{1}{\sqrt{n}}%
\sum_{i\in N_n}a_i^{c}\left\{ \hat{e}_i-e_i\right\}  \label{dec891}
\\
&&+\frac{1}{\sqrt{n}}\sum_{i\in N_n}\left\{ \hat{e}_i-e_i\right\}
\left\{ \hat{a}_i^{c}-a_i^{c}\right\} .  \notag
\end{eqnarray}%
From (\ref{dec67}),%
\begin{equation}
\hat{a}_i^{c}-a_i^{c}=-\frac{\bar{\varepsilon}}{\hat{v}}+\left( \frac{1}{%
	\hat{v}}-\frac{1}{v_n}\right) a_i^{c}v_n,  \label{dec675}
\end{equation}%
and hence%
\begin{equation}
\frac{1}{n}\sum_{i\in N_n}(\hat{a}_i^{c}-a_i^{c})^{2}\leq \frac{2\bar{%
		\varepsilon}^{2}}{\hat{v}^{2}}+2\left( \frac{1}{\hat{v}}-\frac{1}{v_n}%
\right) ^{2} \left(\frac{1}{n}\sum_{i\in N_n}(a_i^{c})^{2} \right) v_n^2.  \label{bd3}
\end{equation}%
By (\ref{st1}) and (\ref{c}), the leading term is $O_{P}(d_{av,n}/n)$. As
for the last term,%
\begin{equation}
\frac{1}{n}\sum_{i\in N_n}\mathbf{E}(a_i^{c})^{2}\leq \frac{1}{n}%
\sum_{i\in N_n}\frac{1}{|N_n\backslash \overline{N}_n(i)|}\sum_{j\in
	N_n\backslash \overline{N}_n(i)}\mathbf{E}e_{j}^{2}\leq \max_{i\in N_n}%
\mathbf{E}e_i^{2}\leq C.  \label{dec892}
\end{equation}%
Applying this and (\ref{bd67}) to (\ref{bd3}), we find that%
\begin{equation}
\frac{1}{n}\sum_{i\in N_n}(\hat{a}_i^{c}-a_i^{c})^{2}=O_{P}\left( 
\frac{d_{av,n}}{n}\right) .  \label{rate4}
\end{equation}%
Now, we bound the last term in (\ref{dec891}) by%
\begin{equation}
\sqrt{n}\sqrt{\frac{1}{n}\sum_{i\in N_n}(\hat{e}_i-e_i)^{2}}\sqrt{%
	\frac{1}{n}\sum_{i\in N_n}(\hat{a}_i^{c}-a_i^{c})^{2}}=O_{P}\left( 
\frac{d_{av,n}}{\sqrt{n}}\right) ,  \label{dec63}
\end{equation}%
using (\ref{st2}) and (\ref{rate4}).

We focus on the leading two terms on the right hand side of (\ref{dec891}).
Note that%
\begin{eqnarray}
\frac{1}{n}\sum_{i\in N_n}a_i^{c}e_i &=&\frac{1}{n}\sum_{i\in
	N_n}\{a_i^{c}e_i-\mathbf{E}[a_i^{c}e_i]\}+\frac{1}{n}\sum_{i\in
	N_n}\mathbf{E}[a_i^{c}e_i]  \label{rate32} \\
&=&\frac{1}{n}\sum_{i\in N_n}\{a_i^{c}e_i-\mathbf{E}%
[a_i^{c}e_i]\}=O_P\left( \frac{d_{av,n}}{n}\right) ,  \notag
\end{eqnarray}%
by (\ref{zer}) and (\ref{eq79}). Now we write (using (\ref{dec675})) 
\begin{eqnarray*}
	\frac{1}{\sqrt{n}}\sum_{i\in N_n}\left( \hat{a}_i^{c}-a_i^{c}\right)
	e_i &=&-\frac{\bar{\varepsilon}}{\hat{v}}\frac{1}{\sqrt{n}}\sum_{i\in
		N_n}e_i+\sqrt{n}v_n\left( \frac{1}{\hat{v}}-\frac{1}{v_n}\right) 
	\frac{1}{n}\sum_{i\in N_n}a_i^{c}e_i \\
	&=&O_{P}\left( \frac{d_{av,n}}{\sqrt{n}}\right) +O_{P}\left( \frac{%
		d_{av,n}^{3/2}}{n}\right) =O_{P}\left( \frac{d_{av,n}}{\sqrt{n}}%
	\right) ,
\end{eqnarray*}%
by (\ref{bd67}) and (\ref{rate32}). Similarly,%
\begin{eqnarray*}
	\frac{1}{\sqrt{n}}\sum_{i\in N_n}a_i^{c}\left( \hat{e}_i-e_i\right)
	&=&-\frac{\bar{\varepsilon}}{\hat{v}}\frac{1}{\sqrt{n}}\sum_{i\in
		N_n}a_i^{c}+\sqrt{n}v_n\left( \frac{1}{\hat{v}}-\frac{1}{v_n}\right) 
	\frac{1}{n}\sum_{i\in N_n}a_i^{c}e_i \\
	&=&O_{P}\left( \frac{d_{av,n}}{\sqrt{n}}\right) ,
\end{eqnarray*}%
by (\ref{bd67}), (\ref{rateV}) and (\ref{rate32}). This completes the proof
of (i).\medskip

\noindent (ii) The proof is similar to that of (i). Indeed, we first write
the sum as%
\begin{eqnarray}
&&\frac{1}{\sqrt{n}}\sum_{i\in N_n}\left\{ \hat{e}_{i,\pi }\hat{a}_{i,\pi
}^{c}-e_{\pi (i)}a_{i,\pi }^{c}\right\}  \label{dec893} \\
&=&\frac{1}{\sqrt{n}}\sum_{i\in N_n}e_{\pi (i)}\left\{ \hat{a}_{i,\pi
}^{c}-a_{i,\pi }^{c}\right\} +\frac{1}{\sqrt{n}}\sum_{i\in N_n}a_{i,\pi
}^{c}\left\{ \hat{e}_{\pi (i)}-e_{\pi (i)}\right\}  \notag \\
&&+\frac{1}{\sqrt{n}}\sum_{i\in N_n}\left\{ \hat{e}_{\pi (i)}-e_{\pi
	(i)}\right\} \left\{ \hat{a}_{i,\pi }^{c}-a_{i,\pi }^{c}\right\} .  \notag
\end{eqnarray}%
From (\ref{dec67}),%
\begin{equation}
\frac{1}{n}\sum_{i\in N_n}(\hat{a}_{i,\pi }^{c}-a_{i,\pi }^{c})^{2}\leq 
\frac{2\bar{\varepsilon}^{2}}{\hat{v}^{2}}+2\left( \frac{1}{\hat{v}}-\frac{1%
}{v_n}\right) ^{2}\frac{1}{n}\sum_{i\in N_n}(a_{i,\pi }^{c})^{2}.
\label{f2}
\end{equation}%
Similarly as before,%
\begin{equation*}
\frac{1}{n}\sum_{i\in N_n}\mathbf{E}[(a_{i,\pi }^{c})^{2}|\pi ]\leq \frac{1%
}{n}\sum_{i\in N_n}\frac{1}{|N_n\backslash \overline{N}_n(i)|}\sum_{j\in
N_n\backslash \overline{N}_n(i)}\mathbf{E}[e_{\pi (j)}^{2}|\pi ]\leq
\max_{i\in N_n}\mathbf{E}e_i^{2}.
\end{equation*}%
Hence we find that%
\begin{equation*}
\frac{1}{n}\sum_{i\in N_n}(\hat{a}_{i,\pi }^{c}-a_{i,\pi
}^{c})^{2}=O_{P}\left( \frac{d_{av,n}}{n}\right) ,\ \Pi _n\text{-unif.}
\end{equation*}%
Using this and noting that $\sum_{i\in N_n}(\hat{e}_{\pi (i)}-e_{\pi
	(i)})^{2}=\sum_{i\in N_n}(\hat{e}_i-e_i)^{2}$, we bound the last term
in (\ref{dec893}) by%
\begin{equation}
\sqrt{n}\sqrt{\frac{1}{n}\sum_{i\in N_n}(\hat{e}_i-e_i)^{2}}\sqrt{%
	\frac{1}{n}\sum_{i\in N_n}(\hat{a}_{i,\pi }^{c}-a_{i,\pi }^{c})^{2}}%
=O_{P}\left( \frac{d_{av,n}}{\sqrt{n}}\right) ,\ \Pi _n\text{-unif.,}
\label{f5}
\end{equation}
as we saw in (\ref{dec63}). Thus the last term in (\ref{dec893}) is $O_{P}(d_{av,n}/\sqrt{n})$.

We write 
\begin{eqnarray*}
	\frac{1}{\sqrt{n}}\sum_{i\in N_n}\left( \hat{a}_{i,\pi }^{c}-a_{i,\pi
	}^{c}\right) e_{\pi (i)} &=&-\frac{\bar{\varepsilon}}{\hat{v}}\frac{1}{\sqrt{%
		n}}\sum_{i\in N_n}e_i+\sqrt{n}v_n\left( \frac{1}{\hat{v}}-\frac{1}{%
	v_n}\right) \frac{1}{n}\sum_{i\in N_n}a_{i,\pi }^{c}e_{\pi (i)} \\
&=&O_{P}\left( \frac{d_{av,n}}{\sqrt{n}}\right) +O_{P}\left( \frac{%
	d_{av,n}^{3/2}}{n}\right) =O_{P}\left( \frac{d_{av,n}}{\sqrt{n}}%
\right) ,
\end{eqnarray*}%
by (\ref{eq81}). Similarly as before, using (\ref{rateV2}),%
\begin{equation*}
\frac{1}{\sqrt{n}}\sum_{i\in N_n}a_{i,\pi }^{c}\left( \hat{e}_{\pi
	(i)}-e_{\pi (i)}\right) =O_{P}\left( \frac{d_{av,n}}{\sqrt{n}}\right)
,\ \Pi _n\text{-unif.}
\end{equation*}%
This completes the proof of (ii). $\blacksquare $\medskip

\noindent \textbf{2.2.2. Asymptotic Linear Representation}\medskip

Recall the definition:%
\begin{equation*}
C_{\pi }(G_n)=\frac{1}{n}\sum_{i\in N_n}\mathbf{E}\left[ e_{\pi
	(i)}a_{i,\pi }|\pi \right] -\frac{1}{n}\sum_{i\in N_n}\mathbf{E}\left[
e_{\pi (i)}a_{i,\pi }^{c}|\pi \right] .
\end{equation*}%
\medskip

\noindent \textbf{Lemma B9:}\textit{\ Suppose that }$d_{mx,n}^2d_{av,n}^{2}/n%
\rightarrow 0,$\textit{\ as }$n\rightarrow \infty $. \textit{Then the
	following holds.}

\noindent (i)%
\begin{equation*}
\sqrt{n}\{\hat{C}(G_n)-C(G_n)\}=\frac{1}{\sqrt{n}}\sum_{i \in N_n}\left(
q_i-\mathbf{E}q_i\right) +O_{P}\left( \frac{d_{mx,n}d_{av,n}}{%
	\sqrt{n}}\right) .
\end{equation*}%
\noindent (ii)%
\begin{equation*}
\sqrt{n}\{\hat{C}_{\pi }(G_n)-C_{\pi }(G_n)\}=\frac{1}{\sqrt{n}}%
\sum_{i \in N_n}\left( q_{i,\pi }-\mathbf{E}\left[ q_{i,\pi }|\pi \right]
\right) +O_{P}\left( \frac{d_{mx,n}d_{av,n}}{\sqrt{n}}\right) ,\ \Pi
_n\text{\textit{-unif.}}
\end{equation*}%
\medskip

\noindent \textbf{Proof:} (i) From (\ref{zer}), $\mathbf{E}\left[
e_ia_i^{c}\right] =0.$ This together with Lemmas B7(i) and B8(i) yields:%
\begin{eqnarray}
\sqrt{n}\{\hat{C}(G_n)-C(G_n)\} &=&\frac{1}{\sqrt{n}}\sum_{i\in
	N_n}\left\{ \hat{e}_i\hat{a}_i-e_ia_i\right\}  \label{dec71} \\
&&+\frac{1}{\sqrt{n}}\sum_{i\in N_n}\left\{ e_ia_i-\mathbf{E}\left[
e_ia_i\right] \right\} +O_{P}\left( \frac{d_{av,n}}{\sqrt{n}}%
\right) .  \notag
\end{eqnarray}%
We focus on the the leading term:%
\begin{eqnarray}
\frac{1}{\sqrt{n}}\sum_{i\in N_n}\left\{ \hat{e}_i\hat{a}%
_i-e_ia_i\right\} &=&\frac{1}{\sqrt{n}}\sum_{i\in N_n}e_i\left\{ 
\hat{a}_i-a_i\right\} +\frac{1}{\sqrt{n}}\sum_{i\in N_n}a_i\left\{ 
\hat{e}_i-e_i\right\}  \label{dec6} \\
&&+\frac{1}{\sqrt{n}}\sum_{i\in N_n}\left\{ \hat{e}_i-e_i\right\}
\left\{ \hat{a}_i-a_i\right\} .  \notag
\end{eqnarray}%
The last term is $O_{P}(d_{av,n}/\sqrt{n})$, by the arguments similar to (%
\ref{bd3}) and (\ref{dec63}). We focus on the leading two terms on the right
hand side of (\ref{dec6}).

From (\ref{dec67}), (\ref{AL2}), (\ref{bd67}) and (\ref{var1}), we deduce that 
\begin{eqnarray*}
	\frac{1}{\sqrt{n}}\sum_{i\in N_n}\left( \hat{a}_i-a_i\right) e_i &=&-%
	\frac{\bar{\varepsilon}}{\hat{v}}\frac{1}{\sqrt{n}}\sum_{i\in N_n}e_i+%
	\sqrt{n}v_n\left( \frac{1}{\hat{v}}-\frac{1}{v_n}\right) \frac{1}{n}%
	\sum_{i\in N_n}a_ie_i \\
	&=&-\left( \frac{1}{2\sqrt{n}}\sum_{i \in N_n}\left( e_i^{2}-\mathbf{E}%
	e_i^{2}\right) \right) \frac{1}{n}\sum_{i\in N_n}\mathbf{E}\left(
	a_ie_i\right) +O_{P}\left( \frac{d_{mx,n}d_{av,n}}{\sqrt{n}}%
	\right) .
\end{eqnarray*}%
Using (\ref{rateV}) and following similar arguments,%
\begin{eqnarray*}
	\frac{1}{\sqrt{n}}\sum_{i\in N_n}a_i\left( \hat{e}_i-e_i\right) &=&-%
	\frac{\bar{\varepsilon}}{\hat{v}}\frac{1}{\sqrt{n}}\sum_{i\in N_n}a_i+%
	\sqrt{n}v_n\left( \frac{1}{\hat{v}}-\frac{1}{v_n}\right) \frac{1}{n}%
	\sum_{i\in N_n}a_ie_i \\
	&=&-\left( \frac{1}{2\sqrt{n}}\sum_{i \in N_n}\left( e_i^{2}-\mathbf{E}%
	e_i^{2}\right) \right) \frac{1}{n}\sum_{i\in N_n}\mathbf{E}\left(
	a_ie_i\right) +O_{P}\left( \frac{d_{mx,n}d_{av,n}}{\sqrt{n}}%
	\right) .
\end{eqnarray*}%
We conclude that 
\begin{equation}
\frac{1}{\sqrt{n}}\sum_{i\in N_n}\left\{ \hat{e}_i\hat{a}%
_i-e_ia_i\right\} =-\frac{1}{\sqrt{n}}\sum_{i \in N_n}\left( e_i^{2}-%
\mathbf{E}e_i^{2}\right) \gamma+O_{P}\left( \frac{%
	d_{mx,n}d_{av,n}}{\sqrt{n}}\right) ,  \label{exp}
\end{equation}%
using the definition $\gamma=\frac{1}{n}\sum_{i\in N_n}\mathbf{E}%
\left( a_ie_i\right) $. Combining this with the second term on the right
hand side of (\ref{dec71}), we obtain the desired result.\medskip

\noindent (ii) From Lemmas B7(ii) and B8(ii), we write%
\begin{eqnarray}
\sqrt{n}\{\hat{C}_{\pi }(G_n)-C_{\pi }(G_n)\} &=&\frac{1}{\sqrt{n}}%
\sum_{i\in N_n}\left\{ \hat{e}_{\pi (i)}\hat{a}_{i,\pi }-e_{\pi
	(i)}a_{i,\pi }\right\}  \label{dec7} \\
&&+\frac{1}{\sqrt{n}}\sum_{i\in N_n}\left\{ e_{\pi (i)}a_{i,\pi }-\mathbf{E%
}\left[ e_{\pi (i)}a_{i,\pi }|\pi \right] \right\}  \notag \\
&&+O_{P}\left( \frac{d_{av,n}}{\sqrt{n}}\right),\textnormal{ } \Pi_n\textnormal{-unif.}  \notag
\end{eqnarray}%
We write the leading term on the right hand side as%
\begin{eqnarray*}
	&&\frac{1}{\sqrt{n}}\sum_{i\in N_n}\left\{ \hat{e}_{i,\pi }\hat{a}_{i,\pi
	}-e_{\pi (i)}a_{i,\pi }\right\} \\
	&=&\frac{1}{\sqrt{n}}\sum_{i\in N_n}e_{\pi (i)}\left\{ \hat{a}_{i,\pi
	}-a_{i,\pi }\right\} +\frac{1}{\sqrt{n}}\sum_{i\in N_n}a_{i,\pi }\left\{ 
	\hat{e}_{\pi (i)}-e_{\pi (i)}\right\} \\
	&&+\frac{1}{\sqrt{n}}\sum_{i\in N_n}\left\{ \hat{e}_{\pi (i)}-e_{\pi
		(i)}\right\} \left\{ \hat{a}_{i,\pi }-a_{i,\pi }\right\} .
\end{eqnarray*}%
Again the last term is $O_{P}(d_{av,n}/\sqrt{n}),$ $\Pi _n$-unif., due to
arguments similar to (\ref{f2}) and (\ref{f5}). Also, similarly as before,
we write%
\begin{equation*}
\frac{1}{\sqrt{n}}\sum_{i\in N_n}\left( \hat{a}_{i,\pi }-a_{i,\pi }\right)
e_{\pi (i)}=-\frac{\bar{\varepsilon}}{\hat{v}}\frac{1}{\sqrt{n}}\sum_{i\in
	N_n}e_i+\sqrt{n}v_n\left( \frac{1}{\hat{v}}-\frac{1}{v_n}\right) 
\frac{1}{n}\sum_{i\in N_n}a_{i,\pi }e_{\pi (i)}.
\end{equation*}%
The leading term is $O_{P}(d_{av,n}/\sqrt{n})$, and the last term is (by
Lemma B3(ii)(a))%
\begin{eqnarray*}
	&&\sqrt{n}v_n\left( \frac{1}{\hat{v}}-\frac{1}{v_n}\right) \left\{ \frac{%
		1}{n}\sum_{i\in N_n}\mathbf{E}\left( a_{i,\pi }e_{\pi (i)}|\pi \right)
	+O_{P}\left( \frac{d_{mx,n}d_{av,n}^{1/2}}{\sqrt{n}}\right) \right\} \\
	&=&-\left( \frac{1}{2\sqrt{n}}\sum_{i \in N_n}\left( e_i^{2}-\mathbf{E}%
	e_i^{2}\right) \right) \left( \frac{1}{n}\sum_{i\in N_n}\mathbf{E}\left(
	a_{i,\pi }e_{\pi (i)}|\pi \right) \right) +O_{P}\left( \frac{%
		d_{mx,n}d_{av,n}}{\sqrt{n}}\right),
\end{eqnarray*}
using (\ref{AL2}), (\ref{bd67}) and (\ref{gmp}).

Similarly,
\begin{eqnarray*}
	\frac{1}{\sqrt{n}}\sum_{i\in N_n}a_{i,\pi }\left( \hat{e}_{\pi (i)}-e_{\pi
		(i)}\right) &=&-\frac{\bar{\varepsilon}}{\hat{v}}\frac{1}{\sqrt{n}}%
	\sum_{i\in N_n}a_{i,\pi}+\sqrt{n}v_n\left( \frac{1}{\hat{v}}-\frac{1}{v_n}%
	\right) \frac{1}{n}\sum_{i\in N_n}a_{i,\pi }e_{\pi (i)} \\
	&=&-\left( \frac{1}{2\sqrt{n}}\sum_{i \in N_n}\left( e_i^{2}-\mathbf{E}%
	e_i^{2}\right) \right) \left( \frac{1}{n}\sum_{i\in N_n}\mathbf{E}\left(
	a_{i,\pi }e_{\pi (i)}|\pi \right) \right) \\
	&&+O_{P}\left( \frac{d_{mx,n}d_{av,n}}{\sqrt{n}}\right) .
\end{eqnarray*}

From (\ref{dec6}), we conclude that 
\begin{eqnarray*}
	\frac{1}{\sqrt{n}}\sum_{i\in N_n}\left\{ \hat{e}_{i,\pi }\hat{a}_{i,\pi
	}-e_{\pi (i)}a_{i,\pi }\right\} &=&-\frac{1}{\sqrt{n}}\sum_{i \in N_n}\left(
	e_i^{2}-\mathbf{E}e_i^{2}\right) \gamma_{\pi}+O_{P}\left( \frac{%
		d_{mx,n}d_{av,n}}{\sqrt{n}}\right) \\
	&=&-\frac{1}{\sqrt{n}}\sum_{i \in N_n}\left( e_{\pi (i)}^{2}-\mathbf{E}\left[
	e_{\pi (i)}^{2}|\pi \right] \right) \gamma_{\pi}+O_{P}\left( \frac{%
		d_{mx,n}d_{av,n}}{\sqrt{n}}\right) .
\end{eqnarray*}%
In view of (\ref{dec7}), we obtain the desired result. $\blacksquare $%
\medskip \medskip

\noindent {\large 2.3. Asymptotic Normality}\medskip

\noindent \textbf{2.3.1. Nondegeneracy of Asymptotic Variance}\medskip

For each $i\in N_n,$ define 
\begin{equation*}
\tilde{q}_i=\frac{1}{|S_n(i)|}\sum_{j\in S_n(i)}q_{j}\text{ and }%
\tilde{q}_{i,\pi }=\frac{1}{|S_n(i)|}\sum_{j\in S_n(i)}q_{j,\pi }\text{.}
\end{equation*}%
Let $\eta _i =q_i-\mathbf{E}[q_i]$,
\begin{eqnarray*}
	\eta _{i,\pi } =q_{i,\pi }-\mathbf{E}\left[ q_{i,\pi }|\pi \right] ,\text{
		and\ }\tilde{\eta}_{i,\pi }=q_{i,\pi }-\mathbf{E}\left[ \tilde{q}_{i,\pi
	}|\pi \right] .
\end{eqnarray*}%
Once $\pi $ is fixed, $\eta _{i,\pi }-\tilde{\eta}_{i,\pi }$ is
nonstochastic.\medskip

\noindent \textbf{Lemma B10:}\textit{\ Suppose that }$d_{mx,n}^2 d_{av} = O(\sqrt{n})$, \textit{\ as }$n\rightarrow \infty $. \textit{Then the
	following holds.}

\noindent (i) 
\begin{equation}
\frac{1}{|\Pi _n|}\sum_{\pi \in \Pi _n}\frac{1}{n}\sum_{i_{1}\in
	N_n}\sum_{i_{2}\in N_n\backslash \{i_{1}\}}\left\vert \mathbf{E}\left[
\eta _{i_{1},\pi }\eta _{i_{2},\pi }|\pi \right] \right\vert =O\left( \frac{d_{av,n}}{\sqrt{n}}\right) \text{\textit{.}}  \label{rate7}
\end{equation}%
\noindent (ii) 
\begin{equation*}
\frac{1}{|\Pi _n|}\sum_{\pi \in \Pi _n}\frac{1}{n}\sum_{i_{1}\in
	N_n}\sum_{i_{2}\in N_{n,3}(i_{1})}\left\vert \mathbf{E}%
\left[ \tilde{\eta}_{i_{1},\pi }\tilde{\eta}_{i_{2},\pi }|\pi \right]
\right\vert =O\left(\frac{d_{av,n}}{\sqrt{n}}\right) .
\end{equation*}%
\medskip

\noindent \textbf{Proof:} (i) Since $\mathbf{E}\left[ \eta _{i_{1},\pi }\eta
_{i_{2},\pi }|\pi \right] =Cov(q_{i_{1},\pi },q_{i_{2},\pi }|\pi ),$ the
result follows from Lemma B6(i) immediately.\medskip

\noindent (ii) Write%
\begin{eqnarray}
&&\frac{1}{n}\sum_{i_{1}\in N_n}\sum_{i_{2}\in N_{n,3}(i_{1})}\mathbf{E}%
\left[ \tilde{\eta}_{i_{1},\pi }\tilde{\eta}_{i_{2},\pi }|\pi \right]
\label{dec82} \\
&=&L_{1,n}(\pi )+L_{2,n}(\pi )+L_{3,n}(\pi )+\frac{1}{n}\sum_{i_{1}\in
	N_n}\sum_{i_{2}\in N_{n,3}(i_{1})}\mathbf{E}\left[ \eta _{i_{1},\pi }\eta
_{i_{2},\pi }|\pi \right] ,  \notag
\end{eqnarray}%
where%
\begin{eqnarray*}
	L_{1,n}(\pi ) &=&\frac{1}{n}\sum_{i_{1}\in N_n}\sum_{i_{2}\in
		N_{n,3}(i_{1})}\mathbf{E}\left[ \left\{ \tilde{\eta}_{i_{1},\pi }-\eta
	_{i_{1},\pi }\right\} \left\{ \tilde{\eta}_{i_{2},\pi }-\eta _{i_{2},\pi
	}\right\} |\pi \right] , \\
	L_{2,n}(\pi ) &=&\frac{1}{n}\sum_{i_{1}\in N_n}\sum_{i_{2}\in
		N_{n,3}(i_{1})}\mathbf{E}\left[ \left\{ \tilde{\eta}_{i_{1},\pi }-\eta
	_{i_{1},\pi }\right\} \eta _{i_{2},\pi }|\pi \right], \textnormal{ and }\\
	L_{3,n}(\pi ) &=&\frac{1}{n}\sum_{i_{1}\in N_n}\sum_{i_{2}\in
		N_{n,3}(i_{1})}\mathbf{E}\left[ \eta _{i_{1},\pi }\left\{ \tilde{\eta}%
	_{i_{2},\pi }-\eta _{i_{2},\pi }\right\} |\pi \right] .
\end{eqnarray*}%
The average of the last term in (\ref{dec82}) over $\pi \in \Pi _n$ is
bounded by (\ref{rate7}), because $N_{n,3}(i_{1})\subset N_n\backslash
\{i_{1}\}$. Note that%
\begin{eqnarray*}
	&&\frac{1}{n}\sum_{i_{1}\in N_n}\sum_{i_{2}\in N_{n,3}(i_{1})}\mathbf{E}%
	\left[ \eta _{i_{1},\pi }\{\eta _{i_{2},\pi }-\tilde{\eta}_{i_{2},\pi
	}\}|\pi \right] \\
	&=&\frac{1}{n}\sum_{i_{1}\in N_n}\sum_{i_{2}\in N_{n,3}(i_{1})}\mathbf{E}%
	\left[ \eta _{i_{1},\pi }|\pi \right] \mathbf{E}[\tilde{q}_{i_{2},\pi
	}-q_{i_{2},\pi }|\pi ]=0,
\end{eqnarray*}%
because $\mathbf{E}\left[ \eta _{i_{1},\pi }|\pi \right] =0$. Using similar
arguments, we obtain that $L_{2,n}(\pi )=L_{3,n}(\pi )=0$.

It suffices to show that%
\begin{equation}
\frac{1}{|\Pi _n|}\sum_{\pi \in \Pi _n}\left\vert L_{1,n}(\pi
)\right\vert =O\left( \frac{d_{mx,n}^2d_{av,n}^2}{n} \right)  = O\left(\frac{d_{av,n}}{\sqrt{n}}\right),
\label{stat2}
\end{equation}
where the last equality follows by the condition that $d_{mx,n}^2 d_{av} = O(\sqrt{n})$.

First, write 
\begin{equation*}
L_{1,n}(\pi )=B_{1,\pi }-B_{2,\pi }-B_{3,\pi }+B_{4,\pi },
\end{equation*}%
where 
\begin{eqnarray*}
	B_{1,\pi } &=&\frac{1}{n}\sum_{i_{1}\in N_n}\sum_{i_{2}\in N_{n,3}(i_{1})}%
	\mathbf{E}[q_{i_{1},\pi }|\pi ]\mathbf{E}[q_{i_{2},\pi }|\pi ] \\
	B_{2,\pi } &=&\frac{1}{n}\sum_{i_{1}\in N_n}\sum_{i_{2}\in N_{n,3}(i_{1})}%
	\mathbf{E}[q_{i_{1},\pi }|\pi ]\mathbf{E}[\tilde{q}_{i_{2},\pi }|\pi ], \\
	B_{3,\pi } &=&\frac{1}{n}\sum_{i_{1}\in N_n}\sum_{i_{2}\in N_{n,3}(i_{1})}%
	\mathbf{E}[\tilde{q}_{i_{1},\pi }|\pi ]\mathbf{E}[q_{i_{2},\pi }|\pi ],\text{
		and} \\
	B_{4,\pi } &=&\frac{1}{n}\sum_{i_{1}\in N_n}\sum_{i_{2}\in N_{n,3}(i_{1})}%
	\mathbf{E}[\tilde{q}_{i_{1},\pi }|\pi ]\mathbf{E}[\tilde{q}_{i_{2},\pi }|\pi
	].
\end{eqnarray*}

We focus on $B_{1,\pi }$. The other terms can be dealt with similarly. Note
that%
\begin{eqnarray}
\left\vert \mathbf{E}[q_{i,\pi }^{2}|\pi ]\right\vert &\leq &2\left\vert 
\mathbf{E}[e_{\pi (i)}^{2}a_{i,\pi }^{2}|\pi ]\right\vert +2\gamma_{\pi
}^{2}\left\vert \mathbf{E}[e_{\pi (i)}^4|\pi ]\right\vert  \label{bdq} \\
&\leq &C_{1}\max_{i,j \in N_n}\mathbf{E}\left[ e_i^{2}e_{j}^{2}\right]
\leq C_{2},  \notag
\end{eqnarray}%
for some constants $C_{1},C_{2}>0$ by (\ref{gmp}). Hence for some $C>0$,
\begin{eqnarray*}
	\frac{1}{|\Pi _n|}\sum_{\pi \in \Pi _n}\left\vert B_{1,\pi }\right\vert
	&\leq &\frac{C}{|\Pi _n|}\sum_{\pi \in \Pi _n}\frac{1}{n}\sum_{i_{1}\in
		N_n}\sum_{i_{2}\in N_{n,3}(i_{1})}\left\vert \mathbf{E}[q_{i_{1},\pi }|\pi
	]\right\vert \\
	&\leq &\frac{Cd_{mx,n}^2}{|\Pi _n|}\sum_{\pi \in \Pi _n}\frac{1}{n}%
	\sum_{i_{1}\in N_n} d_n(i_1)\left\vert \mathbf{E}[q_{i_{1},\pi }|\pi ]\right\vert
	=O\left( \frac{d_{mx,n}^2 d_{av,n}^2}{n}\right) ,
\end{eqnarray*}%
by Lemma B5(ii). Since $d_{av,n} \ge c$ for some $c>0$, we obtain the desired result. $\blacksquare $\medskip

\noindent \textbf{Lemma B11:}\textit{\ Suppose that }$d_{av,n}/n\rightarrow
0,$\textit{\ as }$n\rightarrow \infty $.\textit{\ Then}%
\begin{equation}
\frac{1}{\left\vert \Pi _n\right\vert }\sum_{\pi \in \Pi _n}\left\vert 
\frac{1}{n}\sum_{i\in N_n}\mathbf{E}\left[ \eta _{i,\pi }^{2}-\tilde{\eta}%
_{i,\pi }^{2}|\pi \right] \right\vert =O\left( \frac{d_{av,n}}{n}\right) 
\text{\textit{, as }}n\rightarrow \infty .  \label{rateV5}
\end{equation}%
\medskip

\noindent \textbf{Proof:}\textit{\ }We write%
\begin{eqnarray*}
	\mathbf{E}\left[ \eta _{i,\pi }^2-\tilde{\eta}_{i,\pi }^2|\pi \right]
	&=& \mathbf{E}\left[ (\eta _{i,\pi }-\tilde{\eta}_{i,\pi })(\eta _{i,\pi }+\tilde{\eta}_{i,\pi })|\pi \right]\\
	&=&\left( \mathbf{E}[\tilde{q}_{i,\pi }|\pi ]-\mathbf{E}\left[ q_{i,\pi
	}|\pi \right] \right) \mathbf{E}\left[ 2q_{i,\pi }-\mathbf{E}\left[ q_{i,\pi
}|\pi \right] -\mathbf{E}[\tilde{q}_{i,\pi }|\pi ]|\pi \right] \\
&=&\left( \mathbf{E}[\tilde{q}_{i,\pi }|\pi ]-\mathbf{E}\left[ q_{i,\pi
}|\pi \right] \right) (\mathbf{E}\left[ q_{i,\pi }|\pi \right] -\mathbf{E}[%
\tilde{q}_{i,\pi }|\pi ]) \\
&=&-\left( \mathbf{E}[\tilde{q}_{i,\pi }|\pi ]-\mathbf{E}\left[ q_{i,\pi
}|\pi \right] \right) ^{2}.
\end{eqnarray*}%
Hence the left hand side of (\ref{rateV5}) is bounded by%
\begin{equation*}
\frac{1}{\left\vert \Pi _n\right\vert }\sum_{\pi \in \Pi _n}\frac{1}{n}%
\sum_{i\in N_n}\left( \mathbf{E}[\tilde{q}_{i,\pi }|\pi ]-\mathbf{E}\left[
q_{i,\pi }|\pi \right] \right) ^{2}=O(d_{av,n}/n),
\end{equation*}%
by Lemma B5(ii). $\blacksquare $\medskip

\noindent \textbf{Lemma B12:}\textit{\ Suppose that }$d_{mx,n}/n=O(1),$%
\textit{\ as }$n\rightarrow \infty $.\textit{\ Then}%
\begin{eqnarray*}
	&&\frac{1}{|\Pi _n|}\sum_{\pi \in \Pi _n}\left\{ \frac{1}{n}\sum_{i\in
		N_n}\mathbf{E}[q_{i,\pi }^{2}|\pi ]-\frac{1}{n}\sum_{i\in
		N_n:d_n(i)\geq 1}\frac{1}{d_n^{2}(i)}\sum_{j\in N_n(i)}\mathbf{E}%
	[e_{\pi (i)}^{2}e_{\pi (j)}^{2}|\pi ]\right\} \\
	&=&O\left( \frac{d_{mx,n}}{n}\right) .
\end{eqnarray*}%
\medskip

\noindent \textbf{Proof:}\textit{\ }We bound%
\begin{equation*}
\frac{1}{n}\sum_{i\in N_n}\mathbf{E}[q_{i,\pi }^{2}|\pi ]-\frac{1}{n}%
\sum_{i\in N_n:d_n(i)\geq 1}\frac{1}{d_n^{2}(i)}\sum_{j\in N_n(i)}%
\mathbf{E}[e_{\pi (i)}^{2}e_{\pi (j)}^{2}|\pi ]\leq D_{1n}(\pi )+D_{2n}(\pi
)+D_{3n}(\pi ),
\end{equation*}%
where%
\begin{eqnarray*}
	D_{1n}(\pi ) &=&\frac{1}{n}\sum_{i\in N_n:d_n(i)\geq 1}\frac{1}{%
		d_n^{2}(i)}\sum_{j_{1},j_{2}\in N_n(i):j_{1}\neq j_{2}}\left\vert 
	\mathbf{E}\left[ e_{\pi (i)}^{2}e_{\pi (j_{1})}e_{\pi (j_{2})}|\pi \right]
	\right\vert ,\\
	D_{2n}(\pi ) &=&\frac{2|\gamma_{\pi}|}{n}\sum_{i\in N_n:d_n(i)\geq 1}%
	\frac{1}{d_n(i)}\sum_{j\in N_n(i)}\left\vert \mathbf{E}\left[ e_{\pi
		(i)}^{3}e_{\pi (j)}|\pi \right] \right\vert , \text{ and} \\
	D_{3n}(\pi ) &=&\frac{\gamma_{\pi}^{2}}{n}\sum_{i\in N_n:d_n(i)\geq
		1}\mathbf{E}\left[ e_{\pi (i)}^{4}|\pi \right] .
\end{eqnarray*}%
Note that we have
\begin{eqnarray*}
	\frac{1}{|\Pi _n|}\sum_{\pi \in \Pi _n}D_{3n}(\pi ) &\leq &\max_{i\in
		N_n}\mathbf{E}\left[ e_i^{4}\right] \frac{1}{|\Pi _n|}\sum_{\pi \in
		\Pi _n}\gamma_{\pi}^{2} \\
	&\leq &\max_{i\in N_n}\mathbf{E}\left[ e_i^{4}\right] \frac{1}{|\Pi _n|%
	}\sum_{\pi \in \Pi _n}\frac{1}{n}\sum_{i\in N_n}\left( \mathbf{E}[e_{\pi
	(i)}a_{i,\pi }|\pi ]\right) ^{2}=O\left( \frac{d_{av,n}}{n}\right),
\end{eqnarray*}
by Lemma B5(i). Let us turn to $D_{1n}(\pi )$. Let $\Pi
_{n,i}(j_{1},j_{2})$ be the collection of permutations $\pi \in \Pi _n$
such that either $\pi (j_{1})\pi (j_{2})\in E_n$ or both $\pi (i)\pi
(j_{1})$ and $\pi (i)\pi (j_{2})$ are in $E_n$. If $\pi $ is not in $\Pi
_{n,i}(j_{1},j_{2})$, one of the two vertices $\pi (j_{1})$ and $\pi (j_{2})$ (among the three vertices, $\pi (i)$, $\pi (j_{1})$, and $\pi (j_{2})$) is not adjacent to the other two vertices, and hence%
\begin{equation*}
\mathbf{E}\left[ e_{\pi (i)}^{2}e_{\pi (j_{1})}e_{\pi (j_{2})}|\pi \right]
=0.
\end{equation*}%
Furthermore, 
\begin{equation*}
|\Pi _{n,i}(j_{1},j_{2})|\leq nd_{mx,n}(n-2)!+nd_{mx,n}^{2}(n-3)!,
\end{equation*}%
where the first term correspond to the case $\pi (j_{1})\pi (j_{2})\in E_n$
and the second term to the case both $\pi (i)\pi (j_{1})$ and $\pi (i)\pi
(j_{2})$ are in $E_n$. Therefore,%
\begin{eqnarray*}
	\frac{1}{|\Pi _n|}\sum_{\pi \in \Pi _n}D_{1n}(\pi ) &\leq &\frac{C}{n}%
	\sum_{i\in N_n:d_n(i)\geq 1}\frac{1}{d_n^{2}(i)}\sum_{j_{1},j_{2}\in
		N_n(i):j_{1}\neq j_{2}}\frac{|\Pi _{n,i}(j_{1},j_{2})|}{|\Pi _n|} \\
	&\leq &\frac{C}{n}\sum_{i\in N_n:d_n(i)\geq 1}\frac{1}{d_n^{2}(i)}%
	\sum_{j_{1},j_{2}\in N_n(i):j_{1}\neq j_{2}}\left\{ \frac{d_{mx,n}}{n-1}+%
	\frac{d_{mx,n}^{2}}{(n-1)(n-2)}\right\} .
\end{eqnarray*}%
The last term is bounded by (from some large $n$ on) 
\begin{equation*}
\frac{2Cd_{mx,n}}{n(n-1)}\sum_{i\in N_n:d_n(i)\geq 1}\frac{1}{%
	d_n^{2}(i)}\sum_{j_{1},j_{2}\in N_n(i):j_{1}\neq j_{2}}1\leq \frac{%
	2Cd_{mx,n}}{n-1}.
\end{equation*}%
Therefore,%
\begin{equation*}
\frac{1}{|\Pi _n|}\sum_{\pi \in \Pi _n}D_{1n}(\pi )=O\left( \frac{%
	d_{mx,n}}{n}\right) .
\end{equation*}

Now, let us consider $D_{2n}(\pi )$. Using (\ref{gmp}), we bound it by%
\begin{equation*}
\frac{C}{n}\sum_{i\in N_n:d_n(i)\geq 1}\frac{1}{d_n(i)}\sum_{j\in
	N_n(i):\pi (j)\in N_n(\pi (i))}\left\vert \mathbf{E}\left[ e_{\pi
	(i)}^{3}e_{\pi (j)}|\pi \right] \right\vert .
\end{equation*}%
Let $\Pi _n(i,j)$ be the set of permutations $\pi $ such that vertex $j$ becomes a
neighbor of $i$ after permutation $\pi $. Then%
\begin{equation*}
|\Pi _n(i,j)|\leq nd_n(i)(n-2)!,
\end{equation*}%
which is the number of ways one places $\pi (i)$ in one of the $n$ places
and then places $\pi(j)$ in the neighborhood of $i$ and then permute the remaining $n-2$ vertices.
Hence%
\begin{eqnarray*}
	\frac{1}{|\Pi _n|}\sum_{\pi \in \Pi _n}D_{2n}(\pi ) &\leq &C\frac{%
		\max_{i,j \in N_n}\left\vert \mathbf{E}\left[ e_i^{3}e_{j}\right]
		\right\vert }{n}\sum_{i\in N_n:d_n(i)\geq 1}\frac{1}{d_n(i)}\sum_{j\in
		N_n(i)} \frac{|\Pi
		_n(i,j)|}{|\Pi _n|} \\
	&=&O\left( \frac{d_{av,n}n(n-2)!}{n!}\right) =O\left( \frac{d_{av,n}}{n}%
	\right) .
\end{eqnarray*}%
$\blacksquare $\medskip

Lemma B13 below establishes the nondegeneracy of the variance of $\zeta
_{n,\pi }$. Recall the definition:%
\begin{equation*}
\zeta _{n,\pi }=\frac{1}{\sqrt{n}}\sum_{i\in N_n}\left( q_{i,\pi }-\mathbf{%
	E}[q_{i,\pi }|\pi ]\right) =\frac{1}{\sqrt{n}}\sum_{i\in N_n}\eta _{i,\pi
},
\end{equation*}%
and $h_{n,\pi }^{2}=Var(\zeta _{n,\pi }|\pi ).$\medskip

\noindent \textbf{Lemma B13:}\textit{\ Suppose that }$%
d_{av,n}/\sqrt{n} \rightarrow 0$\textit{, as }$n\rightarrow \infty $%
. \textit{Then},%
\begin{equation*}
\frac{1}{\Pi _n}\sum_{\pi \in \Pi }1\left\{ h_{n,\pi }^{2}>\frac{c}{2}%
\right\} \rightarrow 1,
\end{equation*}%
\textit{as }$n\rightarrow \infty $.\medskip

\noindent \textbf{Proof: }To deal with asymptotically negligible terms, for
each $\varepsilon >0,$ let us define%
\begin{equation*}
\Pi _n(\varepsilon )=\left\{ \pi \in \Pi _n:\left\vert \frac{1}{n}%
\sum_{i_{1}\in N_n}\sum_{i_2 \in N_n \setminus \{i_1\}}\mathbf{E}\left[ \eta
_{i_{1},\pi }\eta _{i_{2},\pi }|\pi \right] \right\vert +\frac{1}{n}%
\sum_{i\in N_n}\left( \mathbf{E}\left[ q_{i,\pi }|\pi \right] \right)
^{2}\leq \varepsilon \right\} .
\end{equation*}%
Since $d_{av,n}/\sqrt{n}\rightarrow 0$, we apply Markov's
inequality to Lemmas B10(i) and B5(ii) and deduce that for each $\varepsilon>0$, $|\Pi
_n(\varepsilon )|/|\Pi _n|\rightarrow 1$ as $n\rightarrow \infty $.

We write%
\begin{eqnarray*}
	h_{n,\pi }^{2} &=&\frac{1}{n}\sum_{i\in N_n}\mathbf{E}\left[ \eta _{i,\pi
	}^{2}|\pi \right] +\frac{1}{n}\sum_{i_{1}\in N_n} \sum_{i_{2} \in N_n \backslash \{i_1\}} \mathbf{E}\left[ \eta _{i_{1},\pi }\eta _{i_{2},\pi }|\pi \right]
	\\
	&=&\frac{1}{n}\sum_{i\in N_n}\mathbf{E}\left[ q_{i,\pi }^{2}|\pi \right] -%
	\frac{1}{n}\sum_{i\in N_n}\left( \mathbf{E}\left[ q_{i,\pi }|\pi \right]
	\right) ^{2}+\frac{1}{n}\sum_{i_{1}\in N_n} \sum_{i_{2} \in N_n \backslash \{i_1\}} \mathbf{E}\left[ \eta _{i_{1},\pi }\eta _{i_{2},\pi }|\pi \right] .
\end{eqnarray*}%
Whenever $\pi \in \Pi _n(\varepsilon )$, the last two terms are bounded by 
$\varepsilon $. We focus on the leading term. By Lemma B12,%
\begin{eqnarray*}
	&&\frac{1}{\Pi _n}\sum_{\pi \in \Pi }\frac{1}{n}\sum_{i\in N_n}\mathbf{E}%
	\left[ q_{i,\pi }^{2}|\pi \right] \\
	&=&\frac{1}{\Pi _n}\sum_{\pi \in \Pi }\frac{1}{n}\sum_{i\in
		N_n:d_n(i)\geq 1}\frac{1}{d_n^{2}(i)}\sum_{j\in N_n(i)}\mathbf{E}%
	\left[ e_{\pi (i)}^{2}e_{\pi (j)}^{2}|\pi \right] +O\left( \frac{d_{mx,n}}{n}%
	\right) \\
	&=&\frac{1}{n}\sum_{i\in N_n:d_n(i)\geq 1}\frac{1}{d_n^{2}(i)}%
	\sum_{j\in N_n(i)}\frac{1}{n(n-1)}\sum_{ij\in \tilde{N}_n}\mathbf{E}%
	\left[ e_i^{2}e_{j}^{2}\right] +O\left( \frac{d_{mx,n}}{n}\right) \\
	&=&\frac{d_{avi,n}}{n(n-1)}\sum_{ij\in \tilde{N}_n}\mathbf{E}\left[
	e_i^{2}e_{j}^{2}\right] +O\left( \frac{d_{mx,n}}{n}\right) .
\end{eqnarray*}%
The leading term is bounded from below by $c>0$ by Assumption 1(ii). Taking $%
\varepsilon \in (0,c/2)$, we obtain the desired result. $\blacksquare $%
\medskip

\noindent \textbf{2.3.2. Asymptotic Normality}\medskip

\noindent \textbf{Lemma B14:}
(i) \textit{Suppose that }$d_{mx,n,3}^4/n
\rightarrow 0$\textit{, as }$n\rightarrow \infty.$\textit{ Then,}
\begin{equation}
\sup_{t\in \mathbf{R}}\left\vert P\left\{ \frac{\zeta _n}{\sqrt{Var(\zeta
		_n)}}\leq t\right\} -\Phi (t)\right\vert \rightarrow 0\text{\textit{, as }}%
n\rightarrow \infty,  \label{sum4}
\end{equation}
\textit{where }$\Phi $\textit{\ is the CDF of} $N(0,1).$\medskip

\noindent (ii) \textit{Suppose that }$d_{mx,n} d_{av,n}/n \rightarrow 0$\textit{, as }$n\rightarrow \infty.$ \textit{For each }$\varepsilon>0$,\textit{ define 
}$\tilde{\Pi}_n(\varepsilon )\subset \Pi _n\times \Pi _n$\textit{\ to
be the collection of }$(\pi _{1},\pi _{2})$\textit{'s such that}%
\begin{equation}
\left\vert Cov(\zeta _{n,\pi _{1}},\zeta _{n,\pi _{2}}|\pi _{1},\pi
_{2})\right\vert <\varepsilon .  \label{bd}
\end{equation}

\noindent \textit{Then for each }$\varepsilon >0$, \textit{we have }$|\tilde{\Pi}%
_n(\varepsilon )|/|\Pi _n|^{2}\rightarrow 1$\textit{\ as} $n\rightarrow
\infty $.\medskip

\noindent (iii) \textit{Suppose that }$d_{mx,n,3}^4/n
\rightarrow 0$\textit{, as }$n\rightarrow \infty.$ \textit{Then for each }$\varepsilon \in (0,c/2)$,\textit{\ there
	exists }$\Pi _{n,2}(\varepsilon )\subset \Pi _n\times \Pi _n$\textit{\ such that }$%
|\Pi _{n,2}(\varepsilon )|/|\Pi _n|^{2}\rightarrow 1$\textit{\ as} $%
n\rightarrow \infty $\textit{, and for any numbers }$b_{1},b_{2}\in \mathbf{R%
}$\textit{\ such that }$b_{1}^{2}+b_{2}^{2}=1$, 
\begin{eqnarray}
&&\max_{(\pi _{1},\pi _{2})\in \Pi _{n,2}(\varepsilon )}\sup_{t\in \mathbf{R}%
}\left\vert P\left\{ \frac{b_{1}\zeta _{n,\pi _{1}}}{h_{n,\pi _{1}}}+\frac{%
b_{2}\zeta _{n,\pi _{2}}}{h_{n,\pi _{2}}}\leq t|\pi _{1},\pi _{2}\right\}
-\Phi (t)\right\vert  \label{sum5} \\
&\leq &\frac{2\varepsilon}{c-2\varepsilon}\sup_{a>0}\phi (a)a+o(1)\text{\textit{%
		, as }}n\rightarrow \infty ,  \notag
\end{eqnarray}%
\textit{where we recall }$h_{n,\pi }^{2}=Var(\zeta _{n,\pi }|\pi )$\textit{\
	and }$\phi $\textit{\ is the PDF of }$N(0,1)$.\medskip

\noindent \textbf{Proof:}\textit{\ }(i) The sum $\zeta _n\ $is a sum of
mean-zero random variables having $G_n'=(N_n,E_n')$
as a dependency graph, where $ij\in E_n'$ if and only if $i$ and $%
j$ are within three edges, i.e., connected through not more than three edges. Furthermore, the maximum degree of $G_n'$
is $d_{mx,n,3}$. Therefore by Theorem 2.4 of Penrose (2003),
p.27, we have%
\begin{eqnarray*}
	&&\left\vert P\left\{ \frac{\zeta _n}{\sqrt{Var(\zeta _n)}}\leq
	t\right\} -\Phi (t)\right\vert  \\
	&\leq &\frac{2(d_{mx,n,3}+1)}{(2\pi )^{1/4}n^{1/4}Var(\zeta _n)^{3/4}}\sqrt{\frac{1}{n}\sum_{i\in N_n}\mathbf{E}\left[
		|q_i-\mathbf{E}q_i|^{3}\right] } \\
	&&+\frac{6(d_{mx,n,3}+1)^{3/2}}{n^{1/2}Var(\zeta _n)}\sqrt{\frac{1}{n} \sum_{i\in N_n}\mathbf{E}\left[ |q_i-\mathbf{E}q_i|^{4}\right] }.
\end{eqnarray*}%
Since $Var(\zeta _n)>c$ by the choice $P\in \mathcal{P}_n(G_n;c,M)$,
the terms on the right hand side are of order%
\begin{equation*}
O\left( \frac{d_{mx,n,3}}{n^{1/4}}\right) =o(1)\text{ and }O\left( \frac{%
	d_{mx,n,3}^{3/2}}{n^{1/2}}\right) =o(1),
\end{equation*}%
uniformly in $t\in \mathbf{R}$. Hence we obtain the desired result.\medskip 

\noindent (ii) Observe that%
\begin{eqnarray*}
	&&\frac{1}{|\Pi _n|^{2}}\sum_{\pi _{1}\in \Pi _n}\sum_{\pi _{2}\in \Pi
		_n}\left\vert Cov(\zeta _{n,\pi _{1}},\zeta _{n,\pi _{2}}|\pi _{1},\pi
	_{2})\right\vert \\
	&\leq &\frac{1}{n}\sum_{i_{1},i_{2}\in N_n}\frac{1}{|\Pi _n|^{2}}%
	\sum_{\pi _{1}\in \Pi _n}\sum_{\pi _{2}\in \Pi _n}\left\vert
	Cov(q_{i_{1},\pi _{1}},q_{i_{2},\pi _{2}}|\pi _{1},\pi _{2})\right\vert .
\end{eqnarray*}%
Hence, by Lemma B6(ii),%
\begin{equation*}
\frac{1}{|\Pi _n|^{2}}\sum_{\pi _{1}\in \Pi _n}\sum_{\pi _{2}\in \Pi
	_n}\left\vert Cov(\zeta _{n,\pi _{1}},\zeta _{n,\pi _{2}}|\pi _{1},\pi
_{2})\right\vert =O\left( \frac{d_{mx,n} d_{av,n}}{n}\right) =o(1).
\end{equation*}%
By Markov's inequality, 
\begin{eqnarray*}
	1-\frac{\tilde{\Pi}_n(\varepsilon )}{|\Pi _n|^{2}} &=&\frac{1}{|\Pi
		_n|^{2}}\sum_{\pi _{1}\in \Pi _n}\sum_{\pi _{2}\in \Pi _n}1\left\{
	\left\vert Cov(\zeta _{n,\pi _{1}},\zeta _{n,\pi _{2}}|\pi _{1},\pi
	_{2})\right\vert >\varepsilon \right\} \\
	&\leq &\frac{1}{\varepsilon |\Pi _n|^{2}}\sum_{\pi _{1}\in \Pi
		_n}\sum_{\pi _{2}\in \Pi _n}\left\vert Cov(\zeta _{n,\pi _{1}},\zeta
	_{n,\pi _{2}}|\pi _{1},\pi _{2})\right\vert =o(1).
\end{eqnarray*}%
\medskip

\noindent (iii) The sum in (\ref{sum5}) is a sum of centered random
variables having $G_{n,\pi _{1},\pi _{2}}=(N_n,E_{n,\pi _{1},\pi _{2}})$
as dependency graph, where $ij\in E_{n,\pi _{1},\pi _{2}}$ if and only if at least one vertex from $%
\{\pi _{1}(i),\pi _{2}(i)\}$ is within three edges from at least one vertex from $\{\pi _{1}(j),\pi
_{2}(j)\}$. In either case, and for any choice of $\pi _{1}$ and $\pi _{2}$,
the maximum degree of $G_{n,\pi _{1},\pi _{2}}$ is again bounded by $%
4d_{mx,n,3}$.

Let $\Pi _n'$ be the
collection of $\pi $'s in $\Pi _n$ such that 
\begin{equation}
h_{n,\pi }^{2}>c/2.  \label{bd2}
\end{equation}%
By Lemma B13, $|\Pi_n'|/|\Pi _n|\rightarrow 1$ as $%
n\rightarrow \infty $. Fix small $\varepsilon \in (0,c/2)$ and take $\tilde{\Pi}_n(\varepsilon )$ as in (ii). Let 
$\Pi _{n,2}(\varepsilon )=\tilde{\Pi}_n(\varepsilon )\cap (\Pi
'_n \times \Pi'_n)$. For given $(\pi _{1},\pi _{2})\in \Pi _n^{2},$
let 
\begin{equation*}
\sigma _{n,\pi _{1},\pi _{2}}^{2}=1+\frac{2b_{1}b_{2}}{h_{n,\pi
		_{1}}h_{n,\pi _{2}}}Cov(\zeta _{n,\pi _{1}},\zeta _{n,\pi _{2}}|\pi _{1},\pi
_{2}).
\end{equation*}%
Since $|b_1b_2| \leq (1/2)(b_1^2+b_2^2) =1/2$, using (\ref{bd2}) and (\ref{bd}), we find that $\left\vert \sigma _{n,\pi
	_{1},\pi _{2}}^{2}-1\right\vert \leq 2\varepsilon /c$ whenever $(\pi
_{1},\pi _{2})\in \Pi _{n,2}(\varepsilon )$. This implies that
\begin{eqnarray}
\label{sigma bound}
|\sigma_{n,\pi_1,\pi_2} - 1| \le \frac{2 \varepsilon /c}{1+\sigma_{n,\pi_1,\pi_2}} \le \frac{2 \varepsilon}{c}.	
\end{eqnarray}
Now for each fixed $\pi_{1},\pi _{2}\in \Pi _{n,2}(\varepsilon ),$%
\begin{eqnarray}
\label{deve}
&& \left\vert P\left\{ \frac{b_{1}\zeta _{n,\pi _{1}}}{h_{n,\pi _{1}}}+%
\frac{b_{2}\zeta _{n,\pi _{2}}}{h_{n,\pi _{2}}}\leq \sigma _{n,\pi _{1},\pi
	_{2}}t|\pi _{1},\pi _{2}\right\} -\Phi (\sigma _{n,\pi _{1},\pi
	_{2}}t)\right\vert 
\\ \notag
&\leq &\left\vert P\left\{ \frac{1}{\sigma _{n,\pi _{1},\pi _{2}}}\left( \frac{%
	b_{1}\zeta _{n,\pi _{1}}}{h_{n,\pi _{1}}}+\frac{b_{2}\zeta _{n,\pi _{2}}}{%
	h_{n,\pi _{2}}}\right) \leq t|\pi _{1},\pi _{2}\right\} -\Phi (t)\right\vert  
+ \vert \Phi(\sigma_{n,\pi_1,\pi_2}t)-\Phi(t) \vert.
\end{eqnarray}
By the mean-value theorem, there exists $\sigma^*$ that lies between $\sigma_{n,\pi_1,\pi_2}$ and $1$ (hence $\sigma^*>0$ by (\ref{sigma bound}) and by $\varepsilon <c/2$) such that the last absolute difference is equal to
\begin{eqnarray*}
	\phi(\sigma^*t)\vert (\sigma_{n,\pi_1,\pi_2} - 1) t \vert
	&\le& \phi(\sigma^*t)\left| \left(\frac{\sigma_{n,\pi_1,\pi_2} - 1}{\sigma^*t}\right) \sigma^*t \right| \\
	&\le& \sup_{a \ge 0}\phi(a)a \left| \left(\frac{\sigma_{n,\pi_1,\pi_2} - 1}{\sigma^*t}\right)  \right|
	\le \sup_{a \ge 0}\phi(a)a \frac{2 \varepsilon/c}{1-2\varepsilon/c}.
\end{eqnarray*}
Similarly as in the proof of (i), by Theorem 2.4 of Penrose (2003), the leading term on the right hand side of (\ref{deve}) is equal to
\begin{eqnarray}
&&\frac{C(d_{mx,n,3}+1)}{n^{1/4}}\sqrt{\frac{1}{n} \sum_{i\in N_n}
	\mathbf{E}\left[ \left\vert \frac{q_{i,\pi_{1}}}{h_{n,\pi
			_{1}}}+\frac{q_{i,\pi _{2}}}{h_{n,\pi _{2}}}\right\vert
	^{3}|\pi_{1},\pi_{2}\right] }  \label{dec562} \\
&&+\frac{C(d_{mx,n,3}+1)^{3/2}}{n^{1/2}}\sqrt{\frac{1}{n}\sum_{i\in N_n}%
	\mathbf{E}\left[ \left\vert \frac{q_{i,\pi _{1}}}{h_{n,\pi
			_{1}}}+\frac{q_{i,\pi_{2}}}{h_{n,\pi _{2}}}\right\vert
	^{4}|\pi _{1},\pi _{2}\right] },  \notag
\end{eqnarray}%
for some constant $C>0$. Since for $\pi \in \Pi_n'$ and for $%
\lambda =3,4,$%
\begin{equation*}
\frac{1}{n} \sum_{i\in N_n}\mathbf{E}\left[ \left\vert \frac{q_{i,\pi }}{h_{n,\pi }}%
\right\vert ^{\lambda }|\pi \right] \leq 2^{\lambda - 1}\left( \frac{c}{2}
\right) ^{-\lambda /2}\max_{i,j \in N_n}\mathbf{E}\left[ \left\vert
e_ie_{j}\right\vert ^{\lambda }\right] \left(1 + \max_{i \in N_n} \mathbf{E}[e_i^{2\lambda}]\right)\leq C,
\end{equation*}%
the last two terms in (\ref{dec562}) are $O(d_{mx,n,3}/n^{1/4})$ and $
O(d_{mx,n,3}^{3/2}/n^{1/2})$, and hence are $o(1),$ uniformly over $\pi _{1},\pi _{2}\in \Pi
_{n,2}(\varepsilon )$, similarly as before. $\blacksquare $\medskip \medskip 

\noindent {\large 2.4. Consistency of Variance Estimators}\medskip

\noindent \textbf{2.4.1. First Order Analysis of Estimation Errors}\medskip

\noindent \textbf{Lemma B15:} (i) \textit{\ Suppose that }$d_{mx,n}^{1/2}d_{av,n}/n%
\rightarrow 0$ \textit{as} $n\rightarrow \infty $. \textit{Then}
\begin{equation}
\frac{1}{n}\sum_{i\in N_n}e_i^{2}\left( \hat{a}_i-a_i\right)
^{2}=O_{P}\left( \frac{d_{mx,n}^{1/2}d_{av,n}}{n}\right) \text{\textit{.}}
\label{st4}
\end{equation}%
\noindent (ii) \textit{Suppose that} $d_{mx,n}^2 d_{av,n} / n 
\rightarrow 0$ \textit{as} $n\rightarrow \infty $. \textit{Then}
\begin{equation*}
\frac{1}{n}\sum_{i\in N_n}\left( \hat{q}_i-q_i\right) ^{2}=O_{P}\left( 
\frac{d_{mx,n}^2 d_{av,n}}{n}\right) \text{\textit{.}}
\end{equation*}%
\medskip

\noindent \textbf{Proof:} (i) The left hand side of (\ref{st4}) is
bounded by%
\begin{eqnarray*}
	\sqrt{\frac{1}{n}\sum_{i\in N_n}e_i^{4}}\sqrt{\frac{1}{n}\sum_{i\in
			N_n}\left( \hat{a}_i-a_i\right) ^{4}} &\leq &O_{P}(1)\sqrt{\frac{1}{n}%
		\sum_{i\in N_n:d_n(i)\geq 1}\frac{1}{d_n(i)}\sum_{j\in N_n(i)}\left( 
		\hat{e}_{j}-e_{j}\right) ^{4}} \\
	&\leq &O_{P}(1)\sqrt{\frac{d_{mx,n}}{n}\sum_{i\in N_n:d_n(i)\geq 1}\left( 
		\hat{e}_{i}-e_{i}\right) ^{4}}\\
	&=& O_{P}\left( \frac{d_{mx,n}^{1/2}d_{av,n}}{n}%
	\right) ,
\end{eqnarray*}%
by (\ref{st2}) in Lemma B2.\medskip

\noindent (ii) We first bound%
\begin{equation}
\frac{1}{n}\sum_{i\in N_n}(\hat{q}_i-q_i)^{2}\leq \frac{2}{n}%
\sum_{i\in N_n}\left( \hat{e}_i-e_i\right) ^{2}\hat{b}_i^{2}+\frac{2%
}{n}\sum_{i\in N_n}e_i^{2}(\hat{b}_i-b_i)^{2},  \label{dec89}
\end{equation}%
where $\hat{b}_i=\hat{a}_i-\hat{\gamma}\hat{e}_i$ and $%
b_i=a_i-\gamma e_i$. The leading term is bounded by 
\begin{equation*}
\frac{4}{n}\sum_{i\in N_n}\left( \hat{e}_i-e_i\right) ^{2}(\hat{b}%
_i-b_i)^{2}+\frac{4}{n}\sum_{i\in N_n}\left( \hat{e}_i-e_i\right)
^{2}b_i^{2}.
\end{equation*}%
It is not hard to check that the first term is asymptotically dominated by
the second term. By bounding $\left( \hat{e}_i-e_i\right) ^{2}$ by two
squares from (\ref{dec67}), we bound the second term by 
\begin{equation*}
\frac{8\bar{\varepsilon}^{2}}{\hat{v}^{2}}\frac{1}{n}\sum_{i\in
	N_n}b_i^{2}+\left( \frac{1}{\hat{v}}-\frac{1}{v_n}\right) ^{2}\frac{8}{%
	n}\sum_{i\in N_n}\varepsilon _i^{2}b_i^{2}=O_{P}\left( \frac{d_{av,n}}{%
	n}\right) ,
\end{equation*}%
because $\bar{\varepsilon}=O_{P}(d_{av,n}^{1/2}/\sqrt{n})$ and $\hat{v}%
^{-1}-v_n^{-1}=O_{P}(d_{av,n}^{1/2}/\sqrt{n})$ by (\ref{bd67}). Also, the last
term in (\ref{dec89}) is bounded by%
\begin{eqnarray*}
	&&\frac{4}{n}\sum_{i\in N_n}e_i^{2}\left( \hat{a}_i-a_i\right) ^{2}+%
	\frac{4}{n}\sum_{i\in N_n}e_i^{2}\left( \hat{\gamma}\hat{e}%
	_i-\gamma e_i\right) ^{2} \\
	&=&O_{P}\left( \frac{d_{mx,n}^{1/2}d_{av,n}}{n}\right) +\frac{4}{n}%
	\sum_{i\in N_n}e_i^{2}\left( \hat{\gamma}\hat{e}_i-\gamma e_i\right) ^{2},
\end{eqnarray*}%
by (i). The last sum is bounded by $A_{1n}+A_{2n}$, where
\begin{eqnarray*}
	A_{1n} &=&\frac{8\hat{\gamma}^{2}}{n}\sum_{i\in N_n}e_i^{2}\left( 
	\hat{e}_i-e_i\right) ^{2}\text{ and} \\
	A_{2n} &=&\frac{8\left( \hat{\gamma}-\gamma\right) ^{2}}{n}%
	\sum_{i\in N_n}e_i^{4}.
\end{eqnarray*}%
Following the proof of Lemma B9(i), we find that
\begin{eqnarray*}
	\hat{\gamma}-\gamma &=& \frac{1}{n}\sum_{i \in N_n}\left( \hat{e}%
	_i\hat{a}_i-\mathbf{E}e_ia_i\right)\\
	&=& \frac{1}{n}\sum_{i \in N_n}\left( q_i - \mathbf{E}q_i \right)
	+ O_{P}\left(\frac{d_{mx,n}d_{av,n}}{n}\right) \text{.}  \label{gm2}
\end{eqnarray*}
The last average of $(q_i - \mathbf{E}q_i)$'s is $O_P(d_{mx,n} d_{av,n}^{1/2}/\sqrt{n})$ by Lemma B4(i). Hence we have 
\begin{equation}
\label{gamma_n}
\hat{\gamma} = \frac{1}{n}\sum_{i \in N_n}\hat{e}_i\hat{a}%
_i
= \gamma+O_{P}\left( \frac{
	d_{mx,n}d_{av,n}^{1/2}}{\sqrt{n}}\right)
=O_{P}(1),
\end{equation}
where the last bound follows by (\ref{gm}). Hence we bound $A_{1n}$ by%
\begin{equation*}
O_{P}(1)\times \sqrt{\frac{1}{n}\sum_{i\in N_n}e_i^{4}}\sqrt{\frac{1}{n}%
	\sum_{i\in N_n}\left( \hat{e}_i-e_i\right) ^{4}}=O_{P}\left( \frac{%
	d_{av,n}}{n}\right),
\end{equation*}
by (\ref{st2}) and (\ref{gamma_n}). We deduce that
\begin{equation*}
A_{2n}=O_{P}\left( \frac{d_{mx,n}^2 d_{av,n}}{n}\right) \cdot \frac{1}{n}%
\sum_{i\in N_n}e_i^{4}=O_{P}\left( \frac{d_{mx,n}^2 d_{av,n}}{n}\right),
\end{equation*}%
because $\frac{1}{n}\sum_{i\in N_n}\mathbf{E}e_i^{4}=O(1)$. Collecting
the results for $A_{1n}$ and $A_{2n}$, we conclude that%
\begin{equation*}
\frac{2}{n}\sum_{i\in N_n}e_i^{2}(\hat{b}_i-b_i)^{2}=O_{P}\left( 
\frac{d_{mx,n}^2 d_{av,n}}{n}\right) ,
\end{equation*}%
and hence from (\ref{dec89}) that 
\begin{equation}
\frac{1}{n}\sum_{i\in N_n}(\hat{q}_i-q_i)^{2}=O_{P}\left( \frac{%
	d_{mx,n}^2 d_{av,n}}{n}\right) .  \label{rate1}
\end{equation}%
$\blacksquare $\medskip

For each $i\in N_n,$ recall $\eta_i=q_i-\mathbf{E}q_i$ and let
\begin{eqnarray*}
	\hat{\eta}_i=\hat{q}_i-\bar{q}_i,
\end{eqnarray*}
where we recall $\bar{q}_i=\frac{1}{|S_n(i)|}\sum_{j\in S_n(i)}\hat{q}%
_{j}$. Define%
\begin{equation}
V_n=\frac{1}{n}\sum_{i_{1}\in N_n}\sum_{i_{2}\in \overline{N}%
	_{n,3}(i_{1})}\left( \hat{\eta}_{i_{1}}\hat{\eta}_{i_{2}}-\eta _{i_{1}}\eta
_{i_{2}}\right) ,  \label{V}
\end{equation}%
where $\overline{N}_{n,3}(i)=N_{n,3}(i)\cup \{i\}$.\medskip

\noindent \textbf{Lemma B16:}\textit{\ Suppose that }$d_{mx,n,3}^{3}d_{mx,n}/n=O(1)$%
\textit{, as }$n\rightarrow \infty $. \textit{Then}%
\begin{equation*}
V_n=O_{P}(d_{mx,n,3}^{3/2}d_{mx,n}^{1/2}/\sqrt{n}).
\end{equation*}%
\medskip

\noindent \textbf{Proof: }We write
\begin{equation*}
V_n=V_{1n}+V_{2n},
\end{equation*}%
where%
\begin{eqnarray*}
	V_{1n} &=&\frac{1}{n}\sum_{i\in N_n}\left( \hat{\eta}_i^{2}-\eta
	_i^{2}\right) \text{ and} \\
	V_{2n} &=&\frac{1}{n}\sum_{i_{1}\in N_n}\sum_{i_{2}\in
		N_{n,3}(i_{1})}\left( \hat{\eta}_{i_{1}}\hat{\eta}_{i_{2}}-\eta _{i_{1}}\eta
	_{i_{2}}\right) .
\end{eqnarray*}

We deal with $V_{1n}$ first. Note that
\begin{equation}
\frac{1}{n}\sum_{i\in N_n}\left\vert \hat{\eta}_i-\eta _i\right\vert
^{2}\leq \frac{2}{n}\sum_{i\in N_n}\left\vert \hat{q}_i-q_i\right\vert
^{2}+\frac{2}{n}\sum_{i\in N_n}\left\vert \bar{q}_i-\mathbf{E}%
q_i\right\vert^{2}  \label{dec77}
\end{equation}%
by the definitions of $\hat{\eta}_i$ and $\eta _i$. The leading term is $%
O_{P}\left( d_{mx,n}^2 d_{av,n}/n\right) $ by Lemma B15(ii). We turn to the
last term in (\ref{dec77}). We bound
\begin{eqnarray}
\frac{1}{n}\sum_{i\in N_n}\left\vert \bar{q}_i-\tilde{q}_i\right\vert
^{2} &\leq &\frac{1}{n}\sum_{i\in N_n}\frac{1}{|S_n(i)|}\sum_{j\in
	S_n(i)}\left\vert \hat{q}_{j}-q_{j}\right\vert ^{2}  \label{arg} \\
& = &\frac{1}{n}\sum_{d\in D_n'}\sum_{i\in N_{n,d}}\frac{1}{%
	|N_{n,d}|}\sum_{j\in N_{n,d}}\left\vert \hat{q}_{j}-q_{j}\right\vert ^{2} 
\notag \\
&=&\frac{1}{n}\sum_{d\in D_n'}\sum_{j\in N_{n,d}}\left\vert \hat{q%
}_{j}-q_{j}\right\vert ^{2}=\frac{1}{n}\sum_{i\in N_n}\left\vert \hat{q}%
_i-q_i\right\vert ^{2}  \notag \\
&=&O_{P}\left( \frac{d_{mx,n}^2 d_{av,n}}{n}\right) \text{,}  \notag
\end{eqnarray}%
where $D_n'\equiv \{d=1,...,d_{mx,n}:|N_{n,d}|\geq 1\}$, the first equality follows because for each $i \in N_{n,d}$, $S_n(i) = N_{n,d}$, and the last
bound follows by Lemma B15(ii) again. Hence%
\begin{equation*}
\frac{1}{n}\sum_{i\in N_n}\left\vert \bar{q}_i-\mathbf{E}%
q_i\right\vert ^{2}\leq \frac{2}{n}\sum_{i\in N_n}\left\vert \tilde{q}%
_i-\mathbf{E}q_i\right\vert ^{2}+O_{P}\left( \frac{d_{mx,n}^2 d_{av,n}}{%
	n}\right) \text{.}
\end{equation*}%
Since $\mathbf{E}q_i=\mathbf{E}\tilde{q}_i$ by Assumption 2(ii), the
expected value of the leading term is bounded by%
\begin{eqnarray}
&&\frac{2}{n}\sum_{d\in D_n'}|N_{n,d}|Var\left( \frac{1}{|N_{n,d}|%
}\sum_{j\in N_{n,d}}q_{j}\right)  \label{dec56} \\
&\leq &\frac{2}{n}\sum_{d\in D_n'}\frac{1}{|N_{n,d}|}\sum_{j\in
	N_{n,d}}Var\left( q_{j}\right) +\frac{2}{n}\sum_{d\in D_n'}\frac{1%
}{|N_{n,d}|}\sum_{j\in N_{n,d}}\sum_{k\in N_{n,d}\cap N_{n,3}(j)}Cov\left(
q_{j},q_{k}\right).  \notag
\end{eqnarray}%
Since $Var\left( q_{j}\right) \leq \mathbf{E}\left[ q_i^{2}\right] \leq C$%
, the leading term is $O(d_{mx,n}/n)$ and the last term is bounded by $O(d_{mx,n}d_{mx,n,3}/n)$  noting that $|D_n'|=O(d_{mx,n})$. Thus we have
\begin{equation}
\frac{1}{n}\sum_{i\in N_n}(\bar{q}_i-\mathbf{E}q_i)^{2}=O_{P}\left( 
\frac{d_{mx,n}d_{mx,n,3}}{n}\right) \text{.}  \label{rate2}
\end{equation}%
From (\ref{dec77}), we deduce that
\begin{equation}
\frac{1}{n}\sum_{i\in N_n}\left\vert \hat{\eta}_i-\eta _i\right\vert
^{2}=O_{P}\left( \frac{d_{mx,n}d_{mx,n,3}}{n}\right) =o_{P}(1)\text{.}
\label{bound}
\end{equation}
Now, using this, we also find that
\begin{equation*}
\frac{1}{n}\sum_{i\in N_n}\left\vert \hat{\eta}_i+\eta_i\right\vert ^{2}\leq \frac{2}{n}\sum_{i\in N_n}\left\vert 2\eta_i\right\vert ^{2}+\frac{2}{n}\sum_{i\in N_n}\left\vert \hat{\eta}_i-%
\eta_i\right\vert ^{2}=O_{P}(1).
\end{equation*}
Therefore,%
\begin{eqnarray*}
	|V_{1n}| &\leq &\frac{1}{n}\sum_{i\in N_n}\left\vert \hat{\eta}_i^{2}-%
	\eta_i^{2}\right\vert \leq \sqrt{\frac{1}{n}\sum_{i\in
			N_n}\left\vert \hat{\eta}_i-\eta_i\right\vert ^{2}}\sqrt{\frac{%
			1}{n}\sum_{i\in N_n}\left\vert \hat{\eta}_i+\eta_i\right\vert
		^{2}} \\
	&=&O_{P}\left( \frac{(d_{mx,n}d_{mx,n,3})^{1/2}}{\sqrt{n}}\right) \text{.}
\end{eqnarray*}

Let us turn to $V_{2n}$. We write%
\begin{eqnarray*}
	&&\frac{1}{n}\sum_{i_{1}\in N_n}\sum_{i_{2}\in N_{n,3}(i_{1})}\left( \hat{%
		\eta}_{i_{1}}\hat{\eta}_{i_{2}}-\eta_{i_1}\eta_{i_2}\right) \\
	&=&\frac{1}{n}\sum_{i_{1}\in N_n}\sum_{i_{2}\in N_{n,3}(i_{1})}\hat{\eta}%
	_{i_{1}}\left( \hat{\eta}_{i_{2}}-\eta_{i_2}\right) +\frac{1}{n}%
	\sum_{i_{1}\in N_n}\sum_{i_{2}\in N_{n,3}(i_{1})}\eta_{i_2}\left( 
	\hat{\eta}_{i_{1}}-\eta_{i_1}\right) .
\end{eqnarray*}%
The leading term on the right hand side is equal to%
\begin{equation}
\frac{1}{n}\sum_{i_{1} \in N_n}\eta_{i_1}\sum_{i_{2}\in
	N_{n,3}(i_{1})}\left( \hat{\eta}_{i_{2}}-\eta_{i_2}\right) +\frac{1}{%
	n}\sum_{i_{1} \in N_n}\left( \hat{\eta}_{i_{1}}-\eta_{i_1}\right)
\sum_{i_{2}\in N_{n,3}(i_{1})}\left( \hat{\eta}_{i_{2}}-\eta_{i_2}\right) .  \label{dec46}
\end{equation}%
The last term is bounded by%
\begin{eqnarray*}
	&&\sqrt{\frac{1}{n}\sum_{i_{1} \in N_n}\left( \hat{\eta}_{i_{1}}-\eta_{i_1}\right) ^{2}}\sqrt{\frac{1}{n}\sum_{i_{1} \in N_n}\left(
		\sum_{i_{2}\in N_{n,3}(i_{1})}\left( \hat{\eta}_{i_{2}}-\eta_{i_2}\right) \right) ^{2}} \\
	&\leq &\frac{d_{mx,n,3}}{n}\sum_{i_{1} \in N_n}\left( \hat{\eta}_{i_{1}}-\eta_{i_1}\right) ^{2}=O_{P}\left( \frac{d_{mx,n}d_{mx,n,3}^2}{n}\right) \text{,}
\end{eqnarray*}%
by (\ref{bound}). The leading term in (\ref{dec46}) is
bounded by%
\begin{eqnarray*}
	&&\sqrt{\frac{1}{n}\sum_{i_{1} \in N_n}\eta_{i_1}^{2}}\sqrt{\frac{1}{%
			n}\sum_{i_{1} \in N_n}\left( \sum_{i_{2}\in N_{n,3}(i_{1})}\left( \hat{\eta}%
		_{i_{2}}-\eta_{i_2}\right) \right) ^{2}} \\
	&&\leq O_{P}(d_{mx,n,3})\sqrt{\frac{1}{n}\sum_{i \in N_n} \left( \hat{\eta}_i-%
		\eta_i\right) ^{2}} = O_P \left(\frac{d_{mx,n,3}^{3/2}d_{mx,n}^{1/2}}{\sqrt{n}}\right),
\end{eqnarray*}%
by applying (\ref{bound}) to the last term. We deduce that%
\begin{equation*}
V_{2n}=\frac{1}{n}\sum_{i_{1}\in N_n}\sum_{i_{2}\in N_{n,3}(i_{1})}\left( 
\hat{\eta}_{i_{1}}\hat{\eta}_{i_{2}}-\eta_{i_1} \eta_{i_2}\right) =O_{P}\left( \frac{d_{mx,n,3}^{3/2}d_{mx,n}^{1/2}}{\sqrt{n}}\right) .
\end{equation*}%
Collecting these results for $V_{1n}$ and $V_{2n},$ we conclude that%
\begin{equation}
V_n=O_{P}\left( \frac{d_{mx,n,3}^{3/2}d_{mx,n}^{1/2}}{\sqrt{n}}\right) \text{.}  \label{V2}
\end{equation}%
$\blacksquare $\medskip

Let%
\begin{equation}
W_n=\frac{1}{n}\sum_{i_{1}\in N_n}\sum_{i_{2}\in \overline{N}_{n,3}(i_{1})}\left( \eta _{i_{1}}\eta _{i_{2}}-\mathbf{E}\left[ \eta
_{i_{1}}\eta _{i_{2}}\right] \right) \text{,}  \label{Wn}
\end{equation}%
where $\overline{N}_{n,3}(i)=N_{n,3}(i)\cup \{i\}$.\medskip

\noindent \textbf{Lemma B17:}\textit{\ Suppose that }$%
d_{mx,n,3}^2d_{av,n}/n=O(1)$\textit{, as }$n\rightarrow \infty $. \textit{%
	Then,}%
\begin{equation*}
W_n=O_{P}\left( \frac{d_{mx,n,3}\sqrt{d_{av,n}}}{\sqrt{n}}\right) \text{.%
}
\end{equation*}%
\medskip

\noindent \textbf{Proof:}\textit{\ }We define $\bar{E}_n=E_n \cup \{ii:i\in N_n\}$ and
\begin{equation}
A_n=\left\{ (i_{1}j_{1},i_{2}j_{2})\in \bar{E}_n\times \bar{E}%
_n:i_{1}j_{1}\sim i_{2}j_{2}\right\} ,  \label{An}
\end{equation}%
which is the set of the pairs of edges that are adjacent. For $\mathbf{i}%
=(i_{1}j_{1},i_{2}j_{2}),$ we define%
\begin{eqnarray}
\xi '(\mathbf{i}) &=&\left[ (e_{i_{1}}e_{j_{1}}-\gamma e_{i_{1}}^{2})-\mathbf{E}(e_{i_{1}}e_{j_{1}}-\gamma e_{i_{1}}^{2})%
\right]  \label{chsi2} \\
&&\times \left[ (e_{i_{2}}e_{j_{2}}-\gamma e_{i_{2}}^{2})-\mathbf{E}%
(e_{i_{2}}e_{j_{2}}-\gamma e_{i_{2}}^{2})\right] ,  \notag
\end{eqnarray}%
and write%
\begin{equation*}
W_n=\frac{1}{n}\sum_{\mathbf{i}\in A_n}\frac{\xi '(\mathbf{i})-%
	\mathbf{E}\left[ \xi '(\mathbf{i})\right] }{%
	d_n^{+}(i_{1})d_n^{+}(i_{2})},
\end{equation*}%
where $d_n^{+}(i)=\max \{d_n(i),1\}$. Now for some $C>0,$%
\begin{eqnarray*}
	Var\left( \frac{1}{n}\sum_{\mathbf{i}\in A_n}\frac{\xi '(\mathbf{i%
		})}{d_n^{+}(i_{1})d_n^{+}(i_{2})}\right) &=&\frac{1}{n^{2}}\sum_{(%
		\mathbf{i},\mathbf{i}')\in A_n\times A_n}\frac{Cov\left( \xi
		'(\mathbf{i}),\xi '(\mathbf{i}')\right) }{%
		d_n^{+}(i_{1})d_n^{+}(i_{2})d_n^{+}(i_{1}^{\prime
		})d_n^{+}(i_{2}')} \\
	&\leq &\frac{C}{n^{2}}\sum_{(\mathbf{i},\mathbf{i}')\in B_n}\frac{%
		1}{d_n^{+}(i_{1})d_n^{+}(i_{2})d_n^{+}(i_{1}^{\prime
		})d_n^{+}(i_{2}')},
\end{eqnarray*}%
using (\ref{gm}), where $B_n=\left\{ (\mathbf{i},\mathbf{i}')\in
A_n\times A_n:\mathbf{i}\sim \mathbf{i}'\right\} .$ Here $%
\mathbf{i}\sim \mathbf{i}'$ means that $\mathbf{i}$ is adjacent to $%
\mathbf{i}'$. The last inequality follows by Assumption 1(iii).
Define $\tilde{N}_{n,3}\subset \tilde{N}_n$ to be such that for any $%
(i_{1},i_{2})\in \tilde{N}_{n,3}$, vertices $i_{1}$ and $i_{2}$ are within
three edges away. Then%
\begin{eqnarray}
&&\sum_{(\mathbf{i},\mathbf{i}')\in B_n}\frac{1}{%
	d_n^{+}(i_{1})d_n^{+}(i_{2})d_n^{+}(i_{1}^{\prime
	})d_n^{+}(i_{2}')}  \label{derV} \\
&=&\sum_{i_{1},i_{2}\in \tilde{N}_{n,3}}\sum_{j_{1}\in \overline{N}%
	_n(i_{1})}\sum_{j_{2}\in \overline{N}_n(i_{2})}\sum_{i_{1}^{\prime
	},i_{2}'\in \tilde{N}_{n,3}}\sum_{j_{1}'\in \overline{N}%
	_n(i_{1}')}\sum_{j_{2}'\in \overline{N}_n(i_{2}')}%
\frac{1\left\{ \mathbf{i}\sim \mathbf{i}'\right\} }{%
	d_n^{+}(i_{1})d_n^{+}(i_{2})d_n^{+}(i_{1}^{\prime
	})d_n^{+}(i_{2}')}  \notag \\
&\leq & 16 \sum_{i_{1},i_{2}\in \tilde{N}_{n,3}}\sum_{j_{1}\in \overline{N}%
	_n(i_{1})}\sum_{j_{2}\in \overline{N}_n(i_{2})}\sum_{i_{1}^{\prime
	},i_{2}'\in \tilde{N}_{n,3}}\sum_{j_{1}'\in \overline{N}%
	_n(i_{1}')}\sum_{j_{2}'\in \overline{N}_n(i_{2}')}%
\frac{1\left\{ i_{1}i_{2}\sim i_{1}'i_{2}'\right\} }{%
	d_n^{+}(i_{1})d_n^{+}(i_{2})d_n^{+}(i_{1}^{\prime
	})d_n^{+}(i_{2}')}.  \notag
\end{eqnarray}%
Note that%
\begin{equation*}
\frac{|\overline{N}_n(i)|}{d_n^{+}(i)}\leq \frac{d_n(i)+1}{d_n^{+}(i)}=%
\frac{d_n(i)}{d_n^{+}(i)}+\frac{1}{d_n^{+}(i)}\leq 2,
\end{equation*}%
and hence the last term in (\ref{derV}) is bounded by%
\begin{equation*}
C\sum_{i_{1},i_{2}\in \tilde{N}_{n,3}}\sum_{i_{1}',i_{2}^{\prime
	}\in \tilde{N}_{n,3}}1\left\{ i_{1}i_{2}\sim i_{1}'i_{2}^{\prime
}\right\} .
\end{equation*}%
The last term is bounded by%
\begin{eqnarray*}
	C\sum_{i_{1},i_{2}\in \tilde{N}_{n,3}}\sum_{i_{1}'\in
		N_n(i_{1})}\sum_{i_{2}'\in N_{n,3}(i_{1}')}1 &\leq
	&Cd_{mx,n,3}\sum_{i_{1},i_{2}\in \tilde{N}_{n,3}}\sum_{i_{1}'\in
		N_n(i_{1})}1=Cd_{mx,n,3}\sum_{i_{1},i_{2}\in \tilde{N}_{n,3}}d_n(i_{1})
	\\
	&=&Cd_{mx,n,3}\sum_{i_{1}\in N_n}d_n(i_{1})\sum_{i_{2}\in
		N_{n,3}(i_{1})}1 \\
	&\leq &Cnd_{mx,n,3}^2\frac{1}{n}\sum_{i_{1}\in
		N_n}d_n(i_{1})=Cnd_{mx,n,3}^2d_{av,n}.
\end{eqnarray*}%
Therefore,%
\begin{equation}
Var\left( \frac{1}{n}\sum_{\mathbf{i}\in A_n}\frac{\xi '(\mathbf{i%
	})}{d_n^+(i_{1})d_n^+(i_{2})}\right) =O\left( \frac{nd_{mx,n,3}^2d_{av,n}}{%
	n^{2}}\right) =O\left( \frac{d_{mx,n,3}^{2}d_{av,n}}{n}\right) .  \label{rb}
\end{equation}%
This gives the desired rate. $\blacksquare $\medskip

\noindent \textbf{Lemma B18:}\textit{\ Suppose that }$d_{mx,n,3}^3 d_{mx,n}/n%
\rightarrow 0$\textit{, as }$n\rightarrow \infty $. \textit{Then}%
\begin{equation*}
\hat{\sigma}_n^{2}=Var\left( \zeta _n\right) +O_{P}\left(\frac{d_{mx,n,3}^{3/2}d_{mx,n}^{1/2}}{\sqrt{
		n}}\right),
\end{equation*}%
\textit{where we recall the definition }$\zeta _n \equiv \frac{1}{\sqrt{n}}\sum_{i\in N_n}\left( q_i-%
\mathbf{E}q_i\right) .$\medskip

\noindent \textbf{Proof :}\textit{\ }We write 
\begin{equation}
Var\left( \zeta _n\right) =\sigma _n^{2}+\frac{1}{n}\sum_{i_{1}\in
	N_n}\sum_{i_{2}\in N_n\backslash \overline{N}_{n,3}(i_{1})}\mathbf{E}\left[ \eta
_{i_{1}}\eta _{i_{2}}\right] ,  \label{dec811}
\end{equation}%
where%
\begin{equation*}
\sigma _n^{2}=\frac{1}{n}\sum_{i_{1}\in N_n}\sum_{i_{2}\in \overline{N}%
	_{n,3}(i_{1})}\mathbf{E}\left[ \eta _{i_{1}}\eta _{i_{2}}\right] .
\end{equation*}%
However, whenever $i_{1}\in N_n$ and $i_{2}\in N_n\backslash
\overline{N}_{n,3}(i_{1})$, $q_{i_{1}}$ and $q_{i_{2}}$ are independent and hence 
\begin{equation*}
\mathbf{E}\left[ \eta _{i_{1}}\eta _{i_{2}}\right] =\mathbf{E}\left[ \eta
_{i_{1}}\right] \mathbf{E}\left[ \eta _{i_{2}}\right] =0.
\end{equation*}%
Thus $Var(\zeta _n)=\sigma _n^{2}$. Now note that%
\begin{equation}
\hat{\sigma}_n^{2}=\sigma _n^{2}+R_n,  \label{var}
\end{equation}%
where $R_n=V_n+W_n$. The desired result follows by Lemmas B16-B17. $%
\blacksquare $\medskip \medskip

\noindent {\Large 2.5. Consistency of Permutation Variance Estimators}%
\medskip

\noindent \textbf{2.5.1 First Order Analysis of Estimation Errors}\medskip

\noindent \textbf{Lemma B19:}(i) \textit{\ Suppose that }$d_{mx,n}^{1/2}d_{av,n}/n%
\rightarrow 0$ \textit{as} $n\rightarrow \infty $. \textit{Then}%
\begin{equation*}
\frac{1}{n}\sum_{i\in N_n}e_{\pi (i)}^{2}\left( \hat{a}_{i,\pi }-a_{i,\pi
}\right) ^{2}=O_{P}\left( \frac{d_{mx,n}^{1/2}d_{av,n}}{n}\right) ,\ \Pi _n%
\text{-unif.}
\end{equation*}

\noindent (ii) \textit{Suppose that} $d_{mx,n}^2 d_{av,n}/n
\rightarrow 0$ \textit{as} $n\rightarrow \infty $. \textit{Then}
\begin{equation*}
\frac{1}{n}\sum_{i\in N_n}\left( \hat{q}_{i,\pi }-q_{i,\pi }\right)
^{2}=O_{P}\left( \frac{d_{mx,n}^2 d_{av,n}}{n}\right) ,\ \Pi _n\text{%
	-unif.}
\end{equation*}%
\medskip

\noindent \textbf{Proof:}\textit{\ }(i) Similarly as before, we bound $\frac{%
	1}{n}\sum_{i\in N_n}e_{\pi (i)}^{2}\left( \hat{a}_{i,\pi }-a_{i,\pi
}\right) ^{2}$ by%
\begin{eqnarray}
&&\sqrt{\frac{1}{n}\sum_{i\in N_n}e_i^{4}}\sqrt{\frac{1}{n}\sum_{i\in
		N_n}\left( \hat{a}_{i,\pi }-a_{i,\pi }\right) ^{4}}  \label{der67} \\
&\leq &O_{P}(1)\sqrt{\frac{1}{n}\sum_{i\in N_n:d_n(i)\geq 1}\frac{1}{%
		d_n(i)}\sum_{j\in N_n(i)}\left( \hat{e}_{\pi (j)}-e_{\pi (j)}\right) ^{4}%
}  \notag \\
&\leq &O_{P}(d_{mx,n}^{1/2})\sqrt{\frac{1}{n}\sum_{i\in N_n}\left( \hat{e}%
	_i-e_i\right) ^{4}}=O_{P}\left( \frac{d_{mx,n}^{1/2}d_{av,n}}{n}\right) ,
\notag
\end{eqnarray}%
by Lemma B2.\medskip

\noindent (ii) Bound%
\begin{equation}
\frac{1}{n}\sum_{i\in N_n}\left\vert \hat{q}_{i,\pi }-q_{i,\pi
}\right\vert ^{2}\leq \frac{2}{n}\sum_{i\in N_n}\left( \hat{e}_{\pi
(i)}-e_{\pi (i)}\right) ^{2}\hat{b}_{i,\pi }^{2}+\frac{2}{n}\sum_{i\in
N_n}e_{\pi (i)}^{2}(\hat{b}_{i,\pi }-b_{i,\pi })^{2},  \label{dec567}
\end{equation}%
where $\hat{b}_{i,\pi }=\hat{a}_{i,\pi }-\hat{\gamma}_{\pi }\hat{e}_{\pi
	(i)}$ and $b_{i,\pi }=a_{i,\pi }-\gamma_{\pi}e_{\pi (i)}$. The leading
term is bounded by 
\begin{equation*}
\frac{4}{n}\sum_{i\in N_n}\left( \hat{e}_{\pi (i)}-e_{\pi (i)}\right) ^{2}(%
\hat{b}_{i,\pi }-b_{i,\pi })^{2}+\frac{4}{n}\sum_{i\in N_n}\left( \hat{e}%
_{\pi (i)}-e_{\pi (i)}\right) ^{2}b_{i,\pi }^{2}.
\end{equation*}%
Again the leading term is dominated by the second term. We focus on the
second term which we bound by (using (\ref{dec67})) 
\begin{eqnarray}
&&\frac{8\bar{\varepsilon}^{2}}{\hat{v}^{2}}\frac{1}{n}\sum_{i\in
	N_n}b_{i,\pi }^{2}+8v_n^{2}\left( \frac{1}{\hat{v}}-\frac{1}{v_n}%
\right) ^{2}\frac{1}{n}\sum_{i\in N_n}e_{\pi (i)}^{2}b_{i,\pi }^{2}
\label{dec43} \\
&=&O_{P}\left( \frac{d_{av,n}}{n}\right) \cdot \frac{1}{n}\sum_{i\in
	N_n}b_{i,\pi }^{2}+8v_n^{2}\left( \frac{1}{\hat{v}}-\frac{1}{v_n}%
\right) ^{2}\frac{1}{n}\sum_{i\in N_n}q_{i,\pi }^{2}.  \notag
\end{eqnarray}%
As for the leading term, observe that
\begin{eqnarray*}
	\frac{1}{n}\sum_{i\in N_n}b_{i,\pi }^{2} &\leq &\frac{2}{n}\sum_{i\in
		N_n}a_{i,\pi }^{2}+\gamma_{\pi}^{2}\frac{2}{n}\sum_{i\in
		N_n}e_i^{2} \\
	&\leq &\frac{2}{n}\sum_{i\in N_n:d_n(i)\geq 1}\frac{1}{d_n(i)}%
	\sum_{j\in N_n(i)}e_{\pi (j)}^{2}+\gamma_{\pi}^{2}\frac{2}{n}%
	\sum_{i\in N_n}e_i^{2}.
\end{eqnarray*}%
The conditional expectation of the leading term given $\pi $ is bounded by%
\begin{equation}
2\max_{i \in N_n}\mathbf{E}\left[ e_i^{2}\right] \leq 2M,  \label{b1}
\end{equation}%
and the last term is $O_{P}(1),$ $\Pi _n$-unif., because its conditional
expectation given $\pi $ is bounded by $2MC^2$ for $C>0$ in (\ref{gmp}). As
for the last term in (\ref{dec43}),%
\begin{equation}
\frac{1}{n}\sum_{i\in N_n}q_{i,\pi }^{2}\leq \frac{2}{n}\sum_{i\in
	N_n}e_{\pi (i)}^{2}a_{i,\pi }^{2}+\gamma_{\pi}^{2}\frac{2}{n}%
\sum_{i\in N_n}e_i^{4}\leq \frac{2}{n}\sum_{i\in N_n}e_{\pi
	(i)}^{2}a_{i,\pi }^{2}+O_{P}(1).  \label{b24}
\end{equation}%
Note that $\mathbf{E}[e_{\pi (i)}^{2}a_{i,\pi}^{2}|\pi ]\leq C$ (see (\ref%
{bd725})). Applying this to (\ref{b24}) and the last term in (\ref{dec43})
and combining with the fact $\hat{v}^{-1}-v_n^{-1}=O_{P}(d_{av,n}^{1/2}/%
\sqrt{n})$ from (\ref{bd67}), we conclude for the leading term in (\ref{dec567})
the following:%
\begin{equation*}
\frac{2}{n}\sum_{i\in N_n}\left( \hat{e}_{\pi (i)}-e_{\pi (i)}\right) ^{2}%
\hat{b}_{i,\pi }^{2}=O_{P}\left( \frac{d_{av,n}}{n}\right) ,\ \Pi _n\text{%
	-unif.}
\end{equation*}%
As for the last term in (\ref{dec567}), it is bounded by%
\begin{eqnarray*}
	&&\frac{4}{n}\sum_{i\in N_n}e_{\pi (i)}^{2}\left( \hat{a}_{i,\pi
	}-a_{i,\pi }\right) ^{2}+\frac{4}{n}\sum_{i\in N_n}e_{\pi (i)}^{2}\left( 
	\hat{\gamma}_{\pi }\hat{e}_{\pi (i)}-\gamma_{\pi}e_{\pi (i)}\right)
	^{2} \\
	&=&O_{P}\left( \frac{d_{mx,n}^{1/2}d_{av,n}}{n}\right) +\frac{4}{n}%
	\sum_{i\in N_n}e_{\pi (i)}^{2}\left( \hat{\gamma}_{\pi }\hat{e}_{\pi
		(i)}-\gamma_{\pi}e_{\pi (i)}\right) ^{2},
\end{eqnarray*}%
by (i) of this lemma. Using Lemmas B5(ii), arguments in the proof of Lemma B9(ii), and Lemma B4(ii),  
\begin{eqnarray*}
	\hat{\gamma}_{\pi }-\gamma_{\pi} &=& \frac{1}{n}\sum_{i\in N_n}\left\{ 
	\hat{e}_{\pi (i)}\hat{a}_{i,\pi }-\mathbf{E}[e_{\pi (i)}a_{i,\pi }|\pi
	]\right\} \\
	&=& O_{P}\left( \frac{d_{mx,n}d_{av,n}^{1/2}}{\sqrt{n}}\right) \text{,%
	}\ \Pi _n\text{-unif.}
\end{eqnarray*}%
Using this, and following similar arguments as in the proof of Lemma B15(ii), we obtain the desired
rate. $\blacksquare $\medskip

Lemmas B20-B21 below are permutation analogues of Lemmas B16-B17. For each $%
i\in N_n,$ let%
\begin{equation*}
\hat{\eta}_{i,\pi }=\hat{q}_{i,\pi }-\bar{q}_{i,\pi },
\end{equation*}%
where $\bar{q}_{i,\pi }=\frac{1}{|S_n(i)|}\sum_{j\in S_n(i)}\hat{q}%
_{j,\pi }$. Define%
\begin{equation*}
V_{n,\pi }=\frac{1}{n}\sum_{i_{1}\in N_n}\sum_{i_{2}\in \overline{N}_{n,3}(i_{1})}\left( \hat{\eta}_{i_{1},\pi }\hat{\eta}_{i_{2},\pi }-\eta_{i_1,\pi } \eta_{i_2,\pi}\right) .
\end{equation*}%
\medskip

\noindent \textbf{Lemma B20:}\textit{\ Suppose that }$d_{mx,n,3}^3 d_{mx,n}/n=O(1)$%
\textit{, as }$n\rightarrow \infty $. \textit{Then} 
\begin{equation*}
V_{n,\pi }=O_{P}\left( \frac{d_{mx,n,3}^{3/2}d_{mx,n}^{1/2}}{\sqrt{n}}\right) ,\ \Pi _n%
\text{-unif.}
\end{equation*}%
\medskip

\noindent \textbf{Proof: }We follow the proof for $V_n$ in Lemma B16
closely. Since we have prepared Lemma B19(ii), a close inspection of the
proof reveals that it suffices to show that%
\begin{equation}
\frac{1}{n}\sum_{i\in N_n}\left\vert \bar{q}_{i,\pi }-\mathbf{E}[\tilde{q}%
_{i,\pi }|\pi ]\right\vert ^{2}=O_{P}\left( \frac{d_{mx,n}d_{mx,n,3}}{n}\right) ,\
\Pi _n\text{-unif.}  \label{bd7}
\end{equation}%
From the arguments in (\ref{arg}),%
\begin{eqnarray*}
	\frac{1}{n}\sum_{i\in N_n}\left\vert \bar{q}_{i,\pi }-\tilde{q}_{i,\pi
	}\right\vert ^{2} &\leq &\frac{1}{n}\sum_{i\in N_n}\left\vert \hat{q}%
	_{i,\pi }-q_{i,\pi }\right\vert ^{2} \\
	&=&O_{P}\left( \frac{d_{mx,n}^2 d_{av,n}}{n}\right) ,\ \Pi _n\text{-unif.,}
\end{eqnarray*}%
by Lemma B19(ii). Hence%
\begin{equation*}
\frac{1}{n}\sum_{i\in N_n}\left\vert \bar{q}_{i,\pi }-\mathbf{E}[\tilde{q}%
_{i,\pi }|\pi ]\right\vert ^{2}\leq \frac{2}{n}\sum_{i\in N_n}\left\vert 
\tilde{q}_{i,\pi }-\mathbf{E}[\tilde{q}_{i,\pi }|\pi ]\right\vert
^{2}+O_{P}\left( \frac{d_{mx,n}^2 d_{av,n}}{n}\right) ,\ \Pi _n\text{%
	-unif.}
\end{equation*}%
The conditional expectation (given $\pi $) of the leading term on the right
hand side is bounded by%
\begin{eqnarray*}
	&&\frac{2}{n}\sum_{d\in D_n'}|N_{n,d}|Var\left( \frac{1}{|N_{n,d}|%
	}\sum_{j\in N_{n,d}}q_{j,\pi }|\pi \right) \\
	&\leq &\frac{2}{n}\sum_{d\in D_n'}\frac{1}{|N_{n,d}|}\sum_{j\in
		N_{n,d}}Var\left( q_{j,\pi }|\pi \right) \\
	&&+\frac{2}{n}\sum_{d\in D_n'}\frac{1}{|N_{n,d}|}\sum_{j\in
		N_{n,d}}\sum_{k\in N_{n,d}\cap N_{n,3}(j)}Cov\left( q_{j,\pi },q_{k,\pi
	}|\pi \right) .
\end{eqnarray*}%
By (\ref{bdq}), the leading term is $O(d_{mx,n}/n), \Pi_n \textnormal{unif.}$, and the last term is
bounded by%
\begin{equation*}
\frac{2d_{mx,n,3}}{n}\sum_{d\in D_n'}\frac{1}{|N_{n,d}|}\sum_{j\in
	N_{n,d}}\max_{\pi \in \Pi _n}\sqrt{\mathbf{E}(q_{1,\pi }^{2}|\pi )}%
\sqrt{\mathbf{E}(q_{j,\pi }^{2}|\pi )}=O\left( \frac{d_{mx,n,3}d_{mx,n}}{n}\right)
,
\end{equation*}%
for some $C>0.$ Thus we have (\ref{bd7}). Following the same arguments in
the proof of (\ref{V2}), we obtain the desired result. $\blacksquare $%
\medskip

Recall the definition%
\begin{equation*}
\tilde{\eta}_{i,\pi }=q_{i,\pi }-\mathbf{E}\left[ \tilde{q}_{i,\pi }|\pi %
\right] \text{ and }\tilde{q}_{i,\pi }=\frac{1}{|S_n(i)|}\sum_{j\in
	S_n(i)}q_{j,\pi }.
\end{equation*}%
Let 
\begin{equation*}
W_{n,\pi }=\frac{1}{n}\sum_{i_{1}\in N_n}\sum_{i_{2}\in \overline{N}_{n,3}(i_{1})}\left( \tilde{\eta}_{i_{1},\pi }\tilde{\eta}_{i_{2},\pi }-%
\mathbf{E}\left[ \tilde{\eta}_{i_{1},\pi }\tilde{\eta}_{i_{2},\pi }|\pi %
\right] \right) .
\end{equation*}%
\medskip

\noindent \textbf{Lemma B21:}\textit{\ Suppose that }$%
d_{mx,n,3}^2d_{av,n}/n=O(1)$\textit{, as }$n\rightarrow \infty $. \textit{%
	Then} 
\begin{equation*}
\frac{1}{|\Pi _n|}\sum_{\pi \in \Pi _n}\left\vert W_{n,\pi }\right\vert
=O_{P}\left( \frac{d_{mx,n,3}\sqrt{d_{av,n}}}{\sqrt{n}}\right).
\end{equation*}%
\medskip

\noindent \textbf{Proof:}\textit{\ }First, we let
\begin{eqnarray}
    \label{W pi 12}
	W_{1n,\pi } &=&\frac{1}{n}\sum_{i_{1}\in N_n}\sum_{i_{2}\in \overline{N}%
		_{n,3}(i_{1})}\left( \tilde{\eta}_{i_{1},\pi }\tilde{\eta}_{i_{2},\pi }-\eta
	_{i_{1},\pi }\eta _{i_{2},\pi }\right) \text{ and} \\
	W_{2n,\pi } &=&\frac{1}{n}\sum_{i_{1}\in N_n}\sum_{i_{2}\in \overline{N}%
		_{n,3}(i_{1})}\mathbf{E}\left[ \tilde{\eta}_{i_{1},\pi }\tilde{\eta}%
	_{i_{2},\pi }-\eta _{i_{1},\pi }\eta _{i_{2},\pi }|\pi \right] .
\end{eqnarray}%
Then we can write%
\begin{equation*}
W_{n,\pi }=W_{1n,\pi }-W_{2n,\pi }+\tilde{W}_{n,\pi },
\end{equation*}%
where%
\begin{equation*}
\tilde{W}_{n,\pi }=\frac{1}{n}\sum_{i_{1}\in N_n}\sum_{i_{2}\in \overline{N}%
	_{n,3}(i_{1})}\left( \eta _{i_{1},\pi }\eta _{i_{2},\pi }-\mathbf{E}\left[
\eta _{i_{1},\pi }\eta _{i_{2},\pi }|\pi \right] \right) .
\end{equation*}%
We show that%
\begin{eqnarray*}
	\frac{1}{|\Pi _n|}\sum_{\pi \in \Pi _n}\mathbf{E}\left[ \left\vert
	W_{1n,\pi }\right\vert |\pi \right] &=&O\left( \frac{d_{av,n}d_{mx,n,3}}{n}\right) ,\text{ and} \\
	\frac{1}{|\Pi _n|}\sum_{\pi \in \Pi _n}\left\vert W_{2n,\pi }\right\vert
	&=&O\left( \frac{d_{av,n}d_{mx,n,3}}{n}\right).
\end{eqnarray*}%
We bound%
\begin{eqnarray}
\left\vert W_{1n,\pi }\right\vert &\leq &\frac{1}{n}\sum_{i_{1}\in
	N_n}\sum_{i_{2}\in \overline{N}_{n,3}(i_{1})}\left\vert \tilde{\eta}_{i_{1},\pi
}(\tilde{\eta}_{i_{2},\pi }-\eta _{i_{2},\pi })\right\vert  \label{dec673} \\
&&+\frac{1}{n}\sum_{i_{1}\in N_n}\sum_{i_{2}\in \overline{N}_{n,3}(i_{1})}\left%
\vert (\tilde{\eta}_{i_{1},\pi }-\eta _{i_{1},\pi })\eta _{i_{2},\pi
}\right\vert .  \notag
\end{eqnarray}%
Consider the leading term. Since $\tilde{\eta}_{i,\pi }-\eta _{i,\pi }=%
\mathbf{E}\left[ q_{i,\pi }|\pi \right] -\mathbf{E}\left[ \tilde{q}%
_{i,\pi }|\pi \right] $ and $\mathbf{E}[|\tilde{\eta}_{i,\pi }||\pi
]\leq C$, we bound 
\begin{eqnarray*}
	&&\frac{1}{|\Pi _n|}\sum_{\pi \in \Pi _n}\frac{1}{n}\sum_{i_{1}\in
		N_n}\sum_{i_{2}\in \overline{N}_{n,3}(i_{1})}\mathbf{E}\left[ \left\vert \tilde{%
		\eta}_{i_{1},\pi }(\tilde{\eta}_{i_{2},\pi }-\eta _{i_{2},\pi })\right\vert
	|\pi \right] \\
	&\leq &\frac{Cd_{mx,n,3}}{|\Pi _n|}\sum_{\pi \in \Pi _n}\frac{1}{n}\sum_{i_{1}\in
		N_n}\frac{1}{\overline{N}_{n,3}(i_{1})}\sum_{i_{2}\in \overline{N}%
		_{n,3}(i_{1})}\left\vert \mathbf{E}\left[ q_{i_{2},\pi }|\pi \right] -%
	\mathbf{E}\left[ \tilde{q}_{i_{2},\pi }|\pi \right] \right\vert.
\end{eqnarray*}
The last term is bounded by
\begin{eqnarray*}
	&&\frac{Cd_{mx,n,3}}{|\Pi _n|}\sum_{\pi \in \Pi _n}\frac{1}{n}\sum_{i_{1}\in
		N_n}\frac{1}{\overline{N}_{n,3}(i_{1})}\sum_{i_{2}\in \overline{N}%
		_{n,3}(i_{1})}\left\vert \mathbf{E}\left[ q_{i_{2},\pi }|\pi \right]
	\right\vert \\
	&&+\frac{Cd_{mx,n,3}}{|\Pi _n|}\sum_{\pi \in \Pi _n}\frac{1}{n}\sum_{i_{1}\in
		N_n}\frac{1}{\overline{N}_{n,3}(i_{1})}\sum_{i_{2}\in \overline{N}_{n,3}(i_{1})}%
	\frac{1}{|S_n(i_{2})|}\sum_{j\in S_n(i_{2})}\left\vert \mathbf{E}\left[
	q_{j,\pi }|\pi \right] \right\vert.
\end{eqnarray*}
By Lemma B5(ii), the leading term is $O(d_{av,n}d_{mx,n,3}/n)$. From the proof of
Lemma B5(ii), it is easily seen that the last term is $O(d_{av,n}d_{mx,n,3}/n)$ as well. Applying the same argument to the last term in (\ref{dec673}), we
find that%
\begin{equation}
\label{W1 pi}
\frac{1}{|\Pi _n|}\sum_{\pi \in \Pi _n}\mathbf{E}\left[ \left\vert
W_{1n,\pi }\right\vert |\pi \right] =O\left( \frac{d_{av,n}d_{mx,n,3}}{n}\right) .
\end{equation}
Applying the similar arguments to $W_{2n,\pi }$, we conclude that%
\begin{eqnarray}
&& \label{conv} \frac{1}{|\Pi _n|}\sum_{\pi \in \Pi _n}\mathbf{E}\left[ \vert
W_{n,\pi }-\tilde{W}_{n,\pi }\vert |\pi \right] 
\\
&& =\frac{1}{|\Pi _n|}
\sum_{\pi \in \Pi _n}\mathbf{E}\left[ \left\vert W_{1n,\pi }-W_{2n,\pi
}\right\vert |\pi \right] =O\left( \frac{d_{av,n}d_{mx,n,3}}{n}\right) .
\end{eqnarray}

Now, we compute the convergence rate for $\tilde{W}_{n,\pi }$ following the
same argument used for $W_n$ in the proof of Lemma B17. For each $\mathbf{i}=(i_{1}j_{1},i_{2}j_{2})$%
, define $\xi _{\pi }'(\mathbf{i})$ as $\xi '(\mathbf{i})$
except that $e_i$ and $\gamma$ are replaced by $e_{\pi (i)}$ and $%
\gamma_{\pi}$. With $A_n$ and $d_n^{+}(i)$ defined in the proof of
Lemma B17, we write%
\begin{equation*}
W_{n,\pi }=\frac{1}{n}\sum_{\mathbf{i}\in A_n}\frac{\xi _{\pi }(\mathbf{i}%
	)-\mathbf{E}\left[ \xi _{\pi }(\mathbf{i})|\pi \right] }{%
	d_n^{+}(i_{1})d_n^{+}(i_{2})}.
\end{equation*}%
For some $C>0,$%
\begin{eqnarray*}
	Var\left( \frac{1}{n}\sum_{\mathbf{i}\in A_n}\frac{\xi _{\pi }(\mathbf{i})%
	}{d_n^{+}(i_{1})d_n^{+}(i_{2})}|\pi \right) &=&\frac{1}{n^{2}}\sum_{(%
	\mathbf{i},\mathbf{i}')\in A_n\times A_n}\frac{Cov\left( \xi
	_{\pi }(\mathbf{i}),\xi _{\pi }(\mathbf{i}')|\pi \right) }{%
	d_n^{+}(i_{1})d_n^{+}(i_{2})d_n^{+}(i_{1}^{\prime
	})d_n^{+}(i_{2}')} \\
&\leq &\frac{C}{n^{2}}\sum_{(\mathbf{i},\mathbf{i}')\in B_{n,\pi }}%
\frac{1}{d_n^{+}(i_{1})d_n^{+}(i_{2})d_n^{+}(i_{1}^{\prime
	})d_n^{+}(i_{2}')},
\end{eqnarray*}%
using (\ref{gmp}), where $B_{n,\pi }=\left\{ (\mathbf{i},\mathbf{i}^{\prime
})\in A_n\times A_n:\pi (\mathbf{i})\sim \pi (\mathbf{i}^{\prime
})\right\} .$ Then%
\begin{eqnarray*}
	&&\sum_{(\mathbf{i},\mathbf{i}')\in B_{n,\pi }}\frac{1}{%
		d_n^{+}(i_{1})d_n^{+}(i_{2})d_n^{+}(i_{1}^{\prime
		})d_n^{+}(i_{2}')} \\
	&=&\sum_{i_{1},i_{2}\in \tilde{N}_{n,3}}\sum_{j_{1}\in \overline{N}%
		_n(i_{1})}\sum_{j_{2}\in \overline{N}_n(i_{2})}\sum_{i_{1}^{\prime
		},i_{2}'\in \tilde{N}_{n,3}}\sum_{j_{1}'\in \overline{N}%
		_n(i_{1}')}\sum_{j_{2}'\in \overline{N}_n(i_{2}')}%
	\frac{1\left\{ \pi (\mathbf{i})\sim \pi (\mathbf{i}')\right\} }{%
		d_n^{+}(i_{1})d_n^{+}(i_{2})d_n^{+}(i_{1}^{\prime
		})d_n^{+}(i_{2}')} \\
	&\leq &C\sum_{i_{1},i_{2}\in \tilde{N}_{n,3}}\sum_{i_{1}^{\prime
		},i_{2}'\in \tilde{N}_{n,3}}1\left\{ \pi (i_{1})\pi (i_{2})\sim \pi
	(i_{1}')\pi (i_{2}')\right\} ,
\end{eqnarray*}%
similarly as in (\ref{derV}). Let $G_{n,\pi }=(N_n,E_{n,\pi })$ be the
graph where $E_{n,\pi }=\{ij\in \tilde{N}_n:\pi (i)\pi (j)\in E_n\}$.
Define%
\begin{equation*}
N_{n,\pi }(i)=\{j\in N_n:\pi (i)\pi (j)\in E_n\},
\end{equation*}%
and let $\overline{N}_{n,\pi }(i)=N_{n,\pi }(i) \cup \{i\}$ and $\overline{N}_{n,3,\pi }(i)$ be the set of vertices that are
within three edges from $i$ in $G_{n,\pi }$ (including the vertex $i$
itself). Also, define $d_{n,\pi }(i)$ to be the degree of $i$ in $G_{n,\pi }$%
. Note that the maximum degree, the maximum 3-degree, and the average degree of $G_{n,\pi }$ are
the same as those of $G_n$, i.e., $d_{mx,n}$, $d_{mx,n,3}$, and $d_{av,n}$. Hence the
last double sum is bounded by
\begin{eqnarray*}
	\sum_{i_{1},i_{2}\in \tilde{N}_{n,3}}\sum_{i_{1}'\in \overline{N}_{n,\pi
		}(i_{1})}\sum_{i_{2}'\in \overline{N}_{n,3,\pi }(i_{1}')}1
	&\leq & d_{mx,n,3}\sum_{i_{1},i_{2}\in \tilde{N}_{n,3}}\sum_{i_{1}^{\prime
		}\in \overline{N}_{n,\pi }(i_{1})}1=d_{mx,n,3}\sum_{i_{1},i_{2}\in \tilde{N}%
		_{n,3}}d_{n,\pi }(i_{1}) \\
	&=&d_{mx,n,3}\sum_{i_{1}\in N_n}d_{n,\pi }(i_{1})\sum_{i_{2}\in \overline{N}%
		_{n,3,\pi }(i_{1})}1 \\
	&\leq &nd_{mx,n,3}^{2}\frac{1}{n}\sum_{i_{1}\in N_n}(d_{n,\pi
	}(i_{1})+1)=O(nd_{mx,n,3}^{2}d_{av,n}).
\end{eqnarray*}%
Therefore,%
\begin{equation*}
Var\left( \frac{1}{n}\sum_{\mathbf{i}\in A_n}\frac{\xi _{\pi }(\mathbf{i})%
}{d_n^+(i_{1})d_n^+(i_{2})}|\pi \right) =O\left( \frac{d_{mx,n,3}^{2}d_{av,n}}{%
n}\right) .
\end{equation*}%
Thus, we obtain that 
\begin{equation*}
\tilde{W}_{n,\pi }=O_{P}\left( \frac{d_{mx,n,3}\sqrt{d_{av,n}}}{\sqrt{n}}%
\right) ,\ \Pi _n\text{-unif.,}
\end{equation*}%
which, combined with (\ref{conv}), gives us the desired result. $%
\blacksquare $\medskip

\noindent \textbf{2.5.2 Consistency of Permutation\ Variance Estimator}%
\medskip

Recall the definitions:%
\begin{equation*}
h_{n,\pi }^{2}=Var\left( \zeta _{n,\pi }|\pi \right) =\mathbf{E}\left[\left( \frac{%
	1}{\sqrt{n}}\sum_{i \in N_n}\eta _{i,\pi } \right) ^{2} \vert \pi \right],
\end{equation*}%
and%
\begin{equation*}
\sigma _{n,\pi }^{2}=\frac{1}{n}\sum_{i_{1}\in N_n}\sum_{i_{2}\in \overline{N}%
	_{n,3}(i_{1})}\mathbf{E}\left[ \tilde{\eta}_{i_{1},\pi }\tilde{\eta}%
_{i_{2},\pi }|\pi \right] .
\end{equation*}%
The following result shows that $h_{n,\pi }^{2}$ and $\hat{\sigma}_{\pi
}^{2} $ are close to each other.\medskip

\noindent \textbf{Lemma B22:}\textit{\ Suppose that }$d_{mx,n,3}^3d_{mx,n}/n%
\rightarrow 0$\textit{, as }$n\rightarrow \infty $. \textit{Then for each }$%
\varepsilon >0$, \textit{there exists }$\Pi _n(\varepsilon )\subset \Pi
_n$\textit{\ such that}$\ |\Pi _n(\varepsilon )|/|\Pi _n|\rightarrow 1$%
\textit{\ as }$n\rightarrow \infty $\textit{, and for all }$\pi \in \Pi
_n(\varepsilon ),$%
\begin{equation*}
\left\vert h_{n,\pi }^{2}-\hat{\sigma}_{\pi }^{2}\right\vert \leq
\varepsilon +O_{P}\left( \frac{d_{mx,n,3}^{3/2}d_{mx,n}^{1/2}}{\sqrt{n}}\right) .
\end{equation*}

\noindent \textbf{Proof :} First observe that
\begin{equation*}
\hat{\sigma}_{\pi }^{2}=\sigma _{n,\pi }^{2}+R_{n,\pi },
\end{equation*}%
where $R_{n,\pi}=V_{n,\pi }+W_{n,\pi}-W_{1n,\pi}$ and $W_{1n,\pi}$ is defined in (\ref{W pi 12}). By Lemmas B20-B21 and (\ref{W1 pi}),
\begin{equation*}
\frac{1}{|\Pi _n|}\sum_{\pi \in \Pi _n}|R_{n,\pi }|=O_{P}\left( \frac{
	d_{mx,n,3}^{3/2}d_{mx,n}^{1/2}}{\sqrt{n}}\right).
\end{equation*}
We now compare $\sigma _{n,\pi }^{2}$ and $h_{n,\pi }^{2}$. First observe
that%
\begin{eqnarray*}
	h_{n,\pi }^{2}-\tilde{h}_{n,\pi }^{2} &=&\frac{1}{n}\sum_{i_{1}\in
		N_n}\sum_{i_{2}\in N_n\backslash \{i_{1}\}}\mathbf{E}\left[ \eta
	_{i_{1},\pi }\eta _{i_{2},\pi }|\pi \right] \text{ and} \\
	\sigma _{n,\pi }^{2}-\tilde{\sigma}_{n,\pi }^{2} &=&\frac{1}{n}%
	\sum_{i_{1}\in N_n}\sum_{i_{2}\in N_{n,3}(i_{1})}\mathbf{E}\left[ \tilde{%
		\eta}_{i_{1},\pi }\tilde{\eta}_{i_{2},\pi }|\pi \right] ,
\end{eqnarray*}%
where 
\begin{equation*}
\tilde{h}_{n,\pi }^{2}=\frac{1}{n}\sum_{i\in N_n}\mathbf{E}\left[ \eta
_{i,\pi }^{2}|\pi \right] \text{ and\ }\tilde{\sigma}_{n,\pi }^{2}=\frac{%
	1}{n}\sum_{i\in N_n}\mathbf{E}\left[ \tilde{\eta}_{i,\pi }^{2}|\pi %
\right] .
\end{equation*}%
By Lemma B10,%
\begin{eqnarray*}
	\frac{1}{|\Pi _n|}\sum_{\pi \in \Pi _n}|h_{n,\pi }^{2}-\tilde{h}_{n,\pi
	}^{2}| &=&O\left( \frac{d_{av,n}}{\sqrt{n}}\right) \text{ and} \\
	\frac{1}{|\Pi _n|}\sum_{\pi \in \Pi _n}|\sigma _{n,\pi }^{2}-\tilde{%
		\sigma}_{n,\pi }^{2}| &=&O\left( \frac{d_{av,n}}{\sqrt{n}}\right) ,
\end{eqnarray*}%
and by Lemma B11, 
\begin{equation*}
\frac{1}{|\Pi _n|}\sum_{\pi \in \Pi _n}\left\vert \tilde{\sigma}_{n,\pi
}^{2}-\tilde{h}_{n,\pi }^{2}\right\vert =O\left( \frac{d_{av,n}}{n}%
\right) .
\end{equation*}%
Hence the desired result follows by taking $\Pi _n(\varepsilon )$ to be
the set of permutations $\pi \in \Pi _n$ such that%
\begin{eqnarray*}
	\left\vert h_{n,\pi }^{2}-\tilde{h}_{n,\pi }^{2}\right\vert &\leq &\frac{%
		\varepsilon }{3}, \\
	\left\vert \sigma _{n,\pi }^{2}-\tilde{\sigma}_{n,\pi }^{2}\right\vert &\leq
	&\frac{\varepsilon }{3},\text{ and }\left\vert \tilde{\sigma}_{n,\pi }^{2}-%
	\tilde{h}_{n,\pi }^{2}\right\vert \leq \frac{\varepsilon }{3},
\end{eqnarray*}%
and applying Markov's inequality. $\blacksquare $\medskip \medskip

\noindent {\large 2.6. Proof of Theorem 1}\medskip

\noindent \textbf{Proof of Theorem 1: }By Lemmas B9, B18, and B14, we find
that%
\begin{equation}
\frac{\sqrt{n}\{\hat{C}(G_n)-C(G_n)\}}{\hat{\sigma}_n}\overset{d}{%
	\rightarrow }N(0,1).  \label{dec4}
\end{equation}%
We turn to the permutation test statistic. For this, we use Lemma A1. First,
we show (C1). Recall the definition:\ $h_{n,\pi }^{2}=Var\left( \zeta
_{n,\pi }|\pi \right) .$ Let $b_{1},b_{2}\in \mathbf{R}$ be such that $%
b_{1}^{2}+b_{2}^{2}=1$. Write for any fixed sequence of permutations $\pi
_{1},\pi _{2}\in \Pi _n,$%
\begin{eqnarray*}
	&&\frac{b_{1}\sqrt{n}\hat{C}_{\pi _{1}}(G_n)}{h_{n,\pi _{1}}}+\frac{b_{2}%
		\sqrt{n}\hat{C}_{\pi _{2}}(G_n)}{h_{n,\pi _{2}}} \\
	&=&\frac{b_{1}\zeta _{n,\pi _{1}}}{h_{n,\pi _{1}}}+\frac{b_{2}\zeta _{n,\pi
			_{2}}}{h_{n,\pi _{2}}}+\frac{b_1}{h_{n,\pi _{1}}\sqrt{n}}\sum_{i\in N_n}%
	\mathbf{E}\left[ q_{i,\pi _{1}}|\pi _{1}\right] \\
	&&+\frac{b_2}{h_{n,\pi _{2}}\sqrt{n}}\sum_{i\in N_n}\mathbf{E}\left[
	q_{i,\pi _{2}}|\pi _{2}\right] +o_{P}(1),
\end{eqnarray*}%
by Lemma B9(ii). For $\varepsilon >0$, we take $\Pi _{1n}(\varepsilon )$ to be
the set of $\pi \in \Pi _n$ such that
\begin{equation*}
\frac{1}{h_{n,\pi }}\left\vert \frac{1}{\sqrt{n}}\sum_{i\in N_n}\mathbf{E}%
\left[ q_{i,\pi }|\pi \right] \right\vert \leq \varepsilon .
\end{equation*}%
Then by Lemmas B13 and B5(ii), $|\Pi _{1n}(\varepsilon )|/|\Pi
_n|\rightarrow 1$ as $n\rightarrow \infty $, because $d_{av,n}/\sqrt{n} \rightarrow 0$. By Lemma B14(iii), there
exists $\tilde{\Pi}_n(\varepsilon )\subset \Pi _n\times \Pi _n$ such
that for each $(\pi _{1},\pi _{2})\in \tilde{\Pi}_n(\varepsilon )$, we
have 
\begin{equation}
\left\vert P\left\{ \frac{b_{1}\zeta _{n,\pi _{1}}}{h_{n,\pi _{1}}}+\frac{%
	b_{2}\zeta _{n,\pi _{2}}}{h_{n,\pi _{2}}}\leq t|\pi _{1},\pi _{2}\right\}
-\Phi (t)\right\vert \leq C\varepsilon +o(1),  \label{cv}
\end{equation}%
and $|\tilde{\Pi}_n(\varepsilon )|/|\Pi _n|^{2}\rightarrow 1$, as $%
n\rightarrow \infty $. Hence for each $(\pi _{1},\pi _{2})\in \tilde{\Pi}%
_n(\varepsilon )\cap \Pi _{1n}^{2}(\varepsilon )$,%
\begin{equation*}
\left\vert P\left\{ \frac{b_{1}\sqrt{n}\hat{C}_{\pi _{1}}(G_n)}{h_{n,\pi
		_{1}}}+\frac{b_{2}\sqrt{n}\hat{C}_{\pi _{2}}(G_n)}{h_{n,\pi _{2}}}\leq
t|\pi _{1},\pi _{2}\right\} -\Phi (t)\right\vert \leq C\varepsilon +o(1),
\end{equation*}%
for some $C>0.$ Hence (C1) of Lemma A1 is fulfilled.

As for (C2), from Lemma B22, for each $\varepsilon >0$, there exists $\Pi
_{2n}(\varepsilon )\subset \Pi _n$ such that$\ |\Pi _{2n}(\varepsilon
)|/|\Pi _n|\rightarrow 1$ as $n\rightarrow \infty $, and for all $\pi \in
\Pi _{2n}(\varepsilon ),$%
\begin{equation*}
\left\vert h_{n,\pi }^{2}-\hat{\sigma}_{\pi }^{2}\right\vert \leq
\varepsilon +O_{P}\left( \frac{d_{mx,n,3}^{3/2}d_{mx,n}^{1/2}}{\sqrt{n}}\right) .
\end{equation*}%
Therefore, (C2) is satisfied. Thus by Lemma A1, we obtain the desired
result. $\blacksquare $

\end{document}